\documentclass[a4paper, 11pt]{article}
\pdfoutput=1 

\usepackage{jheppub,multirow,parskip}

\newcommand{\rf}[1]{ref.~\cite{#1}}
\newcommand{\rfs}[1]{refs.~\cite{#1}}  

\newcommand{\req}[1]{eq.~\eqref{#1}}   
\newcommand{\reqs}[1]{eqs.~\eqref{#1}}

\newcommand{\rfig}[1]{figure~\ref{#1}}
\newcommand{\rfigs}[1]{figures~\ref{#1}}
\newcommand{\rtab}[1]{table~\ref{#1}}
\newcommand{\rtabs}[1]{tables~\ref{#1}}
\newcommand{\rsec}[1]{section~\ref{#1}}
\newcommand{\rsecs}[1]{sections~\ref{#1}}
\newcommand{\rapp}[1]{appendix~\ref{#1}}

\newcommand{\p}[1]{\textsc{#1}}   

\newcommand{\sla}[1]{\ifmmode%
  \setbox0=\hbox{$#1$}%
  \setbox1=\hbox to\wd0{\hss$/$\hss}\else%
  \setbox0=\hbox{#1}%
  \setbox1=\hbox to\wd0{\hss/\hss}\fi%
  #1\hskip-\wd0\box1 }

\newcommand{\sh}{\hat{s}}
\newcommand{\rsh}{\sqrt{\hat{s}}}
\newcommand{\mhiggs}{M_\mathrm{Higgs}}

\newcommand{\calO}{{\cal O}}
\newcommand{\calM}{{\cal M}}

\newcommand{\GeV}{\,\mathrm{GeV}}
 
\newcommand{\ee}[1]{\begin{equation}#1\end{equation}}
\newcommand{\ea}[1]{\begin{align}#1\end{align}}

\newcommand{\es}[1]{\begin{split}#1\end{split}}

\title{\boldmath Higgs interference effects at the one-loop level in the 1-Higgs-Singlet extension of the Standard Model}

\author[a]{Nikolas Kauer,}
\author[a,b]{Alexander Lind,}
\author[c]{Philipp Maierh\"ofer,}
\author[d]{and Weimin Song}
\affiliation[a]{Department of Physics, Royal Holloway, University of London, Egham Hill, Egham TW20 0EX, U.K.}
\affiliation[b]{Department of Physics and Astronomy, University of Sussex, Brighton BN1 9QH, U.K.}
\affiliation[c]{Physikalisches Institut, Albert-Ludwigs-Universit\"at Freiburg, 79104 Freiburg, Germany}
\affiliation[d]{Rutherford Appleton Laboratory, Harwell Oxford, Didcot OX11 0QX, U.K.}
\emailAdd{n.kauer@rhul.ac.uk}
\emailAdd{alexander.lind.2017@live.rhul.ac.uk}
\emailAdd{philipp.maierhoefer@physik.uni-freiburg.de}
\emailAdd{weimin.song@cern.ch}

\abstract{
A detailed study of Higgs interference effects at the one-loop level in the 1-Higgs-Singlet extension of the Standard Model (1HSM) is presented for the $W^+W^-$ and $t\bar{t}$ decay modes with fully leptonic $WW$ decay.  We explore interference effects for benchmark points with a heavy Higgs mass that significantly exceeds $2m_t$.  
In the $WW$ channel, the Higgs signal and the interfering continuum background are loop induced.  In the $t\bar{t}$ channel, which features a tree-level background, we also calculate the interference with the one-loop background, which, due to the appearance of the absorptive part, is found to dominate the normalisation and shape of differential Higgs cross section distributions and should therefore be considered in experimental analyses.  The commonly used geometric average $K$-factor approximation $K_\text{interference}\approx (K_\text{Higgs}K_\text{background})^{1/2}$ is not appropriate.  We calculate with massive top and bottom quarks.  Our 1HSM and SM implementation in \p{Sherpa+OpenLoops} is publicly available and can be used as parton-level integrator or event generator.
}


\preprint{FR-PHENO-2019-004}

\begin{document}
\maketitle
\flushbottom


\section{Introduction}

In 2012, the ATLAS and CMS experiments at the CERN Large Hadron 
Collider (LHC) announced the discovery of a new scalar resonance with a mass of 
approximately 125 GeV \cite{Aad:2012tfa,Chatrchyan:2012ufa}.
The discovered particle is so far consistent with the Higgs 
boson predicted by the Standard Model (SM) Higgs mechanism 
\cite{Higgs:1964ia,Higgs:1964pj,Higgs:1966ev,Englert:1964et,Guralnik:1964eu}, but many extensions to the SM preserve the minimal 
assumptions of an $SU(2)$ doublet which acquires a vacuum expectation value 
thus inducing a physical Higgs boson that couples to fermions and vector bosons in 
proportion to their mass, while also allowing for an expanded Higgs sector with 
additional, heavier Higgs-like scalar particles.

Since a SM-like Higgs boson has been discovered, a theoretically 
consistent search for an additional Higgs boson has to be based on a model 
that is beyond the SM (BSM).  The simplest extension of the Higgs sector of the 
SM introduces an additional real scalar singlet field which is neutral under the 
SM gauge groups. This 1-Higgs-Singlet extension of the SM, abbreviated as 1HSM, 
has been extensively explored in the literature \cite{Binoth:1996au,Schabinger:2005ei,Patt:2006fw,Bowen:2007ia,Barger:2007im,Barger:2008jx,Bhattacharyya:2007pb,Dawson:2009yx,Bock:2010nz,Fox:2011qc,Englert:2011yb,Englert:2011us,Batell:2011pz,Englert:2011aa,Gupta:2011gd,Dolan:2012ac,Batell:2012mj,No:2013wsa,Coimbra:2013qq,Profumo:2014opa,Logan:2014ppa,Chen:2014ask,Costa:2014qga,Falkowski:2015iwa,Martin-Lozano:2015dja,Kanemura:2015fra,Kanemura:2016lkz,Kanemura:2017gbi,Lewis:2017dme,Casas:2017jjg,Altenkamp:2018bcs}.
The remaining viable parameter space of the 1HSM after LHC Run 1 has been 
studied in \rfs{Pruna:2013bma,Robens:2015gla,Robens:2016xkb,Ilnicka:2018def}.

At the LHC, ATLAS and CMS have been conducting searches for heavier Higgs-like bosons in various di-boson channels, in particular $W^+W^-$ \cite{Chatrchyan:2013yoa,Aad:2015agg,Khachatryan:2015cwa,Aaboud:2016okv,Aaboud:2017eta,Aaboud:2017gsl,Aaboud:2017fgj}, and in various di-fermion channels, in particular $t\bar{t}$ \cite{Aaboud:2017hnm,Aaboud:2018mjh}.

\begin{figure}[tbp]
\vspace{0.cm}
\centering
\includegraphics[width=\textwidth, clip=true]{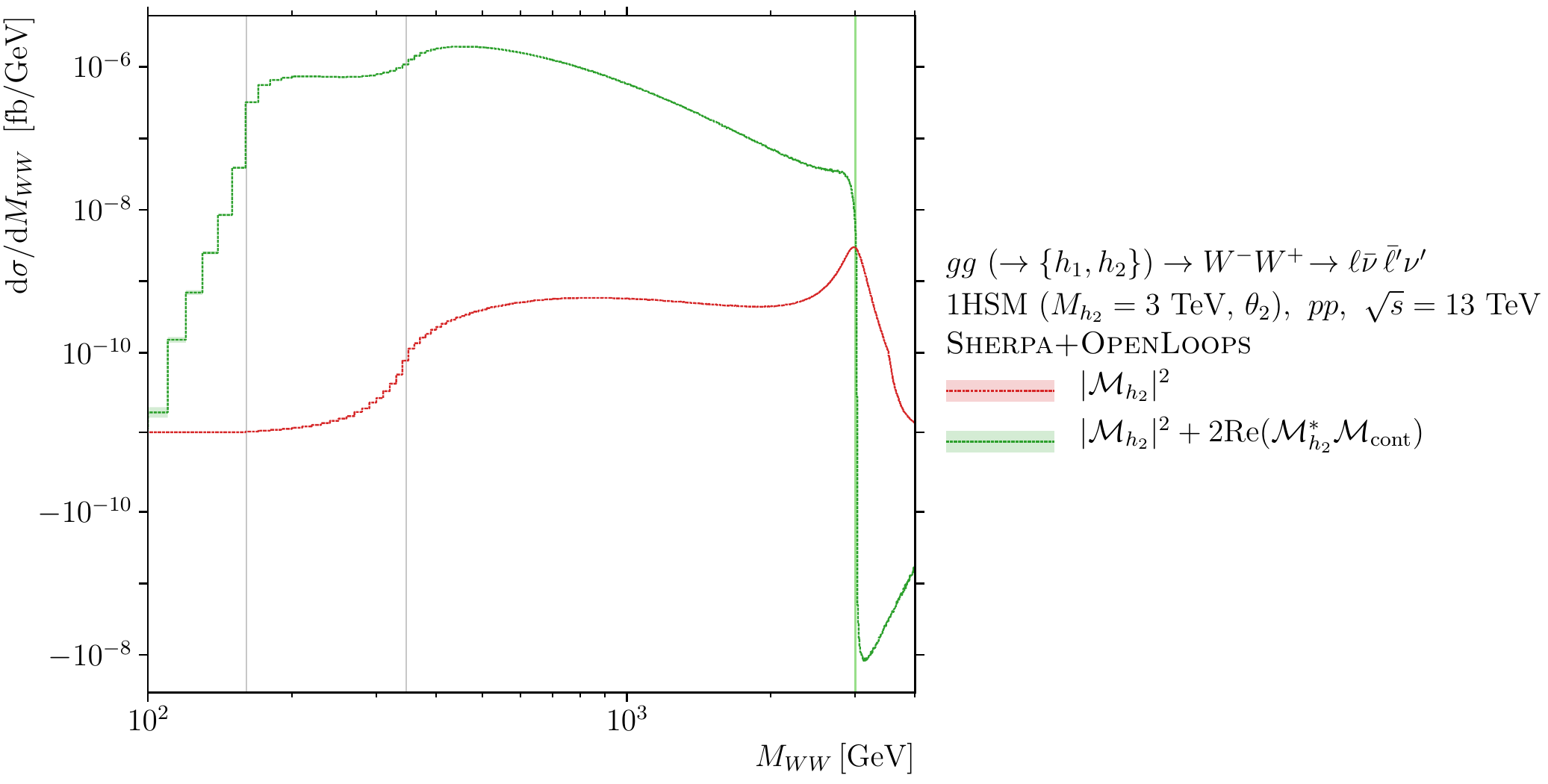}
\caption{\label{fig:lowmass_signal_ww_1hsm8_mass} $M_{WW}$ distribution of the heavy Higgs cross section and including its interference with the continuum background in the 1-Higgs-Singlet Extension of the SM (1HSM) with $M_{h_2}=3$~TeV and mixing angle $\theta_2$ (see \rtab{tab:benchmark}) for $gg\ (\to \{h_1,h_2\}) \to W^-W^+ \!\to \ell\bar{\nu}\,\bar{\ell}^\prime \nu^\prime$ in $pp$ collisions at $\sqrt{s}=13$~TeV.  For details, see \rsecs{sec:model} and \ref{sec:results}.}
\end{figure}

So far, the heavy Higgs searches are geared to establishing a significant excess (``bump'') in the invariant mass spectrum of the final state particles at the position of the heavy resonance.
However, as illustrated in \rfig{fig:lowmass_signal_ww_1hsm8_mass},
%
%
%
the ``bump'' of the heavy resonance can turn out to be a tiny correction to the heavy resonance signal when signal-background interference is taken into account.  Note also that the line shape of the resonance without and with interference has no resemblance.\footnote{%
  For $\sigma(|\calM_{h_2}|^2)$, \rfig{fig:lowmass_signal_ww_1hsm8_mass} shows a shoulder extending from below the heavy Higgs resonance down to $2m_t$.  This significant deviation from the expected Breit-Wigner shape results from the convolution with the strongly rising (for $M_{WW}\to 0$) gluon parton distribution function (PDF).  We note that the shoulder does not effectuate an enhanced experimental sensitivity to the heavy Higgs signal.  This is apparent from the results given in \rsec{sec:results} and \rapp{app:results} and can be understood qualitatively as follows.  Before convolution with the PDF, for the continuum background cross section $d\sigma_\mathrm{cont}\sim \sh^{-1}$ (up to powers of $\log \sh$).  For $\rsh \gg \mhiggs$, the same behaviour applies to $d\sigma_\mathrm{Higgs}$.  However, in the invariant mass region significantly below the Higgs resonance, one has $1/(\sh -\mhiggs^2)^2 \sim \mhiggs^{-4}$, rather than $\sh^{-2}$.  This changes the dependence to $d\sigma_\mathrm{Higgs}\sim \sh$.  In this region, for decreasing $\rsh$, $d\sigma_\mathrm{Higgs}$ decreases while $d\sigma_\mathrm{cont}$ increases.  The background hence outgrows the Higgs cross section when moving further and further below $\mhiggs$.  Convolution with the PDF does not affect this relative change.
}
We emphasise that the signal-background interference is a constituent of the BSM signal and a priori has to be treated on equal footing with the mod-squared BSM amplitude (the ``bump'').  If the BSM amplitude is absent, the interference vanishes.  It is therefore crucial to calculate and study interference effects for heavy resonance searches.  Furthermore, as demonstrated below, in general it is crucial to take one-loop corrections to tree-level amplitudes into account to obtain reliable predictions.

Here, we focus on the case where the 
additional Higgs boson is heavier than the discovered Higgs boson.  In this case,
the BSM signal is affected not only by a sizeable Higgs interference with the 
continuum background, but also by a non-negligible interference between the heavy Higgs boson and the high-mass tail of the light Higgs boson \cite{Kauer:2012hd}, which is fully taken into account in the calculations presented here.

For the $WW$ and $ZZ$ decay modes, interference effects in 2-Higgs models have been studied previously.  In gluon fusion Higgs production, the heavy Higgs-light Higgs interference was studied in the 1HSM in \rfs{Maina:2015ela,Kauer:2015hia,Englert:2015zra} and in 2-Higgs-doublet models (2HDMs) in \rf{Greiner:2015ixr}.%
\footnote{A calculation including full interference effects in a Higgs portal model has 
been carried out in \rf{Englert:2014ffa}.
For Higgs production in vector boson fusion, heavy-light interference in a 
2-Higgs model was studied in \rf{Liebler:2015aka} for an $e^+e^-$ collider
and in more detail including heavy-continuum interference in \rf{Ballestrero:2015jca} for the LHC.}

The peak-dip deformation of the Higgs resonance in $gg\to t\bar{t}$ due to signal-background interference was first studied in the SM in \rf{Dicus:1994bm}.  It was further studied for heavy scalars in 2-Higgs models for masses up to $750$~GeV in \rfs{Bernreuther:2015fts,Hespel:2016qaf}.\footnote{%
See also \rf{Bernreuther:2017yhg}.}
A detailed analytic discussion and illustrative study of the heavy Higgs line shape modification due to signal background interference in $gg\to t\bar{t}$ for scalar masses up to 1~TeV was presented in \rf{Carena:2016npr}.\footnote{%
Loop corrections to the background are not considered in \rf{Carena:2016npr}.} 
Recently, a detailed study of the experimental sensitivity to additional heavy (pseudo)-scalar resonances with mass up to 1~TeV in the singlet model, 2HDM and the hMSSM in $gg\to t\bar{t}$ at the LHC, taking into account signal-background interference effects, was presented in \rf{Djouadi:2019cbm}.  In this paper, we extend the work of \rf{Djouadi:2019cbm} by studying the Higgs signal in the $0.7$~TeV to $3$~TeV mass range for integrated cross sections and differential distributions in $M_{t\bar{t}}$ and various kinematic observables.  Interference effects between all three $gg\to t\bar{t}$ amplitude contributions -- heavy Higgs, continuum background and light Higgs -- are taken into account and illustrated individually.  Significantly, we investigate the impact of higher-order corrections on the Higgs signal by accurately taking into account its interference with the virtual corrections to the $gg\to t\bar{t}$ continuum background.\footnote{In \rf{Djouadi:2019cbm}, the interference is calculated at leading order (LO) and rescaled with the geometric average of inclusive $K$-factors for the signal and QCD background in an attempt to approximate higher-order corrections.  This approach was also used in \rf{Hespel:2016qaf} to obtain approximate next-to-leading order (NLO) results for heavy scalar ($h_2$) production in $gg\to t\bar{t}$+jet with $M_{h_2}=500$ GeV.  In \rf{Bernreuther:2015fts}, for the 2HDM approximate NLO corrections were calculated using the effective $gg(g)H$ vertices obtained in the heavy top quark limit.}  Due to a non-trivial phase, loop-level amplitude contributions can substantially change integrated cross sections and the shapes of differential cross sections \cite{Dixon:2003yb,Kauer:2015dma,Campbell:2017rke}.  Furthermore, it is well-known that ``flat'' inclusive $K$-factors often do not model differential NLO corrections well.
In 2HDMs, signal-background interference effects have also been studied in the context of heavy Higgs searches in the $tbW$ final state \cite{Haisch:2018djm}.
NLO effects in effective field theory fits to $W^+W^-$ production at the LHC have been studied in \rf{Baglio:2018bkm} and the implementation is publicly available through the \p{POWHEG-BOX}.

This paper is organised as follows:  In \rsec{sec:model} we discuss the 1HSM and specify the used benchmark points.  In \rsec{sec:calc} we review the details of our calculation and specify the used input parameters and settings. In \rsec{sec:results}, we present cross sections and distributions for the Higgs signal and its interference in the 1HSM and, for comparison, in the SM for $gg\to$ Higgs $\to WW$ and $t\bar{t}$ with fully leptonic $WW$ decay taking into account tree- and one-loop backgrounds.  In \rsec{sec:discussion}, we discuss our findings.  We conclude in \rsec{sec:sum}.


\section{Model\label{sec:model}}

As a minimal theoretically consistent model with two physical Higgs bosons, we 
consider the 1HSM, i.e.\ the SM with an added real singlet field which is neutral under all SM gauge groups.%
\footnote{Higgs singlet models with an additional $Z_2$ symmetry have generated some interest recently because of the possibility of the additional Higgs boson being a dark matter candidate, but here we consider the most general extension.}
In the following, we give a brief summary of the model.  A more detailed description can be found in \rfs{Chen:2014ask,Heinemeyer:2013tqa}.

The SM Higgs sector is extended by the addition of a new real scalar field, which is a singlet under all the gauge groups of the SM and which also gets a vacuum expectation value (VEV) under electroweak symmetry breaking. The most general gauge-invariant potential can be written as \cite{Schabinger:2005ei,Bowen:2007ia}
\ee{
V = \lambda \left( \Phi^\dagger\Phi - \frac{v^2}{2}\right)^2 + \frac{1}{2} M^2 s^2
+ \lambda_1 s^4 +\lambda_2 s^2 \left( \Phi^\dagger\Phi - \frac{v^2}{2}\right) 
+\mu_1 s^3 + \mu_2 s \left( \Phi^\dagger\Phi - \frac{v^2}{2}\right), 
\label{genpot}
\end{equation}
where $s$ is the real singlet scalar which is allowed to mix with the SM $SU(2)$ Higgs doublet, which in the unitary gauge can be written as
\begin{equation}
\Phi = \begin{pmatrix}
0 \\
(\phi + v)/\sqrt{2}
\end{pmatrix}
\label{capphi}
}
with VEV $v\simeq 246$ GeV.
Here it has already been exploited that (without the $Z_2$ symmetry) shifting the singlet field simply corresponds to a redefinition of the parameter coefficients and 
due to this freedom one can take the VEV of the singlet field to zero, which implies 
$M^2>0$. To avoid vacuum instability the quartic couplings must satisfy 
\begin{equation}
\lambda>0,\quad \lambda_1 > 0,\quad \lambda_2 > -2 \sqrt{\lambda \lambda_1}\,.
\end{equation}
The trilinear couplings $\mu_1$ and $\mu_2$ can have positive or negative sign.
Substituting \req{capphi} into \req{genpot}, one obtains the potential
\ee{
\label{potential}
V = \frac{\lambda}{4}\phi^4 + \lambda v^2 \phi^2 + \lambda v \phi^3 + \frac{1}{2}M^2s^2 + \lambda_1 s^4 + \frac{\lambda_2}{2}\phi^2 s^2 + \lambda_2 v \phi s^2 + \mu_1 s^3 + \frac{\mu_2}{2}\phi^2 s + \mu_2 v \phi s\,.
}
The mass eigenstates can be parametrised in terms of a mixing angle $\theta$ as
\ea{
h_1 &= \phi \cos \theta  - s \sin \theta \,, \label{mixingeq1}\\
h_2 &= \phi \sin \theta + s \cos \theta \,, \label{mixingeq2}
}
where $h_1$ is assumed to be the lighter Higgs boson with a mass of 125 GeV, and
\ee{
\tan 2\theta = \frac{- \mu_2 v}{\lambda v^2 - \frac{1}{2} M^2}
}
with 
\ee{
-\frac{\pi}{4} < \theta < \frac{\pi}{4}
}
under the condition $M^2>2\lambda v^2$.
The model has six independent parameters, which we choose to be $M_{h_1}, M_{h_2}, \theta, \mu_1, \lambda_1$ and $\lambda_2$.  The dependent model parameters are:
\ea{
\lambda &=\frac{\cos\left(2\theta\right)\left(M_{h_1}^2 - M_{h_2}^2\right) + M_{h_1}^2 + M_{h_2}^2 }{4v^2}\,, \label{lambdaeq}\\
M^2 &= \frac{M_{h_2}^2 - M_{h_1}^2 + \sec\left(2\theta\right) \left(M_{h_1}^2 + M_{h_2}^2\right)}{2\sec\left(2\theta\right)}\,, \\
\mu_2 &= -\tan\left(2\theta\right)\ \frac{\lambda v^2 - \frac{1}{2}M^2}{v}\,. 
}

We set $M_{h_1}$ to 125 GeV in accordance with the mass of the observed resonance
and study four values for the mass of the heavy Higgs resonance: $M_{h_2}=700$~GeV, $M_{h_2}=1$~TeV, $M_{h_2}=1.5$~TeV and $M_{h_2}=3$~TeV.
We consider the mixing angles specified in \rtab{tab:benchmark}.
\begin{table}[tbp]
\renewcommand{\arraystretch}{1.2}
\centering 
\begin{tabular}{|l|cccc|}
\hline
$M_{h_2}$ [GeV] & 700 & 1000 & 1500 & 3000 \\ 
\hline
$\theta_1$ &  $\pi/15$  &  $\pi/15$  &  $\pi/22$   &  $\pi/45$  \\
  &  $\approx 0.21$  &  $\approx 0.21$  &  $\approx 0.14$  &  $\approx 0.07$  \\
\hline
$\theta_2$ &  $\pi/8$  &  $\pi/8$  &  $\pi/12$  &  $\pi/24$  \\ 
  &  $\approx 0.39$  &  $\approx 0.39$  &  $\approx 0.26$  &  $\approx 0.13$  \\
\hline
\end{tabular}
\caption{\label{tab:benchmark}Mixing angles $\theta_1$ and $\theta_2$ are given for all considered benchmark points, which are also characterised by $M_{h_1}=125$~GeV and $\mu_{1} = \lambda_{1} = \lambda_{2} = 0$.}
\end{table}
The lower values are consistent with current experimental limits \cite{Robens:2016xkb,Ilnicka:2018def}.%
\footnote{The perturbativity condition $\lambda < 4\pi$ imposes via \req{lambdaeq} the constraint $|\theta| < \theta_0$, which is satisfied for our benchmark points, since $\theta_0\ge 0.42$ for $M_{h_1}=125$~GeV and $200$~GeV~$\lesssim M_{h_2}\le 3$~TeV.  We do not quantitatively consider the RG running of $\lambda$ to high scales, but note that the chosen mixing angles decrease with increasing $M_{h_2}\ge 1$~TeV.  Our $\theta$ values are compatible with the perturbativity constraints shown in refs.~\cite{Robens:2016xkb,Ilnicka:2018def}.}
For each value of $M_{h_2}$, two angles, $\theta_1$ and $\theta_2$, have been chosen to illustrate how the studied interference effects change with the mixing angle. 
Furthermore, we consider model benchmark points with vanishing coupling parameters $\mu_1, \lambda_1$ and $\lambda_2$. ($\lambda_1>0$ is treated as approximately zero.)  We emphasise that this does not imply that the $h_2\to h_1 h_1$ and $h_2\to h_1 h_1 h_1$ decay widths are zero (if kinematically allowed).  This is a consequence of the $\phi$\,-$s$ mixing.  Inverting \reqs{mixingeq1} and \eqref{mixingeq2}, one finds
\ea{
\phi  &= h_1 \cos \theta  + h_2 \sin \theta \,, \label{mixingeqinv1}\\
s  &= - h_1 \sin \theta + h_2 \cos \theta \,.  \label{mixingeqinv2}
}
Substituting \reqs{mixingeqinv1} and \eqref{mixingeqinv2} into the potential $V$, see \req{potential}, more specifically into
\ee{
\frac{\lambda}{4}\phi^4,\ \ \lambda v \phi^3,\ \ \frac{\mu_2}{2}\phi^2 s\,,
}
gives nonvanishing $h_2h_1h_1h_1$ and $h_2h_1h_1$ interactions, since $\lambda$ and $\mu_2$ are non-zero.%
\footnote{The corresponding Feynman rules are given in eqs.\ (331) and (334) in \rf{Heinemeyer:2013tqa}.}  The numerical values of $\Gamma(h_2\to h_1 h_1)$, $\Gamma(h_2\to h_1 h_1 h_1)$ and $\Gamma(h_2\to h_1 h_1 h_1 h_1)$ for our benchmark points are given in \rapp{app:partwidths}.


\section{Calculational details \label{sec:calc}}

We carry out calculations in the 1HSM (signal hypothesis) and the SM (null hypothesis).  As input parameters, we adopt the recommendation of the LHC Higgs Cross Section Working Group in chapter I.1 of \rf{deFlorian:2016spz} with $M_V^{pole}$ and $\Gamma_V^{pole}$ as given by eq.\ (I.1.7).\footnote{For $m_t$, we use the world average of $m_t^\text{OS}$.}  We employ the $G_\mu$ scheme, where 
\ee{
\label{gmuscheme}
\cos\theta_W = \frac{M_W}{M_Z}\,,\ \ \ 
\alpha = \frac{\sqrt{2}}{\pi} G_F M_W^2 \sin^2\theta_W\,.
}
More specifically, we use $G_F = 1.1663787\times 10^{-5}$ GeV$^{-2}$, $M_W = 80.35797$ GeV, $M_Z = 91.15348$ GeV, $\Gamma_W  = 2.08430$ GeV, $\Gamma_Z  = 2.49427$ GeV, $m_t = 173.2$ GeV, $\Gamma_t = 1.369$ GeV and $m_b =  4.92$ GeV. Via \req{gmuscheme}, we have $1/\alpha\approx 132.36$ and $\sin^2\theta_W \approx 0.222838$.

The PDF set \texttt{PDF4LHC15\_nlo\_mc} \cite{Butterworth:2015oua} with default $\alpha_s$ is used, and the CKM matrix is approximated by the identity matrix.%
\footnote{In this context, the associated error is smaller than 0.01\% \cite{Kauer:2012ma}.}
 The renormalisation and factorisation scales are set to $M_{WW}/2$ for $WW$ production and $M_{t\bar{t}}/2$ for $t\bar{t}$ production.  The $pp$ collision energy is $\sqrt{s}=13$ TeV.  Finite top and bottom quark mass effects are fully taken into account.  Lepton masses are neglected.
As unstable particle states arise in the considered processes, the prescription of the complex-mass scheme \cite{Denner:2006ic,Denner:1999gp} is applied to all scattering amplitudes.

SM Higgs widths have been calculated using \p{HDECAY} \cite{Djouadi:1997yw,Djouadi:2018xqq} and \p{Prophecy4f} \cite{Bredenstein:2006rh,Bredenstein:2006nk,Bredenstein:2006ha}.  For the SM Higgs with $M_H = 125$ GeV, one obtains $\Gamma_H=4.087\times 10^{-3}$ GeV.
The Higgs boson widths in the 1HSM are calculated as follows:
\ea{
\Gamma_{h_1} &= \cos^2\theta\ \Gamma_{H}(M_{h_1}), \\
\Gamma_{h_2} &= \sin^2\theta\ \Gamma_{H}(M_{h_2}) + \Gamma(h_2\to n \times h_1), \label{nh1eq}
}
where $\Gamma_{H}(M)$ denotes the width of a SM Higgs boson with mass $M$.%
\footnote{For $M=3$~TeV, $\Gamma_H(M)$ cannot be obtained using \p{HDECAY} due to numerical problems for decay modes with $b$ quark loops caused by the tiny value of the running $b$ quark mass.  For $M=3$~TeV, we therefore approximate $\Gamma_H(M)\approx \Gamma(H\to WW) + \Gamma(H\to ZZ)$.}  We take into account decay modes, where $h_2$ decays into up to four $h_1$ bosons, i.e.\ $2\le n\le 4$ in \req{nh1eq}.  
A custom implementation of the 1HSM in \p{FeynRules} \cite{Alloul:2013bka,Christensen:2008py} and the \p{UFO} \cite{Degrande:2011ua} interface with \p{MadGraph5\_aMC@NLO} \cite{Alwall:2014hca} was used to calculate $\Gamma(h_2 \to h_1 h_1)$, $\Gamma(h_2 \to h_1 h_1 h_1)$ and $\Gamma(h_2 \to h_1 h_1 h_1 h_1)$.  The resulting partial decay widths are given in \rtab{tab:partwidths} in \rapp{app:partwidths}.
The results displayed in \rtab{tab:partwidths} demonstrate that the $h_2$ width contributions from higher $h_1$ multiplicities are suppressed for all considered benchmark points.  
The resulting values for $\Gamma_{h_1}$ and $\Gamma_{h_2}$ and the corresponding $\Gamma/M$ ratios are given in \rtab{tab:widths}.
\begin{table}[tbp]
\renewcommand{\arraystretch}{1.2}
\centering
\small
\begin{tabular}{cc|cccc|}
\hline
\multicolumn{1}{|c|}{$\theta$} & $M_{h_2}$ [GeV] & $700$ & $1000$ & $1500$ & $3000$ \\ 
\hline
\multicolumn{1}{|c|}{\multirow{4}{*}{$\theta_1$}} & $\Gamma_{h_1}$ [GeV] &  $3.910(5) \times 10^{-3}$  &  $3.910(5) \times 10^{-3}$  &  $4.004(5)\times 10^{-3}$  &  $4.067(5)\times 10^{-3}$  \\
\multicolumn{1}{|c|}{} & $\Gamma_{h_1}/M_{h_1}$ &  $3.1283(4) \times 10^{-5}$  & $3.1283(4) \times 10^{-5}$  &  $3.2034(4) \times 10^{-5}$  &  $3.2537(4) \times 10^{-5}$  \\ 
\cline{2-6}
\multicolumn{1}{|c|}{} & $\Gamma_{h_2}$ [GeV] &  10.780(3)  &  34.295(3)  &  79.52(2)  &  86.70(3)  \\
\multicolumn{1}{|c|}{} & $\Gamma_{h_2}/M_{h_2}$ &  0.015400(4) & 0.034295(3) &  0.053013(7)  &  0.028902(9)  \\ 
\hline
\multicolumn{1}{|c|}{\multirow{4}{*}{$\theta_2$}} & $\Gamma_{h_1}$ [GeV] &  $3.488(5)\times 10^{-3}$  &  $3.488(5)\times 10^{-3}$  & $3.813(5)\times 10^{-3}$   &  $4.017(5)\times 10^{-3}$  \\
\multicolumn{1}{|c|}{}  &  $\Gamma_{h_1}/M_{h_1}$ &  $2.7908(4) \times 10^{-5}$  &  $2.7908(4) \times 10^{-5}$  &  $3.0506(4) \times 10^{-5}$  &  $3.2139(4) \times 10^{-5}$  \\
\cline{2-6}
\multicolumn{1}{|c|}{} & $\Gamma_{h_2}$ [GeV] &  33.903(8)  &  116.37(4)  &  273.6(2)  &  322.5(2)  \\
\multicolumn{1}{|c|}{}  &  $\Gamma_{h_2}/M_{h_2}$ &  0.04843(2)  &  0.11637(4)  &  0.18240(8)   &  0.10751(5)  \\ \hline
\end{tabular}
\caption{\label{tab:widths}Decay widths and $\Gamma/M$ ratios of the two physical Higgs bosons $h_{1}$ and $h_{2}$ in the $1$-Higgs-Singlet extension of the SM for the considered benchmark points.  Details as in \rtab{tab:benchmark}.  The error due to rounding and numerical integration is given in brackets.}
\end{table}

We study Higgs boson production in gluon fusion at the LHC for the $WW$ and $t\bar{t}$ decay modes with subsequent fully-leptonic $W$ boson decays in the 1HSM:
\ea{
&g g\ (\to \{h_1,h_2\}) \to W^-W^+ \!\to \ell\bar{\nu}\,\bar{\ell}^\prime \nu^\prime,\\
&g g\ (\to \{h_1,h_2\}) \to t\bar{t} \to b\bar{b}\,\ell\bar{\nu}\,\bar{\ell}^\prime \nu^\prime. \label{ttprocess}
}
The results presented in \rsec{sec:results} have been calculated at LO unless otherwise noted and are given for a single combination of different lepton flavours, for instance $\ell=e^-$, $\ell^\prime=\mu^-$.

Representative Feynman graphs for the light 
and heavy Higgs and interfering continuum background processes in the 1HSM are shown in \rfigs{fig:WWdiagrams} and \ref{fig:ttdiagrams}.
\begin{figure}[tbp]
\centering
\includegraphics[height=4cm, clip=true]{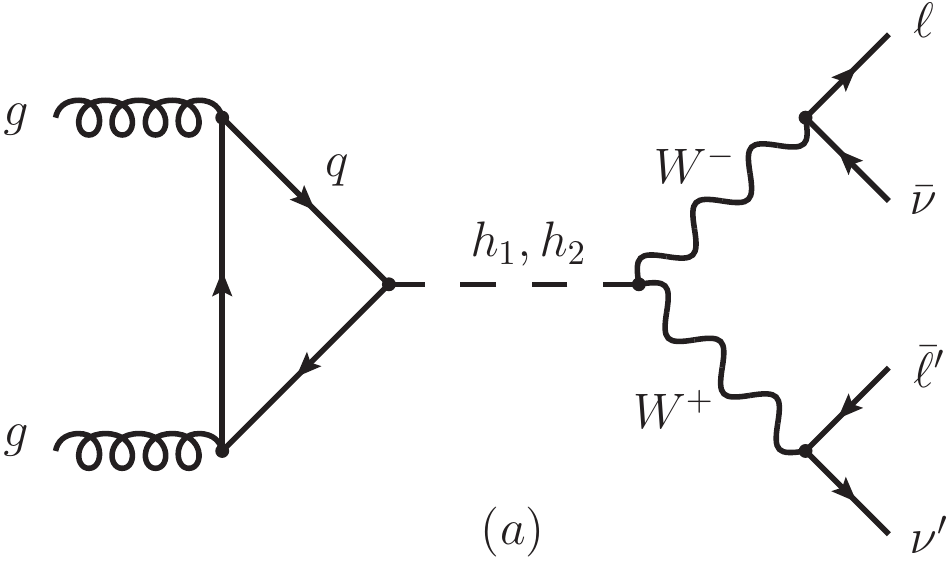}\hfil
\includegraphics[height=4cm, clip=true]{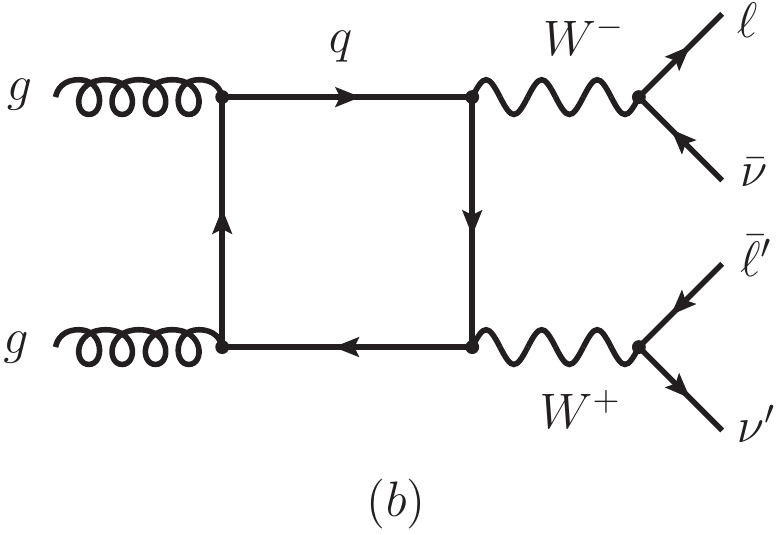}
\caption{\label{fig:WWdiagrams}Representative Feynman graphs for $g g\ (\to \{h_1,h_2\}) \to WW \to 4$ leptons in the SM extended with a real scalar singlet field. The light ($h_1$) and heavy Higgs ($h_2$) production graphs (a) interfere with each other and the gluon-induced continuum background graphs (b).}
\end{figure}
\begin{figure}[tbp]
\centering
\includegraphics[height=6cm, clip=true]{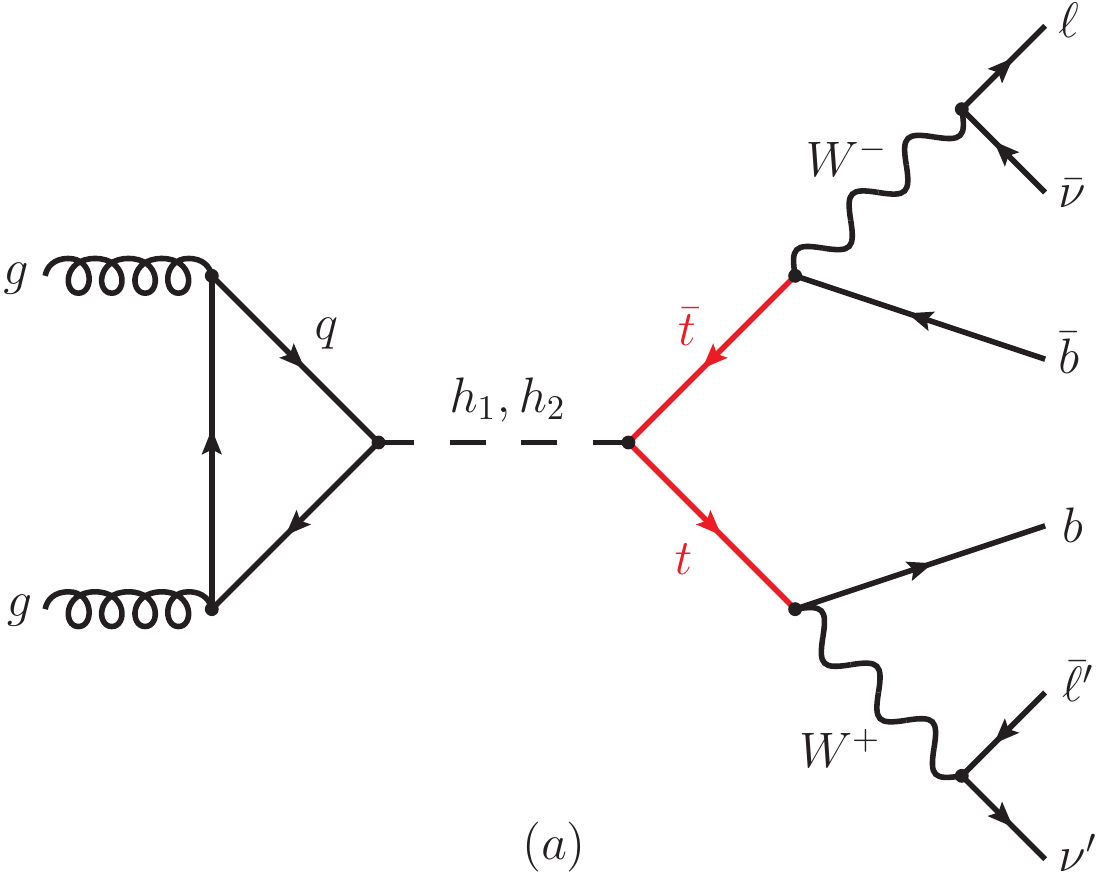}\hfil
\includegraphics[height=6cm, clip=true]{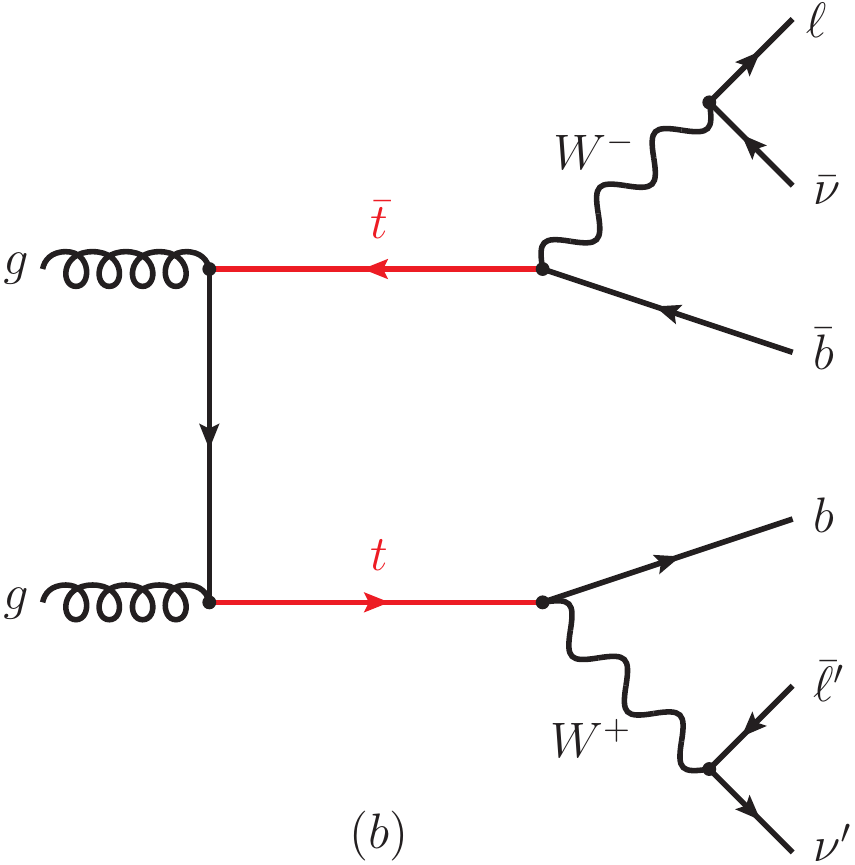}\\
\includegraphics[height=6cm, clip=true]{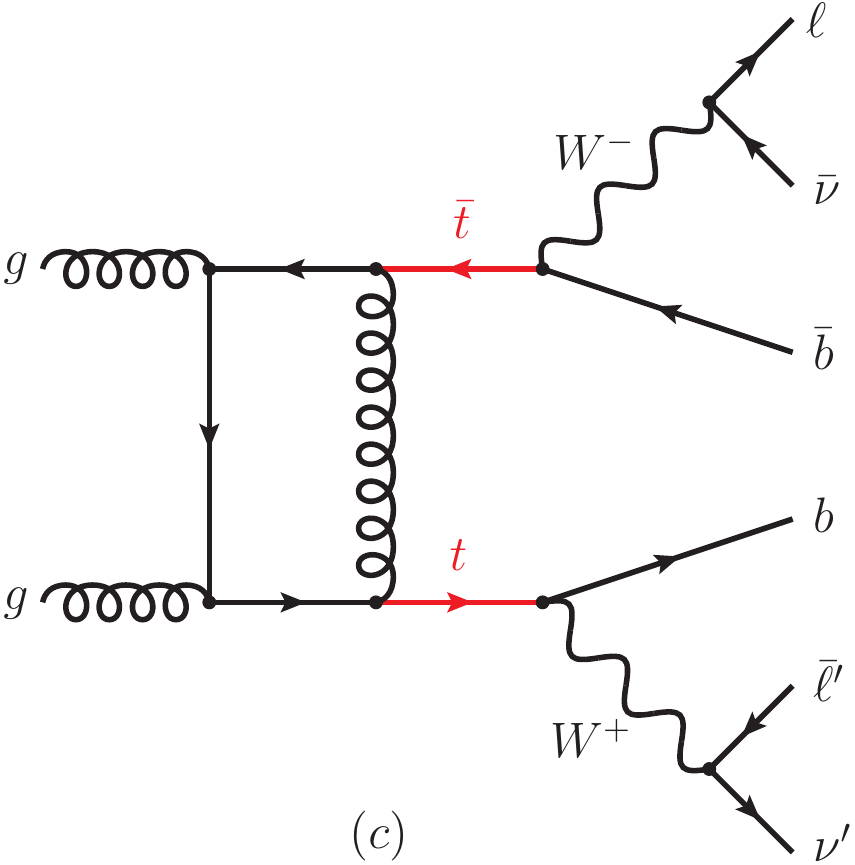}\hfil
\includegraphics[height=6cm, clip=true]{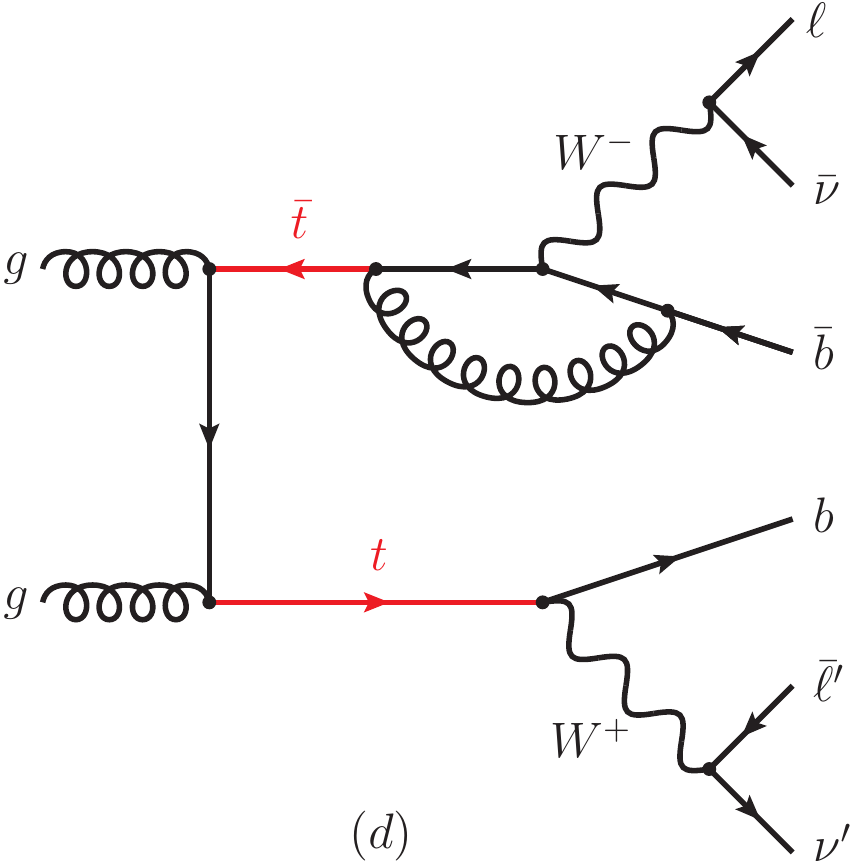}
\caption{\label{fig:ttdiagrams}Representative Feynman graphs for $g g\ (\to \{h_1,h_2\}) \to t\bar{t} \to b\bar{b} + 4$ leptons in the SM extended with a real scalar singlet field. The light ($h_1$) and heavy Higgs ($h_2$) production graphs (a) interfere with each other and the gluon-induced LO continuum background graphs (b).  Interference with the gluon-induced continuum background at the one-loop level, for which representative graphs are shown in (c) and (d), is also considered.  The process is calculated in double pole approximation with a pair of on-shell top quarks (red).}
\end{figure}
All considered amplitudes are at the one-loop level, except for the $gg\to t\bar{t}$ continuum background at LO.
For the $t\bar{t}$ process shown in \rfig{fig:ttdiagrams}, we note that continuum background graphs with $s$-channel gluon propagator do not interfere with the Higgs graphs, which have a colour singlet initial and final state.  For the depicted continuum background graphs, only the colour singlet configurations -- occurring with probability $1/(N_c^2-1)$ -- contribute to the signal-background interference.
The corresponding SM graphs are obtained by substituting $h_1$ with $H$ and discarding $h_2$ contributions.  The amplitudes are calculated using customised \p{OpenLoops} \cite{Cascioli:2011va,Denner:2016kdg} code, which is interfaced to the \p{Sherpa} Monte Carlo (MC) event generator \cite{Gleisberg:2008ta,Krauss:2001iv} and \p{LHAPDF} \cite{Buckley:2014ana}.  Since \p{Sherpa}-2.2.5 does not automatically generate phase space integrators for loop-induced processes, a customised approach is used.  Full spin correlations are taken into account for all considered processes.  
For the top pair process, \req{ttprocess}, since our study focuses on the region with $2m_t < M_{t\bar{t}} < M_{h_2}$, the double pole approximation \cite{Denner:2000bj} with a pair of on-shell top quark states --- shown in red in \rfig{fig:ttdiagrams} --- is applied to simplify our calculations.
It has been shown that higher-order corrections to interference can be larger than the interference at LO \cite{Dixon:2003yb,Kauer:2015dma}.  In the case where LO involves tree-level amplitudes, this can be understood as follows: the relative phase that induces large interference arises primarily through the absorptive part of loop graphs.  We therefore also calculate the interference between the LO Higgs amplitude --- $H$ in the SM and $h_1$ as well as $h_2$ in the 1HSM, see figure \rfig{fig:ttdiagrams}(a) --- and the interfering continuum background amplitude at the one-loop level, see \rfigs{fig:ttdiagrams}(c) and \ref{fig:ttdiagrams}(d).%
\footnote{
The NLO contribution from interference of 2-loop virtual corrections to the loop-induced Higgs amplitude with the tree-level continuum background amplitude is not taken into account. We note that this contribution includes multiscale 2-loop diagrams of the non-factorisable type shown in the centre of figure 9 in \rf{Hespel:2016qaf}, for which results are not yet available.  We believe this tree-2-loop contribution is small compared to the 1-loop-1-loop contribution we compute, because tree-1-loop interference was found to be small compared to 1-loop-1-loop interference in similar processes \cite{Campbell:2014gua,Kauer:2015dma}, but the non-factorisable contribution may be enhanced due to the lifted colour singlet final state restriction.
}
Since the top quark states are treated in narrow-width approximation (NWA), factorising production and decay, nonfactorisable corrections are neglected.%
\footnote{In the inclusive case, nonfactorisable corrections are suppressed by $\Gamma_t/m_t$, i.e.\ $\sim 1\%$ \cite{Fadin:1993kt}.}

The one-loop continuum background amplitudes are affected by ultraviolet (UV) and infrared (IR) singularities, which are treated with conventional dimensional regularisation.  \p{OpenLoops} uses the on-shell scheme to renormalise all masses.  For all sufficiently inclusive transition probabilities (``IR-safe'' observables), the IR poles cancel when the virtual corrections, represented by \rfigs{fig:ttdiagrams}(c) and \ref{fig:ttdiagrams}(d), are combined with the real emission corrections \cite{Kinoshita:1962ur,Lee:1964is} and the collinear counterterms, which, taken together, constitute the full next-to-leading order (NLO) corrections to the continuum background subprocess in \req{ttprocess}.\footnote{A description of the structure of NLO calculations can be found in \rf{Binoth:2010xt}.  In \p{OpenLoops}, the coefficient defined in eq.\ (2.6) of \rf{Binoth:2010xt} is chosen according to eq.\ (2.7) therein.}  In our calculations, we do not take into account the real emission corrections to the LO continuum background amplitude, i.e.\ to \rfig{fig:ttdiagrams}(b), because they do not interfere with the LO signal amplitude, see \rfig{fig:ttdiagrams}(a).  We note that they would have to be included in a full NLO calculation of the signal-background interference, together with the real emission corrections to the LO signal amplitude.%
\footnote{At full NLO, also $gq$ and $q\bar{q}$ subprocesses \cite{Grazzini:2018owa}, which are quark-PDF suppressed at the LHC, formally contribute to the signal-background and $h_1$-$h_2$ interference, as illustrated for $0\to gq\bar{q}ZZ$ in figure 2 of \rf{Campbell:2014gua}.}  
A full NLO calculation of the signal-background interference is beyond the scope of this work.

We note that our 1HSM and SM implementation in \p{Sherpa+OpenLoops} is included in the \href{http://arxiv.org/}{arXiv} submission as ancillary file \path{sherpa_openloops_code.tar.bz2}.


\section{Results\label{sec:results}}

To take into account the fiducial selection at the LHC, we employ a simplified version of the experimental leptonic cuts used in \rf{Aaboud:2018jqu} and standard jet selection criteria \cite{Sirunyan:2019twz}.  More precisely, we apply:\footnote{The $b$ and $\bar{b}$ quark in the final state are not jet-clustered in our LO study.}
\ea{
\label{cuts}
\es{
&p_{T\ell_1} > 22\GeV,\ \ p_{T\ell_2} > 15\GeV,\ \ |\eta_\ell|<2.5,\ \ M_{\ell\bar{\ell}^\prime} > 10\GeV,\ \ \sla{p}_T > 20\GeV,\\
&p_{Tj} > 30\GeV,\ \ |\eta_j|<4.7,\ \ \Delta R_{j\ell} > 0.4\,.
}
}

Integrated results for the SM and all considered 1HSM benchmark points (see \rtab{tab:benchmark}) are shown in \rtabs{tab:res_ww}--\ref{tab:res_tt_h2}.  Mod-squared amplitude contributions are specified using the abbreviations defined in \rtab{tab:abbrev}.  For reference, a nonredundant complete set of integrated results is given in \rapp{app:results}.

To explore the differential dependence, various distributions have been calculated.  In addition to the Higgs invariant mass distribution ($M_{WW}$, $M_{t\bar{t}}$), we have also studied the transverse mass distribution of the $WW$ system ($M_{T,WW}$), the dilepton invariant mass ($M_{\ell\bar{\ell}^\prime}$) and the angular observables $\Delta\eta_{\ell\bar{\ell}^\prime}$, $\Delta\phi_{\ell\bar{\ell}^\prime}$ and $\Delta R_{\ell\bar{\ell}^\prime}$.\footnote{%
$M_{T,WW}$ is defined as in eq.~(3.6) in \rf{Kauer:2012hd}.}

Differential cross section distributions in the 1HSM for $WW$ production and the benchmark point with $M_{h_2}=1500$~GeV and mixing angle $\theta_1$ are displayed in \rfigs{fig:ww_1hsm3_mass}--\ref{fig:ww_1hsm3_deltaRll} and for $t\bar{t}$ production and the benchmark point with $M_{h_2}=700$~GeV and mixing angle $\theta_1$ in \rfigs{fig:tt_1hsm1_mass}--\ref{fig:tt_1hsm1_deltaRll}.  For reference, differential distributions in the SM are displayed in \rapp{app:sm_figs}.

For $WW$ and $t\bar{t}$ production, invariant mass distributions of the relative deviation $\delta=R-1$ of the Higgs cross section including its interference with the background in the 1HSM with $M_{h_2}=\{700,1000,1500,3000\}$~GeV and mixing angles $\theta_1$ and $\theta_2$ compared to the SM are shown in \rfigs{fig:1hsm_ww_si_h12-1256-sm_ww_si_-delta}--\ref{fig:1hsm_tt_si_h12bbv-3478-sm_tt_si_bbv-delta}.  More specifically, $R$ is the ratio of $\sigma(\text{$h_{1+2}$+I(C)})$ to $\sigma(\text{$H$+I(C)})$ or for $t\bar{t}$ production also $\sigma(\text{$h_{1+2}$+I(C$_{+\circlearrowleft}$)})$ to $\sigma(\text{$H$+I(C$_{+\circlearrowleft}$)})$, i.e.\ including the virtual corrections to the continuum background.

Furthermore, $M_{t\bar{t}}$, $M_{\ell\bar{\ell}^\prime}$, $\Delta\eta_{\ell\bar{\ell}^\prime}$ and $\Delta\phi_{\ell\bar{\ell}^\prime}$ distributions of the relative deviation $\delta=R-1$ of the Higgs interference with the background without and with the virtual corrections ($\mathcal{M}_\mathrm{cont,loop}$) in the SM and 1HSM with $M_{h_2}=\{700,1000\}$~GeV and mixing angles $\{\theta_1,\theta_2\}$ for $gg\ (\to \{h_1,h_2,H\}) \to t\bar{t} \to b\bar{b}\,\ell\bar{\nu}\,\bar{\ell}^\prime \nu^\prime$ are displayed in \rfigs{fig:tt_1hsm1256_mass-i_Hh12bv-div-i_Hh12b}--\ref{fig:tt_1hsm1256_deltaPhill-i_Hh12bv-div-i_Hh12b}.  Here, $R$ is the ratio of $\sigma(\text{I($H$,C$_{+\circlearrowleft}$)})$ to $\sigma(\text{I($H$,C)})$ and $\sigma(\text{I($h_{1+2}$,C$_{+\circlearrowleft}$)})$ to $\sigma(\text{I($h_{1+2}$,C)})$ in the SM and 1HSM, respectively.

Supplementary figures with distributions for all studied quantities, models and benchmark points are available at this URL:\\
\url{http://users.hepforge.org/~nkauer/arXiv/plots_08May2019.pdf} \cite{interference_1hsm_plots}.

\begin{table}[tbp]
\vspace{0.cm}
\centering
\small
\renewcommand{\arraystretch}{1.2}
\begin{tabular}{|l|l|}
\hline
Label & $|\mathcal{M}|^2$ \\
\hline
Sq(C) & $|\mathcal{M}_\mathrm{cont}|^2$ \\
Sq($H$) & $|\mathcal{M}_H|^2$ \\
I($H$,C) & $2\mathrm{Re}(\mathcal{M}^\ast_H\mathcal{M}_\mathrm{cont})$ \\
$H$+I(C) & $|\mathcal{M}_H|^2+2\mathrm{Re}(\mathcal{M}^\ast_{{H}}\mathcal{M}_\mathrm{cont})$ \\
I($H$,C$_\circlearrowleft$) & $2\mathrm{Re}(\mathcal{M}^\ast_{{H}}\mathcal{M}_\mathrm{cont,loop})$ \\
I($H$,C$_{+\circlearrowleft}$) & $2\mathrm{Re}(\mathcal{M}^\ast_{{H}}(\mathcal{M}_\mathrm{cont}+\mathcal{M}_\mathrm{cont,loop}))$ \\
$H$+I(C$_{+\circlearrowleft}$) & $|\mathcal{M}_H|^2+2\mathrm{Re}(\mathcal{M}^\ast_{{H}}(\mathcal{M}_\mathrm{cont}+\mathcal{M}_\mathrm{cont,loop}))$ \\
Sq($h_1$) & $|\mathcal{M}_{h_1}|^2$ \\
Sq($h_2$) & $|\mathcal{M}_{h_2}|^2$ \\
I($h_1$,$h_2$) & $2\mathrm{Re}(\mathcal{M}^\ast_{h_1}\mathcal{M}_{h_2})$ \\
$h_2$+I($h_1$) & $|\mathcal{M}_{h_2}|^2+2\mathrm{Re}(\mathcal{M}^\ast_{h_1}\mathcal{M}_{h_2})$ \\
I($h_1$,C) & $2\mathrm{Re}(\mathcal{M}^\ast_{h_1}\mathcal{M}_\mathrm{cont})$ \\
I($h_2$,C) & $2\mathrm{Re}(\mathcal{M}^\ast_{h_2}\mathcal{M}_\mathrm{cont})$ \\
Sq($h_{1+2}$) & $|\mathcal{M}_{h_1}+\mathcal{M}_{h_2}|^2$ \\
I($h_{1+2}$,C) & $2\mathrm{Re}((\mathcal{M}^\ast_{h_1}+\mathcal{M}^\ast_{h_2})\mathcal{M}_\mathrm{cont})$ \\
$h_{1+2}$+I(C) & $|\mathcal{M}_{h_1}+\mathcal{M}_{h_2}|^2+2\mathrm{Re}((\mathcal{M}^\ast_{h_1}+\mathcal{M}^\ast_{h_2})\mathcal{M}_\mathrm{cont})$ \\
I($h_2$,C+$h_1$) & $2\mathrm{Re}(\mathcal{M}^\ast_{h_2}(\mathcal{M}_\mathrm{cont}+\mathcal{M}_\mathrm{h_1}))$ \\
$h_2$+I(C+$h_1$) & $|\mathcal{M}_{h_2}|^2+2\mathrm{Re}(\mathcal{M}^\ast_{h_2}(\mathcal{M}_\mathrm{cont}+\mathcal{M}_\mathrm{h_1}))$ \\
I($h_1$,C+$h_2$) & $2\mathrm{Re}(\mathcal{M}^\ast_{h_1}(\mathcal{M}_\mathrm{cont}+\mathcal{M}_\mathrm{h_2}))$ \\
I($h_1$,C$_\circlearrowleft$) & $2\mathrm{Re}(\mathcal{M}^\ast_{h_1}\mathcal{M}_\mathrm{cont,loop})$ \\
I($h_2$,C$_\circlearrowleft$) & $2\mathrm{Re}(\mathcal{M}^\ast_{h_2}\mathcal{M}_\mathrm{cont,loop})$ \\
$h_{1+2}$+I(C$_{+\circlearrowleft}$) & $|\mathcal{M}_{h_1}+\mathcal{M}_{h_2}|^2+2\mathrm{Re}((\mathcal{M}^\ast_{h_1}+\mathcal{M}^\ast_{h_2})(\mathcal{M}_\mathrm{cont}+\mathcal{M}_\mathrm{cont,loop}))$ \\
I($h_2$,C$_{+\circlearrowleft}$+$h_1$) & $2\mathrm{Re}(\mathcal{M}^\ast_{h_2}(\mathcal{M}_\mathrm{cont}+\mathcal{M}_\mathrm{cont,loop}+\mathcal{M}_\mathrm{h_1}))$ \\
$h_2$+I(C$_{+\circlearrowleft}$+$h_1$) & $|\mathcal{M}_{h_2}|^2+2\mathrm{Re}(\mathcal{M}^\ast_{h_2}(\mathcal{M}_\mathrm{cont}+\mathcal{M}_\mathrm{cont,loop}+\mathcal{M}_\mathrm{h_1}))$ \\
I($h_1$,C$_{+\circlearrowleft}$+$h_2$) & $2\mathrm{Re}(\mathcal{M}^\ast_{h_1}(\mathcal{M}_\mathrm{cont}+\mathcal{M}_\mathrm{cont,loop}+\mathcal{M}_\mathrm{h_2}))$ \\
\hline  
\end{tabular}\\[0cm] 
\caption{\label{tab:abbrev}
Abbreviations used in tables with integrated cross sections and the corresponding mod-squared amplitude expressions.}
\end{table}
%
%
%
\begin{table}[tbp]
\vspace{0.cm}
\centering
\renewcommand{\arraystretch}{1.2}
\begin{tabular}{|c|c|c|c|}
\cline{2-4}
\multicolumn{1}{c|}{} & \multicolumn{3}{|c|}{$gg\ (\to \{H,h_1,h_2\}) \to W^-W^+ \!\to \ell\bar{\nu}\,\bar{\ell}^\prime \nu^\prime$} \\ 
\multicolumn{1}{c|}{} &   \multicolumn{3}{|c|}{$\sigma$ [fb], $pp$, $\sqrt{s}=13$ TeV} \\ 
\multicolumn{1}{c|}{} &   \multicolumn{3}{|c|}{SM and 1HSM (see \rtabs{tab:benchmark} and \ref{tab:abbrev})} \\
\cline{2-4}\noalign{\vspace{0.075cm}}\cline{2-4}
\multicolumn{1}{c|}{} & \multirow{2}{*}{SM} & \multicolumn{2}{|c|}{$|\mathcal{M}|^2$} \\
\cline{3-4}
\multicolumn{1}{c|}{} &  & Sq($H$) & $H$+I(C) \\
\cline{2-4}
\multicolumn{1}{c|}{} & $\sigma$ & $13.689(4)$ & $10.420(5)$ \\
\multicolumn{1}{c|}{} & ratio & $1$ & $0.7612(5)$ \\
\cline{2-4}\noalign{\vspace{0.075cm}}\cline{2-4}
\multicolumn{1}{c|}{} & 1HSM & \multicolumn{2}{|c|}{$|\mathcal{M}|^2$} \\
\cline{2-4}
\multicolumn{1}{c|}{} & $M_{h_2}$ [GeV]  & Sq($h_{1+2}$) & $h_{1+2}$+I(C) \\ \hline
\multirow{8}{*}{$\theta_1$}
 & 700 & $13.090(4)$ & $10.012(5)$ \\
 & ratio & $1$ & $0.7649(5)$ \\ \cline{2-4}
 & 1000 & $13.032(4)$ & $9.934(5)$ \\
 & ratio & $1$ & $0.7623(5)$ \\ \cline{2-4}
 & 1500 & $13.387(4)$ & $10.189(5)$ \\
 & ratio & $1$ & $0.7611(5)$ \\ \cline{2-4}
 & 3000 & $13.619(4)$ & $10.368(5)$ \\
 & ratio & $1$ & $0.7613(5)$ \\
\hline
\multirow{8}{*}{$\theta_2$}
 & 700 & $11.715(4)$ & $9.095(5)$ \\
 & ratio & $1$ & $0.7763(5)$ \\ \cline{2-4}
 & 1000 & $11.503(4)$ & $8.813(5)$ \\
 & ratio & $1$ & $0.7662(5)$ \\ \cline{2-4}
 & 1500 & $12.681(4)$ & $9.651(4)$ \\
 & ratio & $1$ & $0.7611(4)$ \\ \cline{2-4}
 & 3000 & $13.435(4)$ & $10.215(5)$ \\
 & ratio & $1$ & $0.7603(5)$ \\
\hline
\end{tabular}\\[0cm] 
\caption{\label{tab:res_ww}
Cross sections for $gg\ (\to \{H,h_1,h_2\}) \to W^-W^+ \!\to \ell\bar{\nu}\,\bar{\ell}^\prime \nu^\prime$ in $pp$ collisions at $\sqrt{s}=13$~TeV in the Standard Model with $M_H = 125$~GeV and its 1-Higgs-Singlet Extension with $M_{h_1} = 125$~GeV, $M_{h_2} = 700, 1000, 1500, 3000$~GeV and mixing angles $\theta_1$ and $\theta_2$ (see \rtab{tab:benchmark}).  Mod-squared amplitude contributions are specified using the abbreviations defined in \rtab{tab:abbrev}.  The ratio $\sigma/\sigma(\mathrm{Sq}(h_{1+2}))$ is also given.  The selection cuts in \eqref{cuts} are applied.  Cross sections are given for a single lepton flavour combination.  The integration error is displayed in brackets.} 
\end{table}
%
%
%
\begin{table}[tbp]
\vspace{0.cm}
\centering
\renewcommand{\arraystretch}{1.2}
\begin{tabular}{|c|c|c|c|c|}
\cline{2-5}
\multicolumn{1}{c|}{} & \multicolumn{4}{|c|}{$gg\ (\to \{h_1,h_2\}) \to W^-W^+ \!\to \ell\bar{\nu}\,\bar{\ell}^\prime \nu^\prime$} \\ 
\multicolumn{1}{c|}{} &   \multicolumn{4}{|c|}{$\sigma$ [fb], $pp$, $\sqrt{s}=13$ TeV} \\ 
\multicolumn{1}{c|}{} &   \multicolumn{4}{|c|}{1HSM (see \rtabs{tab:benchmark} and \ref{tab:abbrev})} \\ 
\cline{2-5}
\multicolumn{1}{c|}{} & $M_{h_2}$ & \multicolumn{3}{|c|}{$|\mathcal{M}|^2$} \\
\cline{3-5}
\multicolumn{1}{c|}{} &  [GeV]  & Sq($h_2$) & $h_2$+I($h_1$) & $h_2$+I(C+$h_1$) \\
\hline
\multirow{8}{*}{$\theta_1$}
 & 700 & $0.07810(2)$ & $0.04113(4)$ & $0.09591(7)$ \\
 & ratio & $1$ & $0.5266(6)$ & $1.2280(9)$ \\ \cline{2-5}
 & 1000 & $0.010824(2)$ & $-0.01621(2)$ & $0.01780(3)$ \\
 & ratio & $1$ & $-1.498(2)$ & $1.644(2)$ \\ \cline{2-5}
 & 1500 & $0.00027818(5)$ & $-0.005749(2)$ & $0.001214(3)$ \\
 & ratio & $1$ & $-20.668(8)$ & $4.36(1)$ \\ \cline{2-5}
 & 3000 & $5.3026(9)\times 10^{-7}$ & $-0.00032008(7)$ & $3.46(2)\times 10^{-5}$ \\
 & ratio & $1$ & $-603.6(2)$ & $65.2(2)$ \\
\hline
\multirow{8}{*}{$\theta_2$}
 & 700 & $0.27776(5)$ & $0.1737(2)$ & $0.3502(2)$ \\
 & ratio & $1$ & $0.6252(4)$ & $1.2606(7)$ \\ \cline{2-5}
 & 1000 & $0.035182(6)$ & $-0.03845(3)$ & $0.06833(5)$ \\
 & ratio & $1$ & $-1.0928(9)$ & $1.942(2)$ \\ \cline{2-5}
 & 1500 & $0.0008885(2)$ & $-0.016227(5)$ & $0.005293(7)$ \\
 & ratio & $1$ & $-18.262(6)$ & $5.957(8)$ \\ \cline{2-5}
 & 3000 & $2.3605(4)\times 10^{-6}$ & $-0.0010870(3)$ & $0.0001561(4)$ \\
 & ratio & $1$ & $-460.5(2)$ & $66.1(2)$ \\
\hline
\end{tabular}\\[0cm] 
\caption{\label{tab:res_ww_h2}
Cross sections for $gg\ (\to \{h_1,h_2\}) \to W^-W^+ \!\to \ell\bar{\nu}\,\bar{\ell}^\prime \nu^\prime$ in $pp$ collisions at $\sqrt{s}=13$~TeV in the 1HSM with focus on heavy Higgs ($h_2$) production.  The ratio $\sigma/\sigma(\mathrm{Sq}(h_2))$ is also given.  Other details as in \rtab{tab:res_ww}.}
\end{table}
%
%
%
\begin{table}[tbp]
\vspace{0.cm}
\centering
\renewcommand{\arraystretch}{1.2}
\begin{tabular}{|c|c|c|c|c|}
\cline{2-5}
\multicolumn{1}{c|}{} & \multicolumn{4}{|c|}{$gg\ (\to \{H,h_1,h_2\}) \to t\bar{t} \to b\bar{b}\,\ell\bar{\nu}\,\bar{\ell}^\prime \nu^\prime$} \\ 
\multicolumn{1}{c|}{} &   \multicolumn{4}{|c|}{$\sigma$ [fb], $pp$, $\sqrt{s}=13$ TeV} \\ 
\multicolumn{1}{c|}{} &   \multicolumn{4}{|c|}{SM and 1HSM (see \rtabs{tab:benchmark} and \ref{tab:abbrev})} \\
\cline{2-5}\noalign{\vspace{0.075cm}}\cline{2-5}
\multicolumn{1}{c|}{} & \multirow{2}{*}{SM} & \multicolumn{3}{|c|}{$|\mathcal{M}|^2$} \\
\cline{3-5}
\multicolumn{1}{c|}{} &  & Sq($H$) & $H$+I(C) & $H$+I(C$_{+\circlearrowleft}$) \\
\cline{2-5}
\multicolumn{1}{c|}{} & $\sigma$ & $0.13367(4)$ & $-4.984(2)$ & $10.984(5)$ \\
\multicolumn{1}{c|}{} & ratio & $1$ & $-37.28(2)$ & $82.17(5)$ \\
\cline{2-5}\noalign{\vspace{0.075cm}}\cline{2-5}
\multicolumn{1}{c|}{} & 1HSM & \multicolumn{3}{|c|}{$|\mathcal{M}|^2$} \\
\cline{2-5}
\multicolumn{1}{c|}{} & $M_{h_2}$ [GeV]  & Sq($h_{1+2}$) & $h_{1+2}$+I(C) & $h_{1+2}$+I(C$_{+\circlearrowleft}$) \\ \hline
\multirow{8}{*}{$\theta_1$}
 & 700 & $0.12834(4)$ & $-4.779(2)$ & $11.203(5)$ \\
 & ratio & $1$ & $-37.23(2)$ & $87.29(5)$ \\ \cline{2-5}
 & 1000 & $0.11820(4)$ & $-4.739(2)$ & $10.605(5)$ \\
 & ratio & $1$ & $-40.10(2)$ & $89.72(5)$ \\ \cline{2-5}
 & 1500 & $0.12735(4)$ & $-4.874(2)$ & $10.759(5)$ \\
 & ratio & $1$ & $-38.27(2)$ & $84.48(5)$ \\ \cline{2-5}
 & 3000 & $0.13228(4)$ & $-4.957(2)$ & $10.932(5)$ \\
 & ratio & $1$ & $-37.47(2)$ & $82.64(5)$ \\
\hline
\multirow{8}{*}{$\theta_2$}
 & 700 & $0.12576(4)$ & $-4.317(2)$ & $11.797(5)$ \\
 & ratio & $1$ & $-34.33(2)$ & $93.80(5)$ \\ \cline{2-5}
 & 1000 & $0.08696(3)$ & $-4.195(2)$ & $9.846(5)$ \\
 & ratio & $1$ & $-48.24(2)$ & $113.23(6)$ \\ \cline{2-5}
 & 1500 & $0.11365(3)$ & $-4.635(2)$ & $10.278(5)$ \\
 & ratio & $1$ & $-40.78(2)$ & $90.44(5)$ \\ \cline{2-5}
 & 3000 & $0.12895(4)$ & $-4.896(2)$ & $10.796(5)$ \\
 & ratio & $1$ & $-37.97(2)$ & $83.72(5)$ \\
\hline
\end{tabular}\\[0cm] 
\caption{\label{tab:res_tt}
Cross sections for $gg\ (\to \{H,h_1,h_2\}) \to t\bar{t} \to b\bar{b}\,\ell\bar{\nu}\,\bar{\ell}^\prime \nu^\prime$ in $pp$ collisions at $\sqrt{s}=13$~TeV in the SM and 1HSM.  Virtual corrections ($\circlearrowleft$) to the continuum background are taken into account (see main text).  Other details as in \rtab{tab:res_ww}.}
\end{table}
%
%
%
\begin{table}[tbp]
\vspace{0.cm}
\centering
\renewcommand{\arraystretch}{1.2}
\begin{tabular}{|c|c|c|c|c|c|}
\cline{2-6}
\multicolumn{1}{c|}{} & \multicolumn{5}{|c|}{$gg\ (\to \{h_1,h_2\}) \to t\bar{t} \to b\bar{b}\,\ell\bar{\nu}\,\bar{\ell}^\prime \nu^\prime$} \\ 
\multicolumn{1}{c|}{} &   \multicolumn{5}{|c|}{$\sigma$ [fb], $pp$, $\sqrt{s}=13$ TeV} \\ 
\multicolumn{1}{c|}{} &   \multicolumn{5}{|c|}{1HSM (see \rtabs{tab:benchmark} and \ref{tab:abbrev})} \\ 
\cline{2-6}
\multicolumn{1}{c|}{} & $M_{h_2}$ & \multicolumn{4}{|c|}{$|\mathcal{M}|^2$} \\
\cline{3-6}
\multicolumn{1}{c|}{} &  [GeV]  & Sq($h_2$) & $h_2$+I($h_1$) & $h_2$+I(C+$h_1$) & $h_2$+I(C$_{+\circlearrowleft}$+$h_1$) \\
\hline
\multirow{8}{*}{$\theta_1$}
 & 700 & $0.015207(4)$ & $0.00607(1)$ & $-0.00744(2)$ & $0.6966(6)$ \\
 & ratio & $1$ & $0.3990(7)$ & $-0.489(2)$ & $45.80(4)$ \\ \cline{2-6}
 & 1000 & $0.0012148(4)$ & $-0.004079(3)$ & $0.03194(3)$ & $0.09891(8)$ \\
 & ratio & $1$ & $-3.358(2)$ & $26.30(3)$ & $81.42(7)$ \\ \cline{2-6}
 & 1500 & $1.2910(4)\times 10^{-5}$ & $-0.0009172(3)$ & $0.009049(4)$ & $-0.00278(2)$ \\
 & ratio & $1$ & $-71.05(3)$ & $700.9(4)$ & $-2.15(1)\times 10^{2}$ \\ \cline{2-6}
 & 3000 & $7.858(3)\times 10^{-9}$ & $-4.655(2)\times 10^{-5}$ & $0.0005783(2)$ & $-0.0007648(7)$ \\
 & ratio & $1$ & $-5.923(3)\times 10^{3}$ & $7.359(3)\times 10^{4}$ & $-9.733(9)\times 10^{4}$ \\
\hline
\multirow{8}{*}{$\theta_2$}
 & 700 & $0.05395(2)$ & $0.02842(3)$ & $-0.04930(9)$ & $2.436(2)$ \\
 & ratio & $1$ & $0.5268(5)$ & $-0.914(2)$ & $45.16(4)$ \\ \cline{2-6}
 & 1000 & $0.004151(2)$ & $-0.010379(6)$ & $0.07329(9)$ & $0.4855(4)$ \\
 & ratio & $1$ & $-2.501(2)$ & $17.66(3)$ & $117.0(1)$ \\ \cline{2-6}
 & 1500 & $5.566(2)\times 10^{-5}$ & $-0.0026682(8)$ & $0.02068(1)$ & $0.03554(2)$ \\
 & ratio & $1$ & $-47.94(2)$ & $371.6(3)$ & $638.6(4)$ \\ \cline{2-6}
 & 3000 & $8.503(3)\times 10^{-8}$ & $-0.00015896(5)$ & $0.0017632(6)$ & $-0.001246(3)$ \\
 & ratio & $1$ & $-1869.5(7)$ & $2.0736(9)\times 10^{4}$ & $-1.465(3)\times 10^{4}$ \\
\hline
\end{tabular}\\[0cm] 
\caption{\label{tab:res_tt_h2}
Cross sections for $gg\ (\to \{h_1,h_2\}) \to t\bar{t} \to b\bar{b}\,\ell\bar{\nu}\,\bar{\ell}^\prime \nu^\prime$ in $pp$ collisions at $\sqrt{s}=13$~TeV in the 1HSM with focus on heavy Higgs production.  Other details as in \rtab{tab:res_ww_h2}.}
\end{table}

\begin{figure}[tbp]
\vspace{0.cm}
\centering
\includegraphics[width=\textwidth, clip=true]{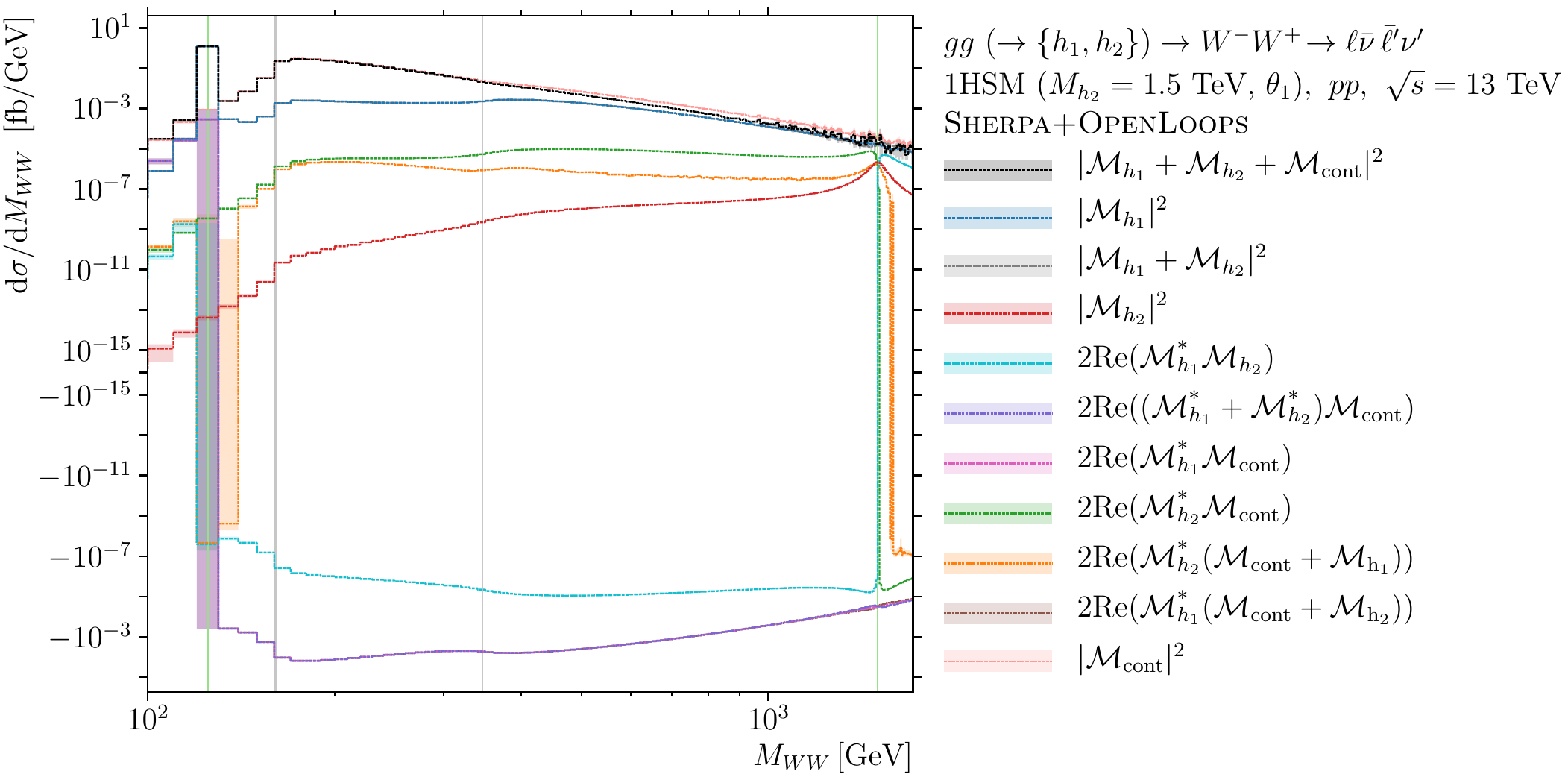}
\caption{\label{fig:ww_1hsm3_mass} $M_{WW}$ distributions for the signal process $gg\ (\to \{h_1,h_2\}) \to W^-W^+ \!\to \ell\bar{\nu}\,\bar{\ell}^\prime \nu^\prime$ in the 1-Higgs-Singlet Extension of the SM (1HSM) with $M_{h_2}=1500$~GeV and mixing angle $\theta_1$ (see \rtab{tab:benchmark}) including its interference with the background in $pp$ collisions at $\sqrt{s}=13$~TeV.  Vertical lines indicate the position of the $WW$ and $t\bar{t}$ thresholds (grey) and of the Higgs resonances (green).  The selection cuts in \eqref{cuts} are applied.  Cross sections are given for a single lepton flavour combination.  The bands show the MC integration error estimate (90\% confidence interval).}
\end{figure}
%
%
%
\begin{figure}[tbp]
\vspace{0.cm}
\centering
\includegraphics[width=\textwidth, clip=true]{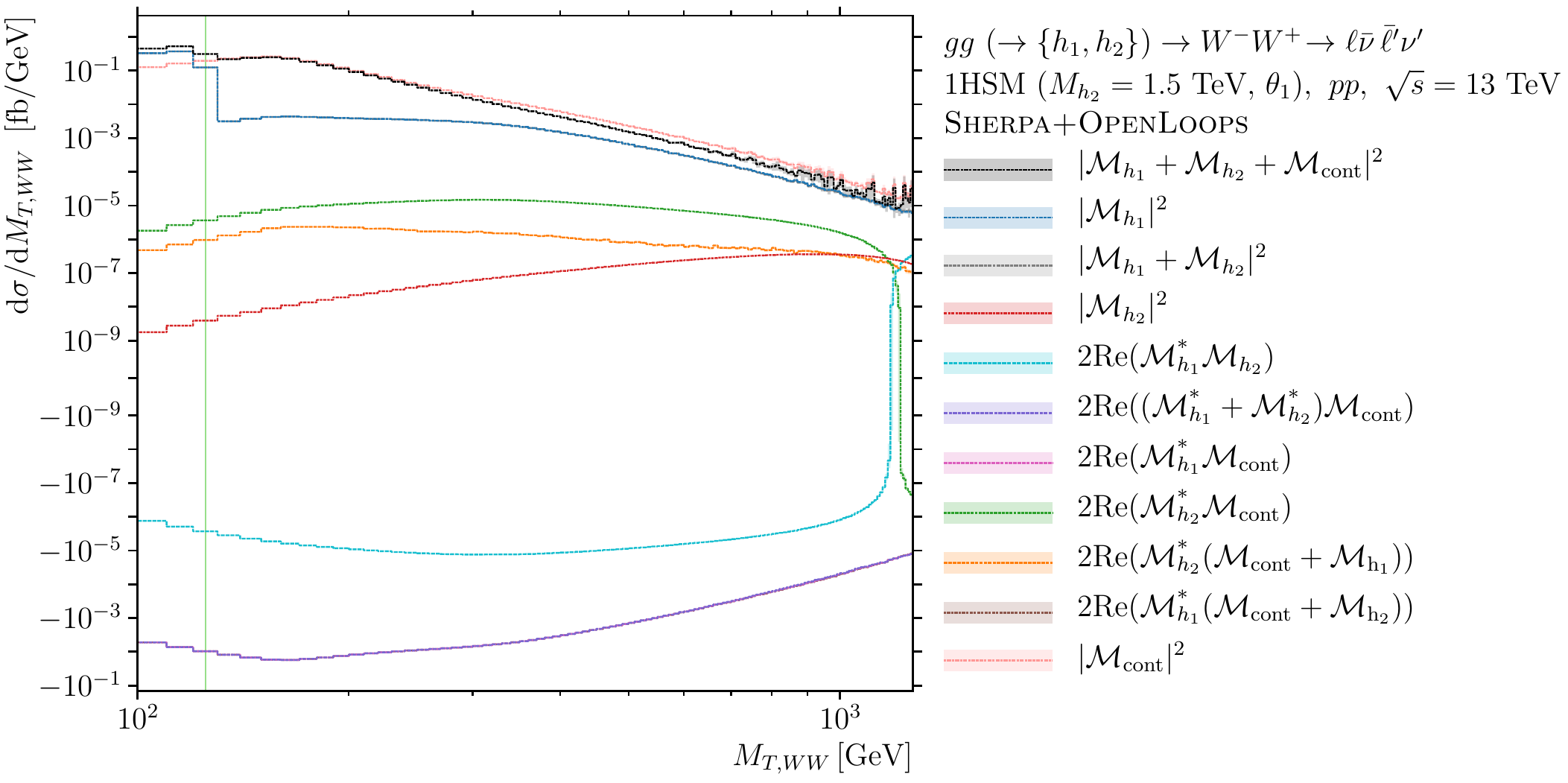}
\caption{\label{fig:ww_1hsm3_mass_trans1} $M_{T,WW}$ distributions for the signal process $gg\ (\to \{h_1,h_2\}) \to W^-W^+ \!\to \ell\bar{\nu}\,\bar{\ell}^\prime \nu^\prime$ in the 1HSM ($M_{h_2}=1500$~GeV, $\theta_1$) including its interference with the background in $pp$ collisions at $\sqrt{s}=13$~TeV.  Other details as in \rfig{fig:ww_1hsm3_mass}.}
\end{figure}
%
%
%
\begin{figure}[tbp]
\vspace{0.cm}
\centering
\includegraphics[width=\textwidth, clip=true]{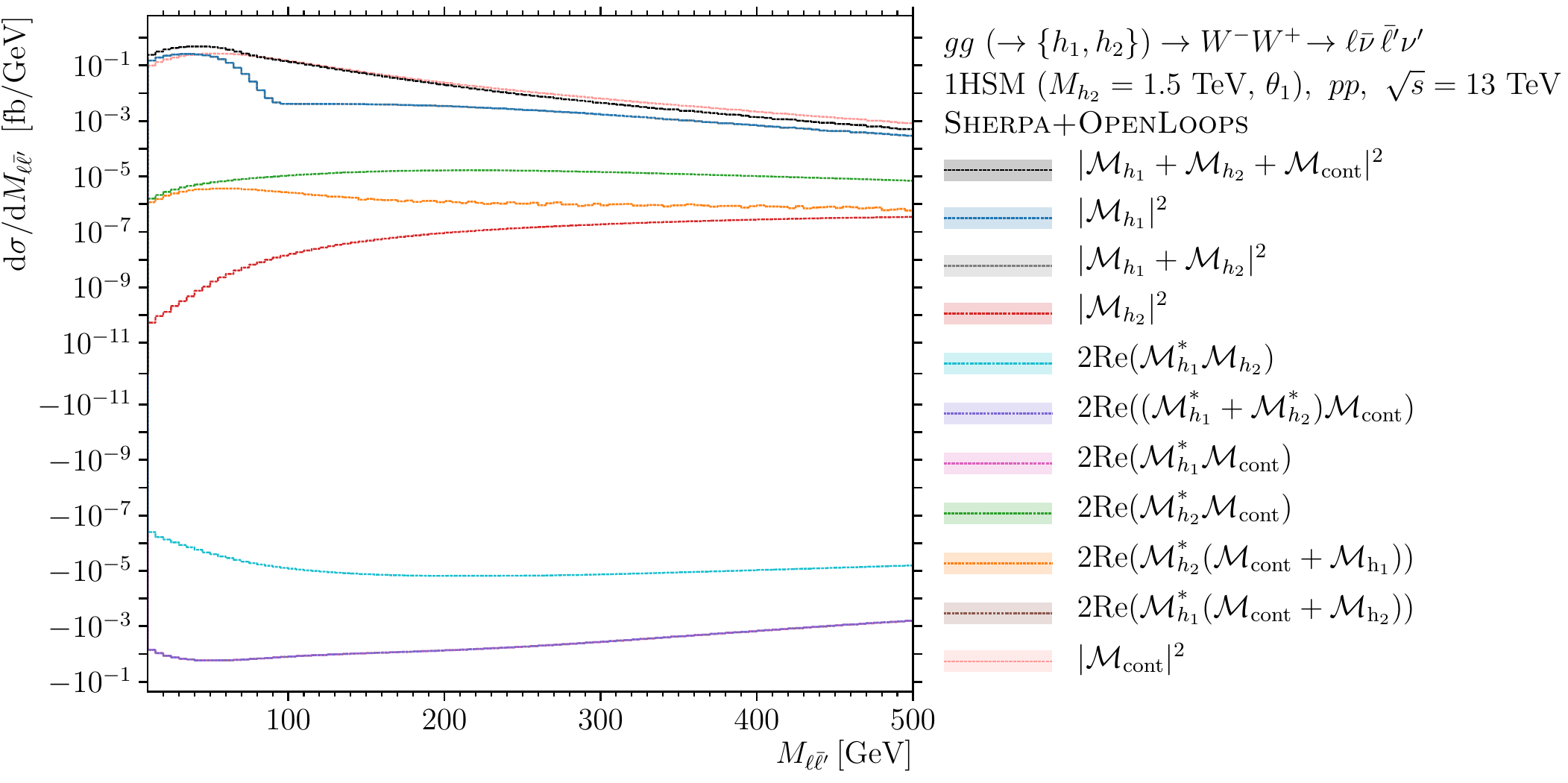}
\caption{\label{fig:ww_1hsm3_mll} $M_{\ell\bar{\ell}^\prime}$ distributions for the signal process $gg\ (\to \{h_1,h_2\}) \to W^-W^+ \!\to \ell\bar{\nu}\,\bar{\ell}^\prime \nu^\prime$ in the 1HSM ($M_{h_2}=1500$~GeV, $\theta_1$) including its interference with the background in $pp$ collisions at $\sqrt{s}=13$~TeV.  Other details as in \rfig{fig:ww_1hsm3_mass}.}
\end{figure}
%
%
%
\begin{figure}[tbp]
\vspace{0.cm}
\centering
\includegraphics[width=\textwidth, clip=true]{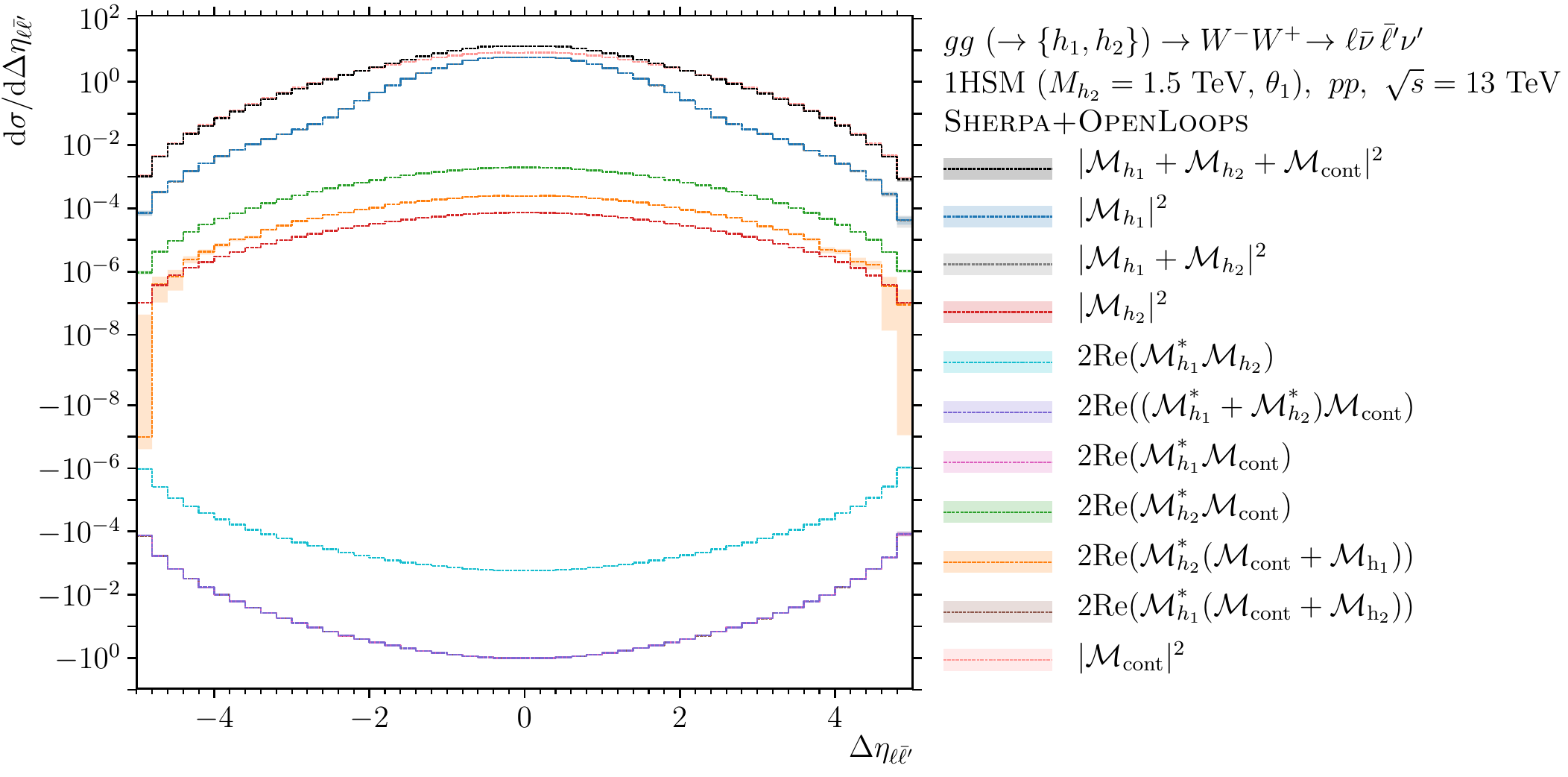}
\caption{\label{fig:ww_1hsm3_deltaEtall} $\Delta\eta_{\ell\bar{\ell}^\prime}$ distributions for the signal process $gg\ (\to \{h_1,h_2\}) \to W^-W^+ \!\to \ell\bar{\nu}\,\bar{\ell}^\prime \nu^\prime$ in the 1HSM ($M_{h_2}=1500$~GeV, $\theta_1$) including its interference with the background in $pp$ collisions at $\sqrt{s}=13$~TeV.  Other details as in \rfig{fig:ww_1hsm3_mass}.}
\end{figure}
%
%
%
\begin{figure}[tbp]
\vspace{0.cm}
\centering
\includegraphics[width=\textwidth, clip=true]{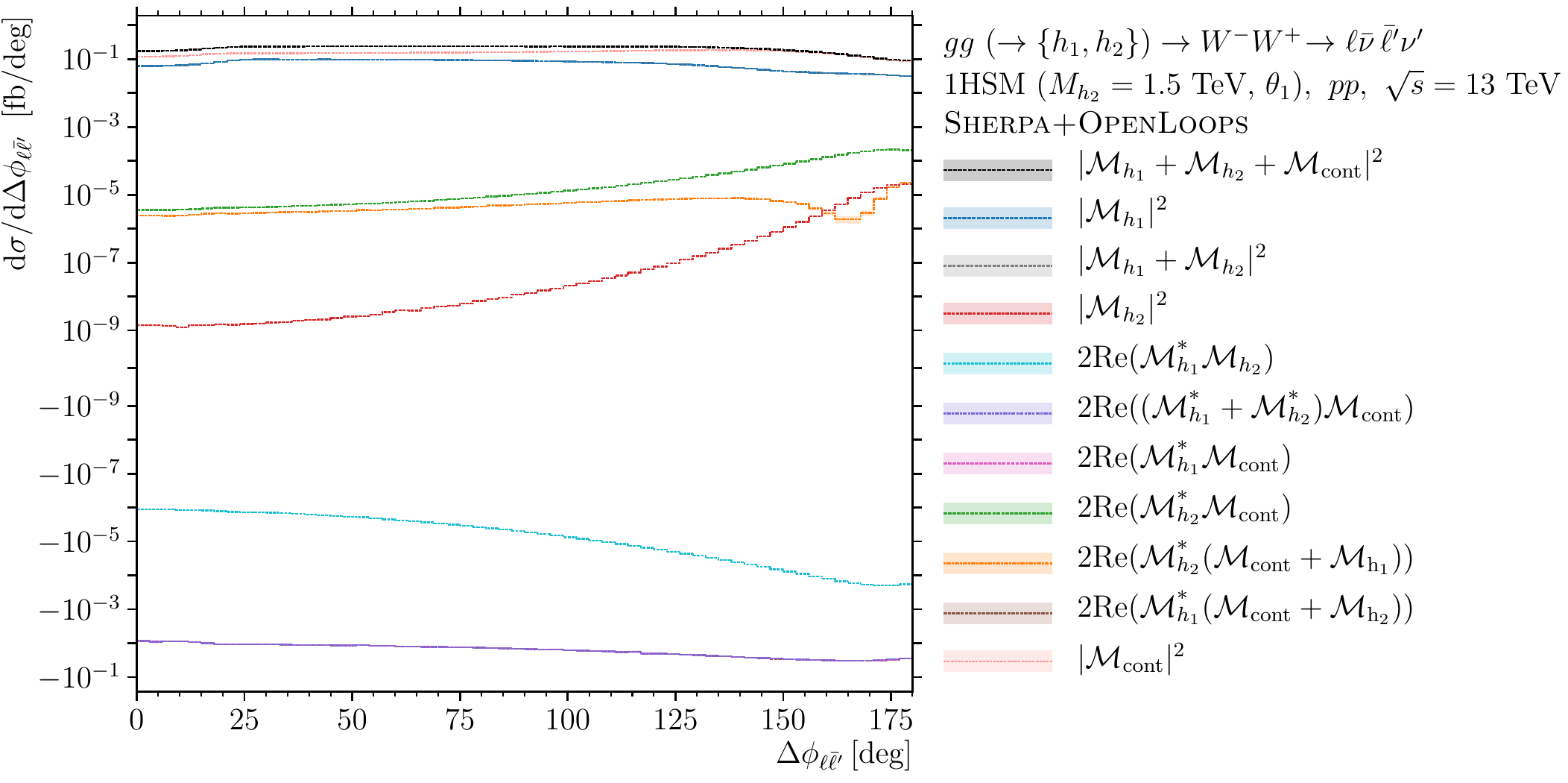}
\caption{\label{fig:ww_1hsm3_deltaPhill} $\Delta\phi_{\ell\bar{\ell}^\prime}$ distributions for the signal process $gg\ (\to \{h_1,h_2\}) \to W^-W^+ \!\to \ell\bar{\nu}\,\bar{\ell}^\prime \nu^\prime$ in the 1HSM ($M_{h_2}=1500$~GeV, $\theta_1$) including its interference with the background in $pp$ collisions at $\sqrt{s}=13$~TeV.  Other details as in \rfig{fig:ww_1hsm3_mass}.}
\end{figure}
%
%
%
\begin{figure}[tbp]
\vspace{0.cm}
\centering
\includegraphics[width=\textwidth, clip=true]{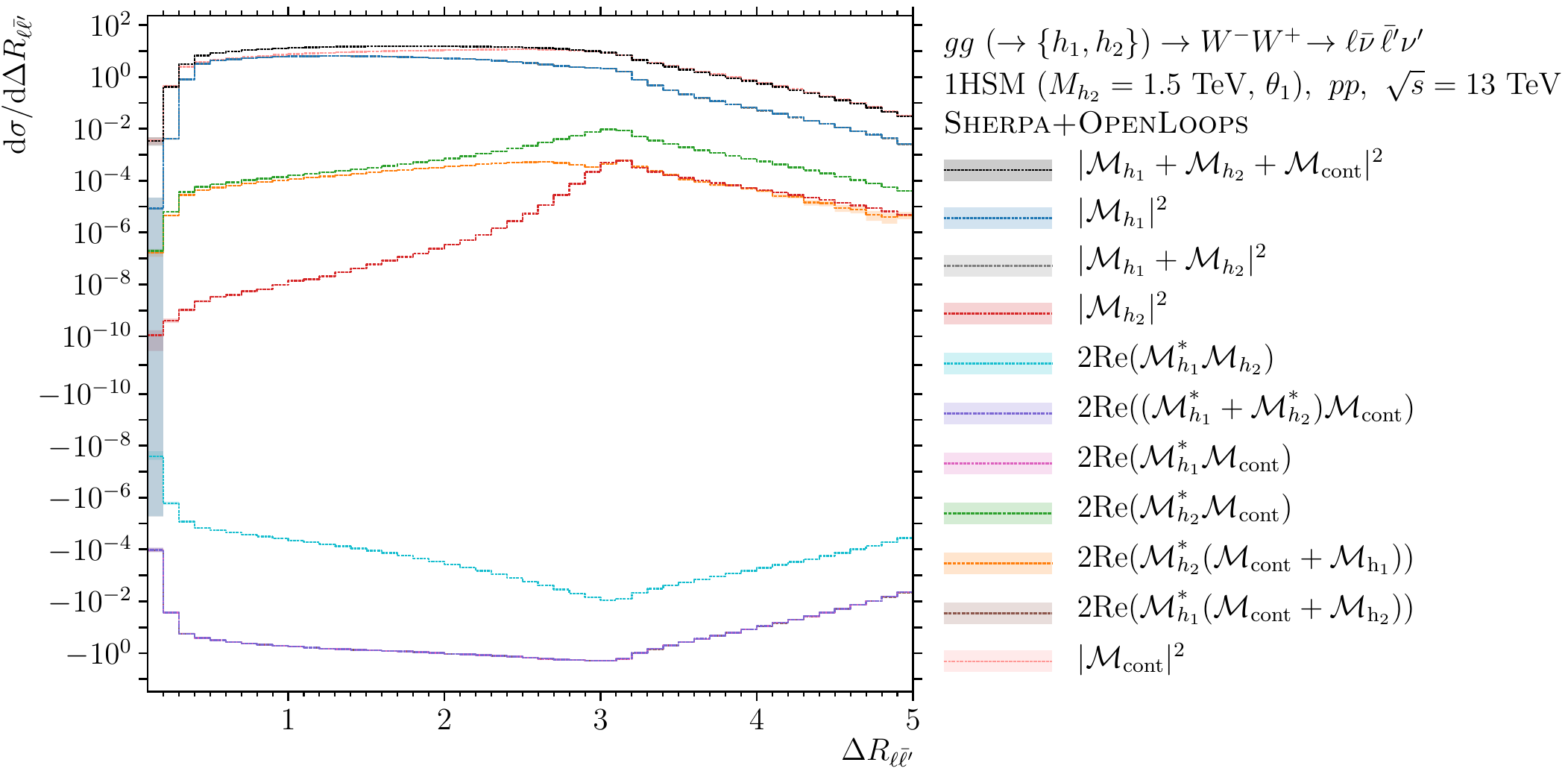}
\caption{\label{fig:ww_1hsm3_deltaRll} $\Delta R_{\ell\bar{\ell}^\prime}$ distributions for the signal process $gg\ (\to \{h_1,h_2\}) \to W^-W^+ \!\to \ell\bar{\nu}\,\bar{\ell}^\prime \nu^\prime$ in the 1HSM ($M_{h_2}=1500$~GeV, $\theta_1$) including its interference with the background in $pp$ collisions at $\sqrt{s}=13$~TeV.  Other details as in \rfig{fig:ww_1hsm3_mass}.}
\end{figure}
%
%
%
%
%
%
\begin{figure}[tbp]
\vspace{0.cm}
\centering
\includegraphics[width=\textwidth, clip=true]{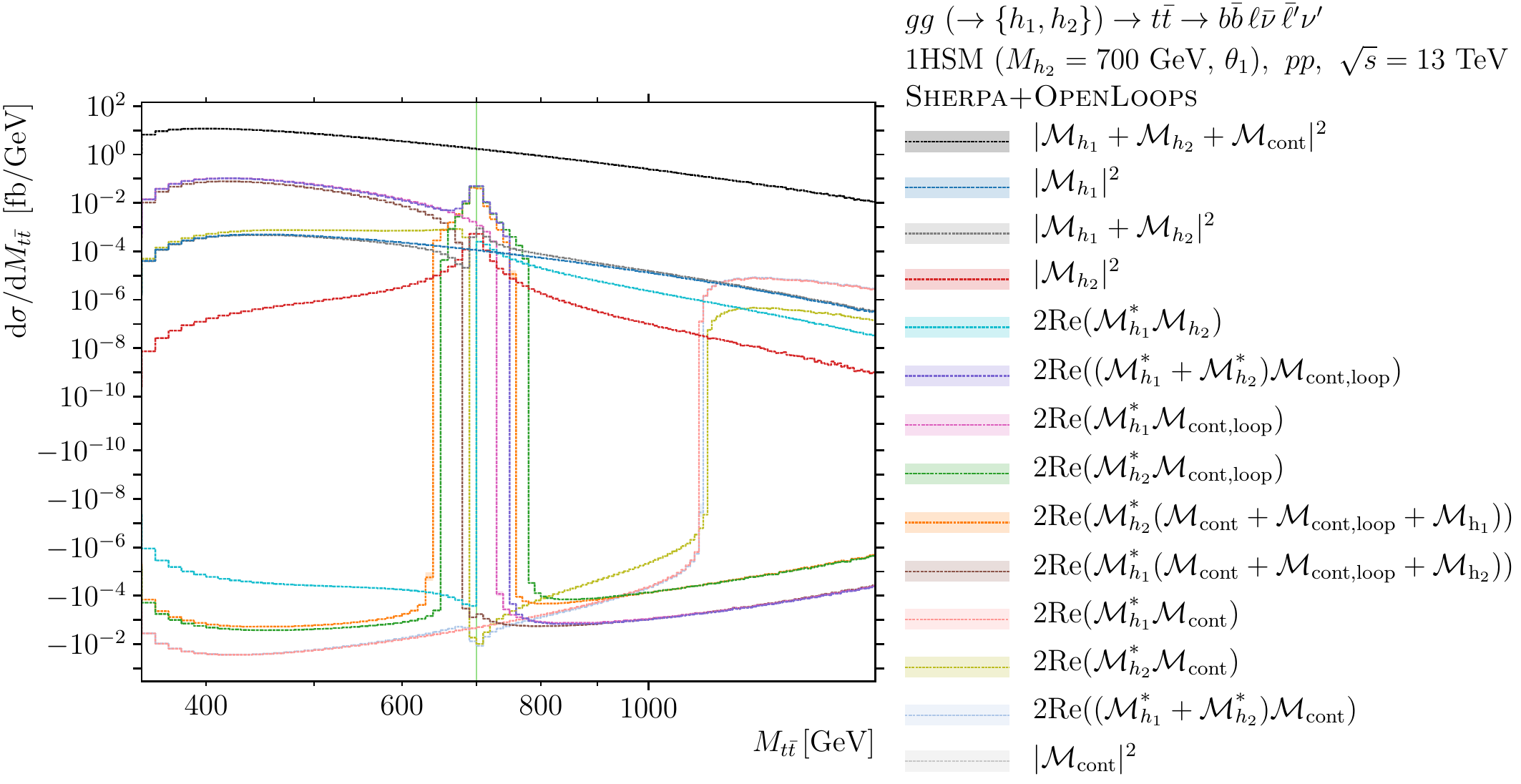}
\caption{\label{fig:tt_1hsm1_mass} $M_{t\bar{t}}$ distributions for the signal process $gg\ (\to \{h_1,h_2\}) \to t\bar{t} \to b\bar{b}\,\ell\bar{\nu}\,\bar{\ell}^\prime \nu^\prime$ in the 1HSM ($M_{h_2}=700$~GeV, $\theta_1$) including its interference with the background in $pp$ collisions at $\sqrt{s}=13$~TeV.  Other details as in \rfig{fig:ww_1hsm3_mass}.}
\end{figure}
%
%
%
\begin{figure}[tbp]
\vspace{0.cm}
\centering
\includegraphics[width=\textwidth, clip=true]{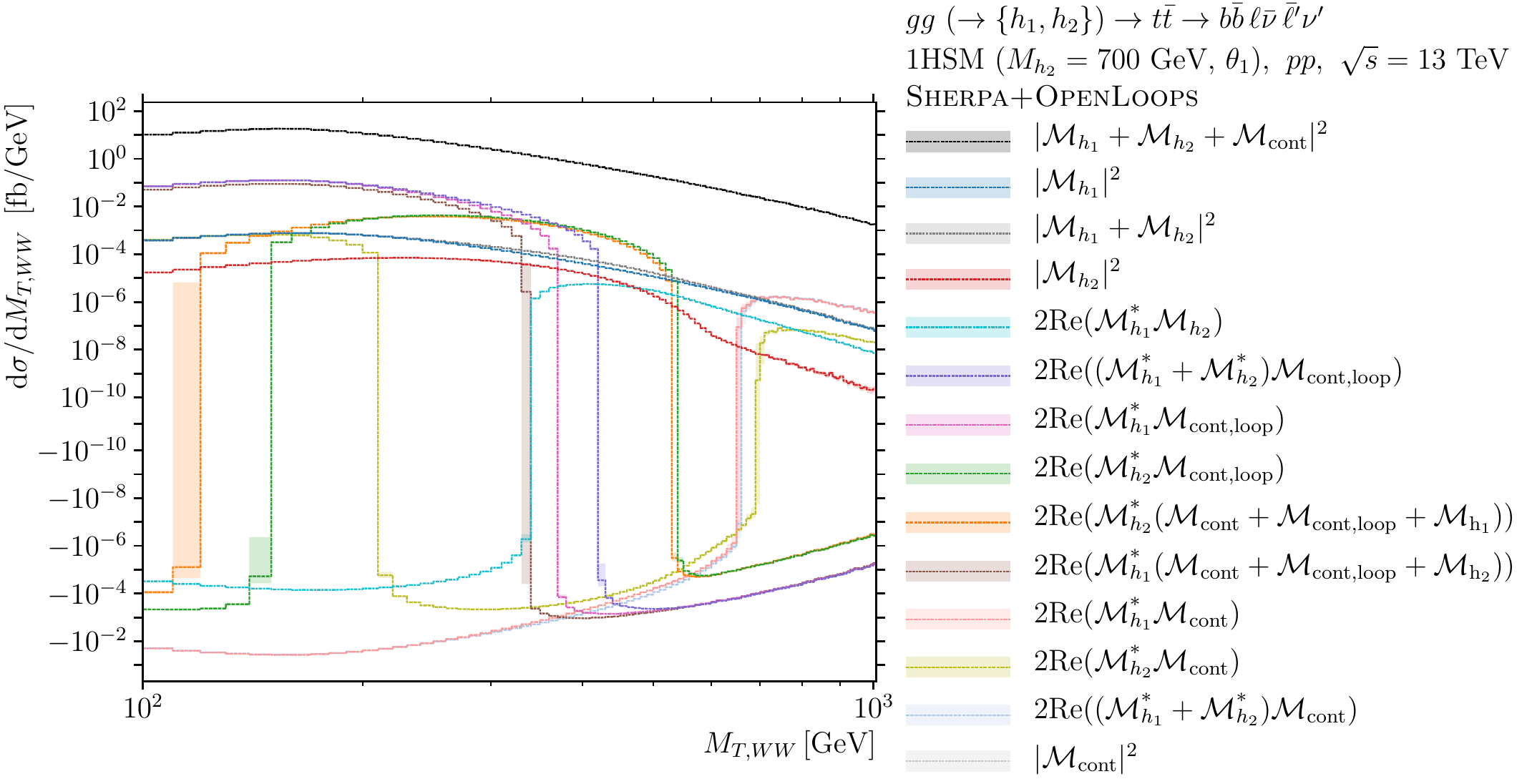}
\caption{\label{fig:tt_1hsm1_mass_trans1} $M_{T,WW}$ distributions for the signal process $gg\ (\to \{h_1,h_2\}) \to t\bar{t} \to b\bar{b}\,\ell\bar{\nu}\,\bar{\ell}^\prime \nu^\prime$ in the 1HSM ($M_{h_2}=700$~GeV, $\theta_1$) including its interference with the background in $pp$ collisions at $\sqrt{s}=13$~TeV.  Other details as in \rfig{fig:ww_1hsm3_mass}.}
\end{figure}
%
%
%
\begin{figure}[tbp]
\vspace{0.cm}
\centering
\includegraphics[width=\textwidth, clip=true]{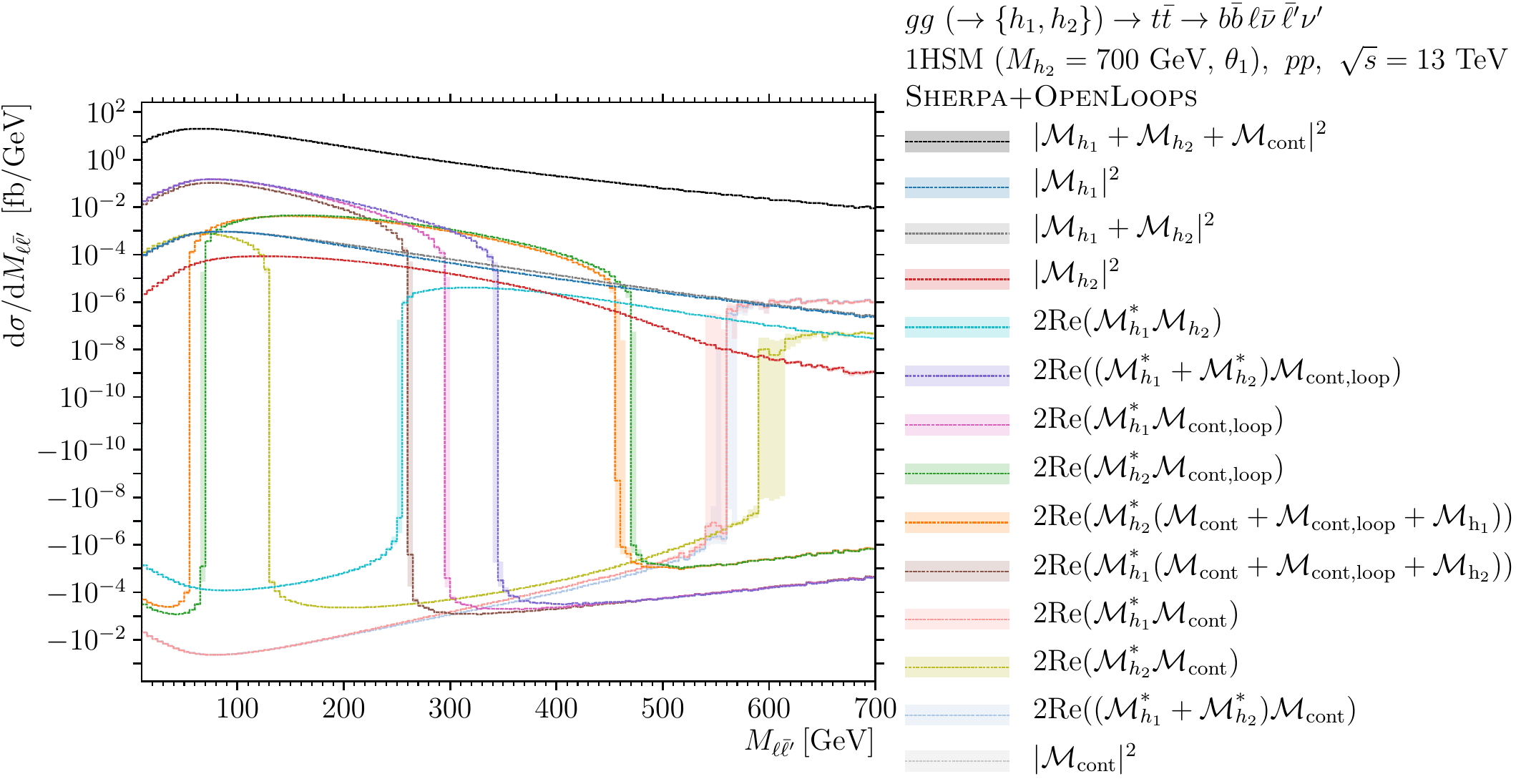}
\caption{\label{fig:tt_1hsm1_mll} $M_{\ell\bar{\ell}^\prime}$ distributions for the signal process $gg\ (\to \{h_1,h_2\}) \to t\bar{t} \to b\bar{b}\,\ell\bar{\nu}\,\bar{\ell}^\prime \nu^\prime$ in the 1HSM ($M_{h_2}=700$~GeV, $\theta_1$) including its interference with the background in $pp$ collisions at $\sqrt{s}=13$~TeV.  Other details as in \rfig{fig:tt_1hsm1_mass}.}
\end{figure}
%
%
%
\begin{figure}[tbp]
\vspace{0.cm}
\centering
\includegraphics[width=\textwidth, clip=true]{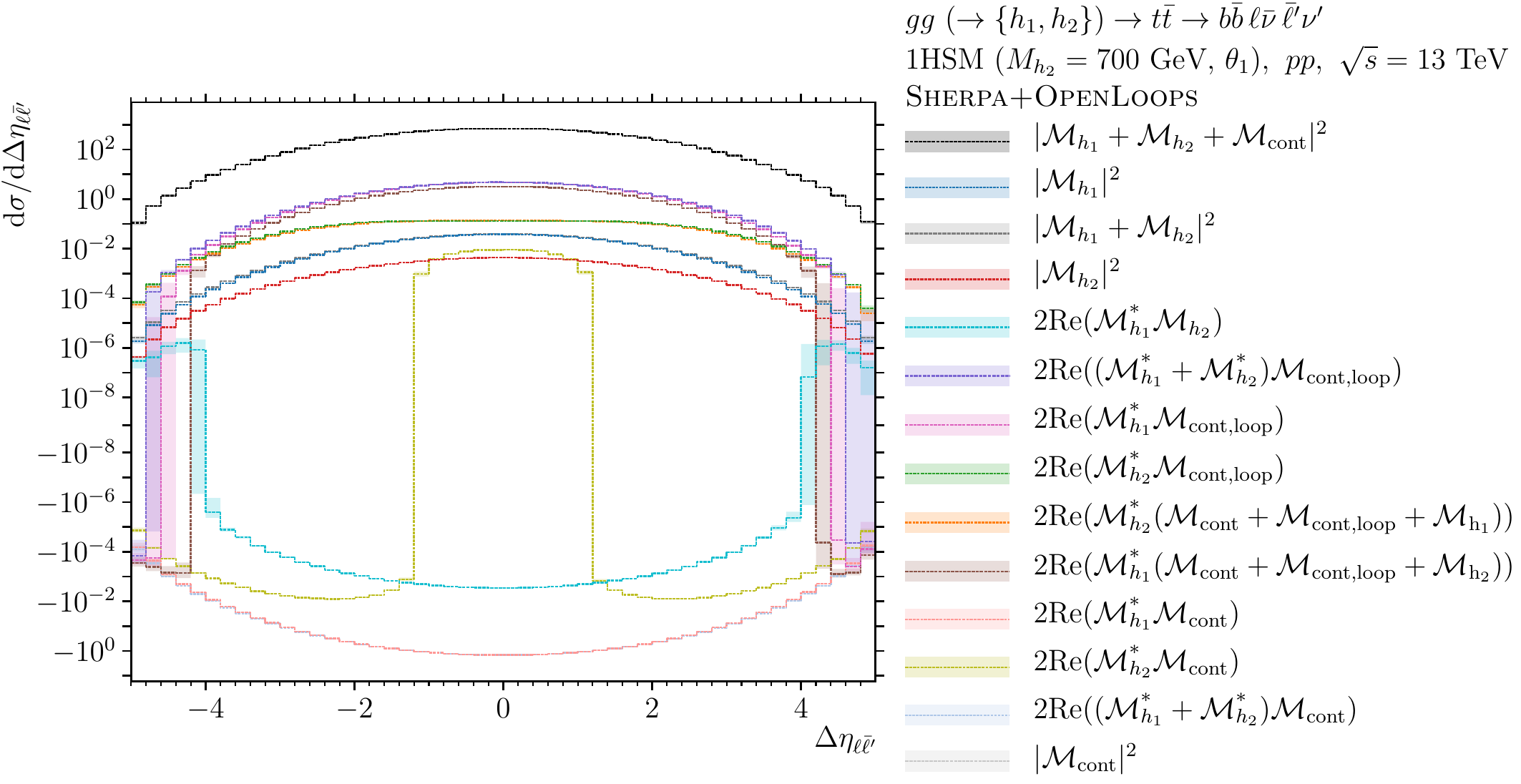}
\caption{\label{fig:tt_1hsm1_deltaEtall} $\Delta\eta_{\ell\bar{\ell}^\prime}$ distributions for the signal process $gg\ (\to \{h_1,h_2\}) \to t\bar{t} \to b\bar{b}\,\ell\bar{\nu}\,\bar{\ell}^\prime \nu^\prime$ in the 1HSM ($M_{h_2}=700$~GeV, $\theta_1$) including its interference with the background in $pp$ collisions at $\sqrt{s}=13$~TeV.  Other details as in \rfig{fig:tt_1hsm1_mass}.}
\end{figure}
%
%
%
\begin{figure}[tbp]
\vspace{0.cm}
\centering
\includegraphics[width=\textwidth, clip=true]{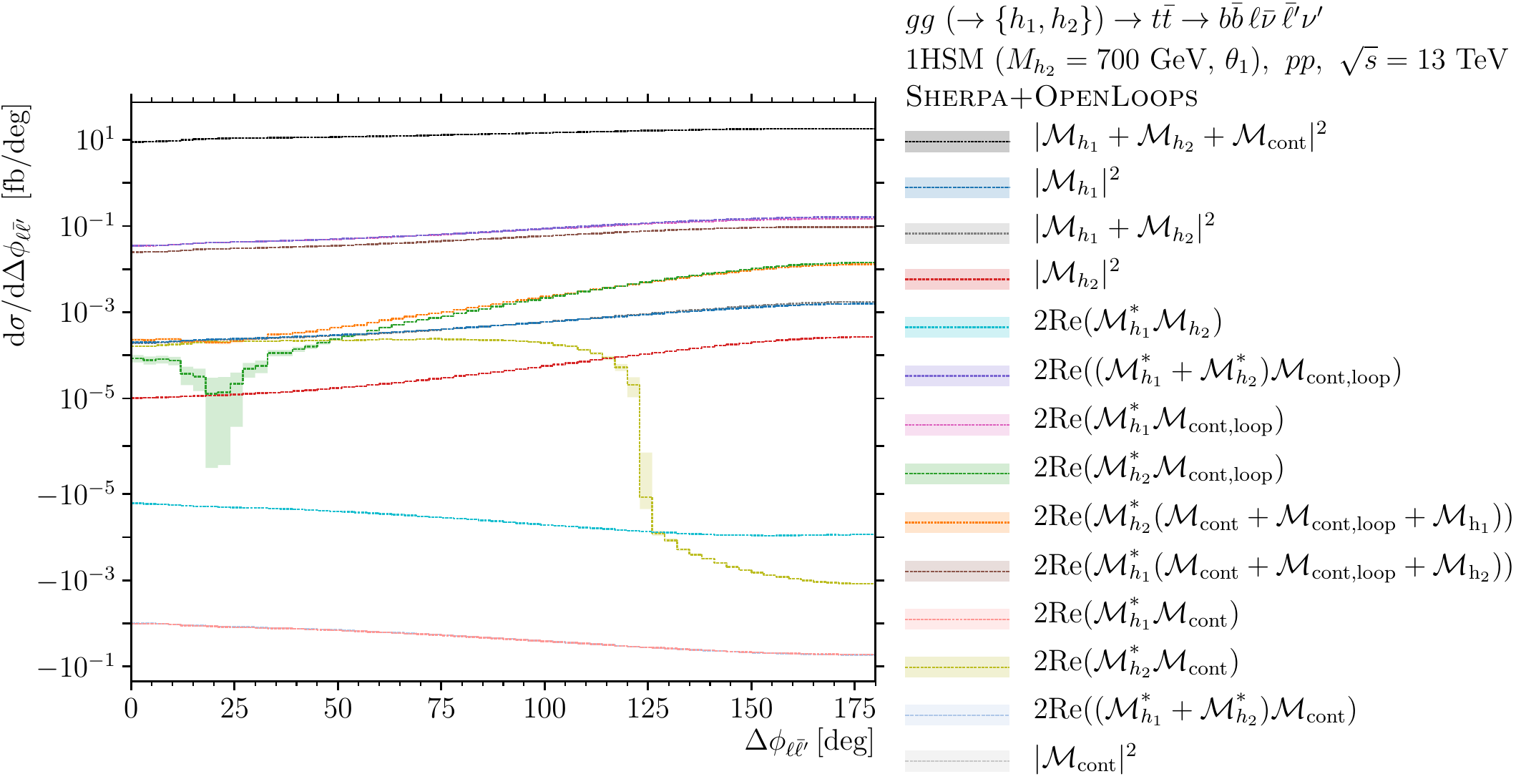}
\caption{\label{fig:tt_1hsm1_deltaPhill} $\Delta\phi_{\ell\bar{\ell}^\prime}$ distributions for the signal process $gg\ (\to \{h_1,h_2\}) \to t\bar{t} \to b\bar{b}\,\ell\bar{\nu}\,\bar{\ell}^\prime \nu^\prime$ in the 1HSM ($M_{h_2}=700$~GeV, $\theta_1$) including its interference with the background in $pp$ collisions at $\sqrt{s}=13$~TeV.  Other details as in \rfig{fig:tt_1hsm1_mass}.}
\end{figure}
%
%
%
\begin{figure}[tbp]
\vspace{0.cm}
\centering
\includegraphics[width=\textwidth, clip=true]{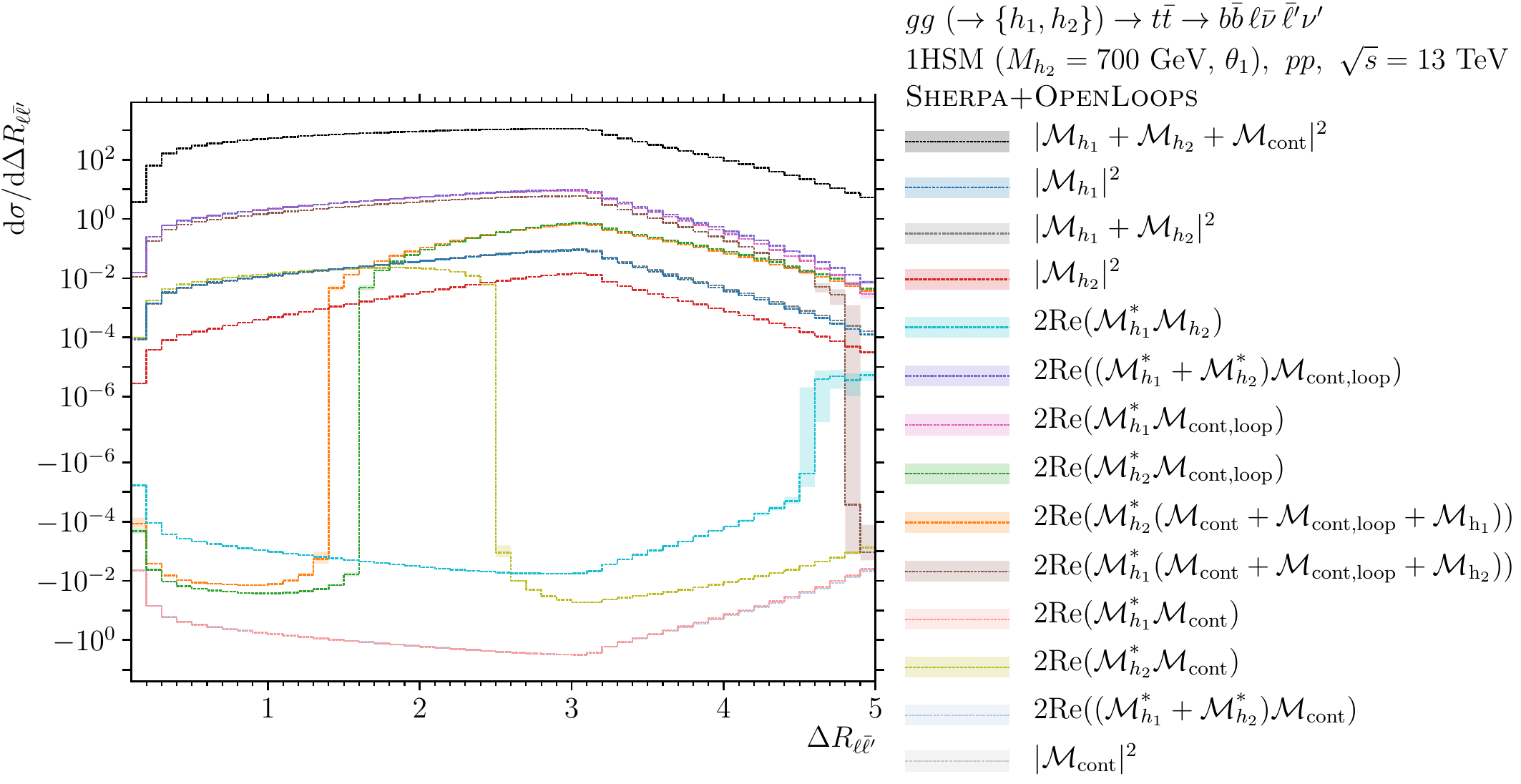}
\caption{\label{fig:tt_1hsm1_deltaRll} $\Delta R_{\ell\bar{\ell}^\prime}$ distributions for the signal process $gg\ (\to \{h_1,h_2\}) \to t\bar{t} \to b\bar{b}\,\ell\bar{\nu}\,\bar{\ell}^\prime \nu^\prime$ in the 1HSM ($M_{h_2}=700$~GeV, $\theta_1$) including its interference with the background in $pp$ collisions at $\sqrt{s}=13$~TeV.  Other details as in \rfig{fig:tt_1hsm1_mass}.}
\end{figure}
\clearpage
%
%
%
\begin{figure}[tbp]
\vspace{0.cm}
\centering
\includegraphics[width=\textwidth, clip=true]{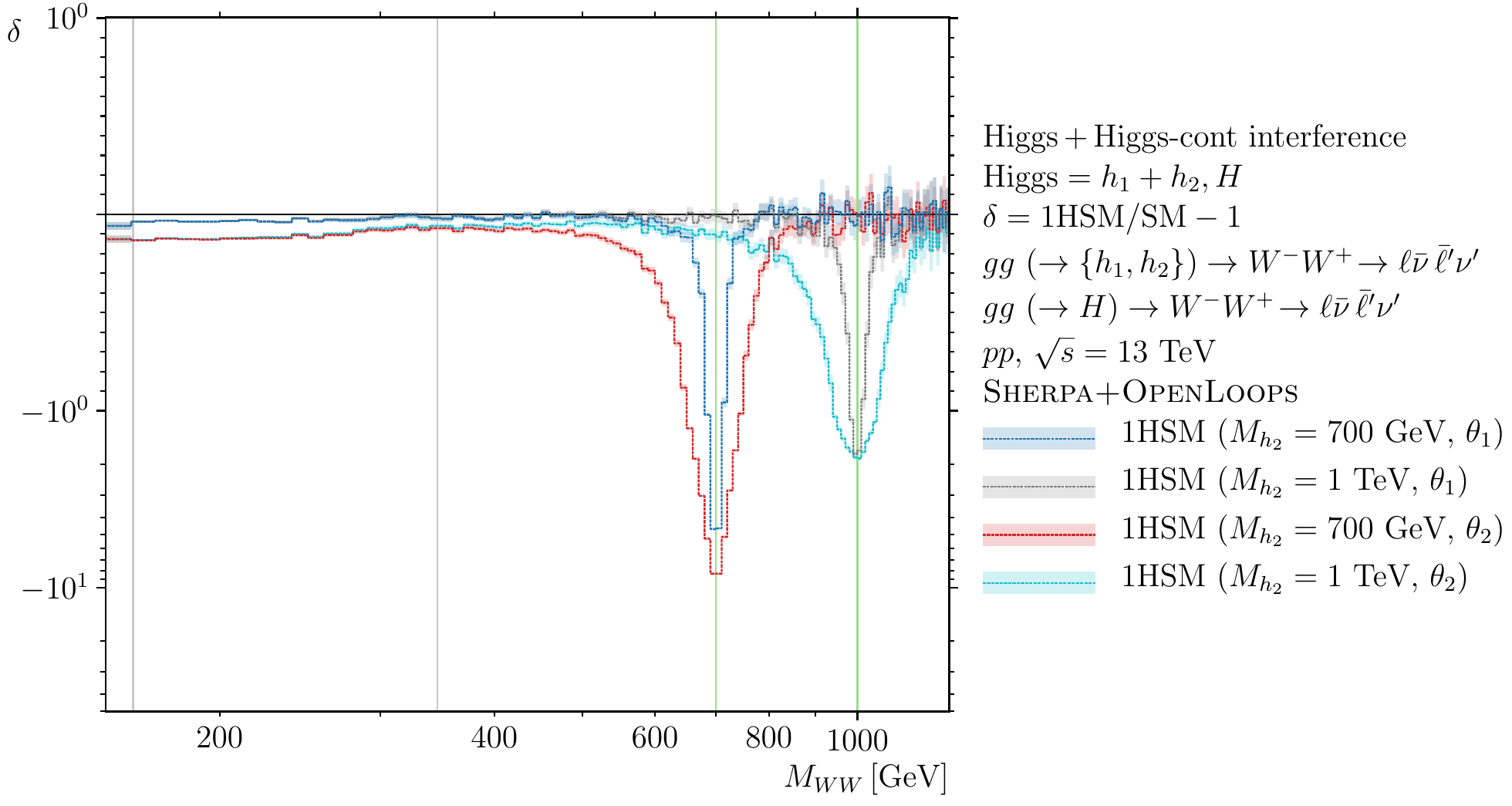}
\caption{\label{fig:1hsm_ww_si_h12-1256-sm_ww_si_-delta} $M_{WW}$ distributions of the relative deviation $\delta=R-1$ of the Higgs cross section including its interference with the background in the 1HSM ($M_{h_2}=\{700,1000\}$~GeV, $\{\theta_1,\theta_2\}$) compared to the SM for $gg\ (\to \{h_1,h_2,H\}) \to W^-W^+ \!\to \ell\bar{\nu}\,\bar{\ell}^\prime \nu^\prime$ in $pp$ collisions at $\sqrt{s}=13$~TeV.  $R$ is the ratio of $\sigma(\text{$h_{1+2}$+I(C)})$ to $\sigma(\text{$H$+I(C)})$.  Other details as in \rfig{fig:ww_1hsm3_mass}.}
\end{figure}
%
%
%
\begin{figure}[tbp]
\vspace{0.cm}
\centering
\includegraphics[width=\textwidth, clip=true]{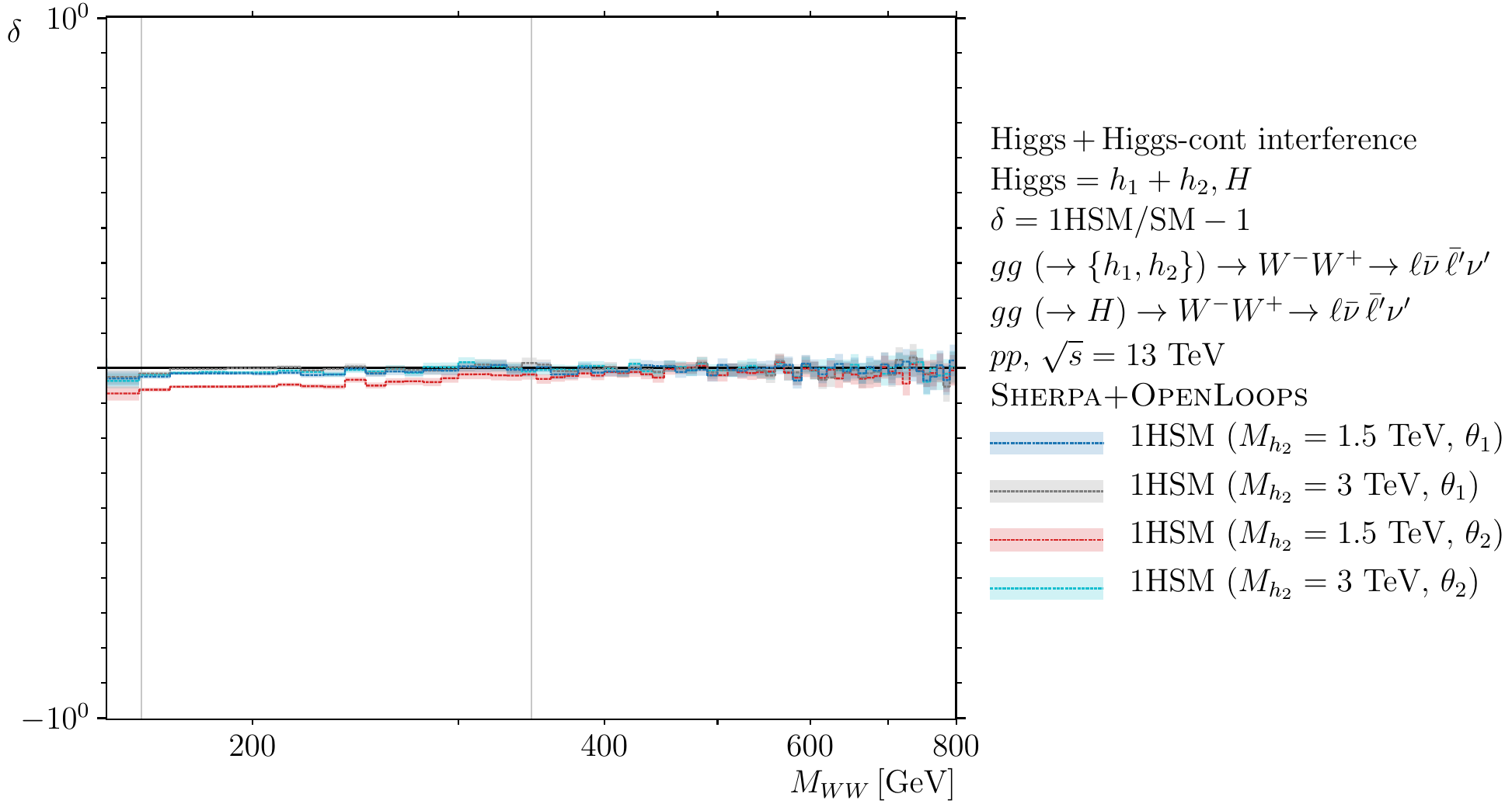}
\caption{\label{fig:1hsm_ww_si_h12-3478-sm_ww_si_-delta} $M_{WW}$ distributions of the relative deviation $\delta=R-1$ of the Higgs cross section including its interference with the background in the 1HSM ($M_{h_2}=\{1.5,3\}$~TeV, $\{\theta_1,\theta_2\}$) compared to the SM for $gg\ (\to \{h_1,h_2,H\}) \to W^-W^+ \!\to \ell\bar{\nu}\,\bar{\ell}^\prime \nu^\prime$ in $pp$ collisions at $\sqrt{s}=13$~TeV.  $R$ is the ratio of $\sigma(\text{$h_{1+2}$+I(C)})$ to $\sigma(\text{$H$+I(C)})$.  Other details as in \rfig{fig:ww_1hsm3_mass}.}
\end{figure}
%
%
%

\begin{figure}[tbp]
\vspace{0.cm}
\centering
\includegraphics[width=\textwidth, clip=true]{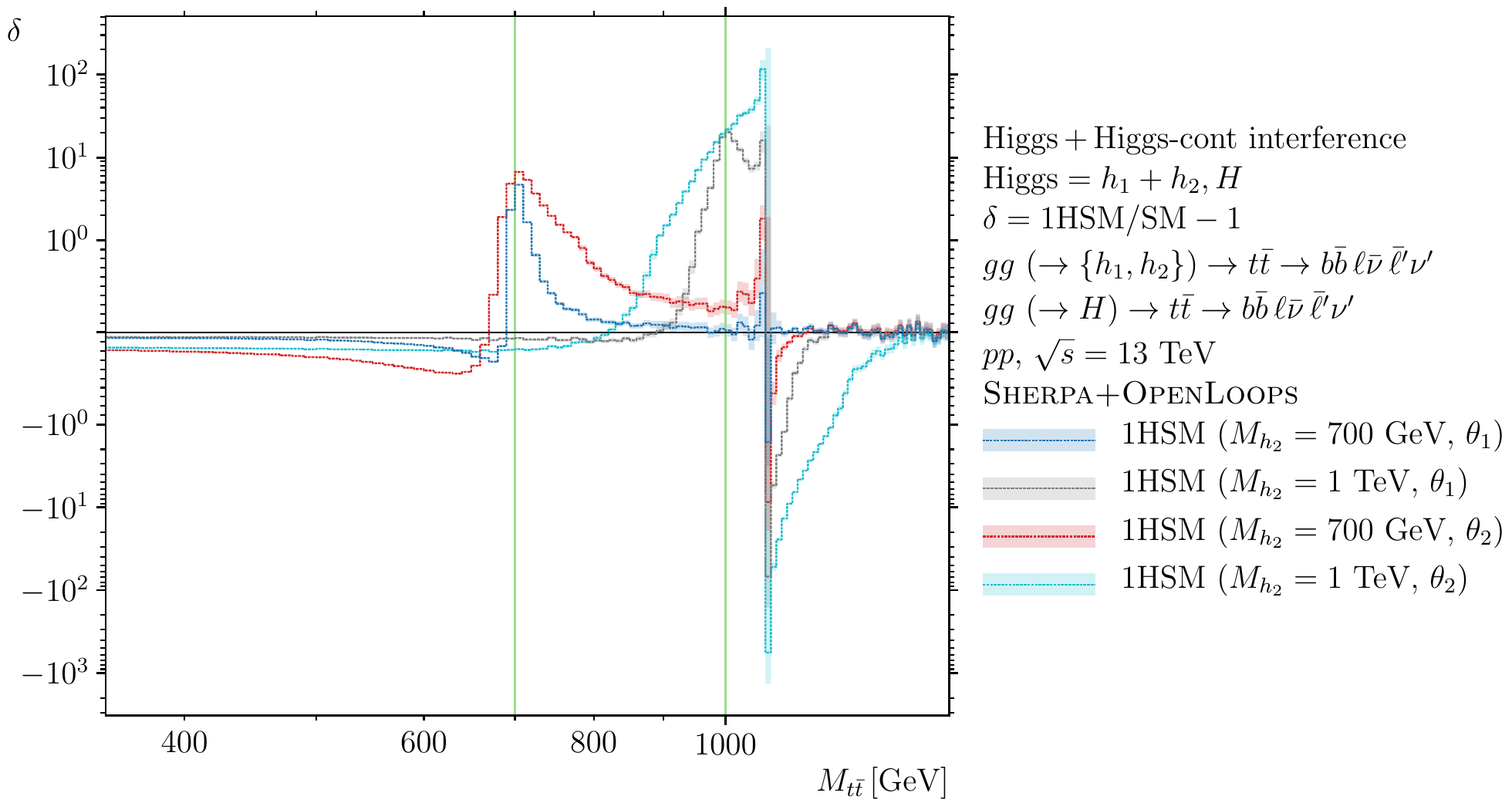}
\caption{\label{fig:1hsm_tt_si_h12-1256-sm_tt_si_-delta} $M_{t\bar{t}}$ distributions of the relative deviation $\delta=R-1$ of the Higgs cross section including its interference with the background in the 1HSM ($M_{h_2}=\{700,1000\}$~GeV, $\{\theta_1,\theta_2\}$) compared to the SM for $gg\ (\to \{h_1,h_2,H\}) \to t\bar{t} \to b\bar{b}\,\ell\bar{\nu}\,\bar{\ell}^\prime \nu^\prime$ in $pp$ collisions at $\sqrt{s}=13$~TeV.  $R$ is the ratio of $\sigma(\text{$h_{1+2}$+I(C)})$ to $\sigma(\text{$H$+I(C)})$.  Other details as in \rfig{fig:ww_1hsm3_mass}.}
\end{figure}
%
%
%
\begin{figure}[tbp]
\vspace{0.cm}
\centering
\includegraphics[width=\textwidth, clip=true]{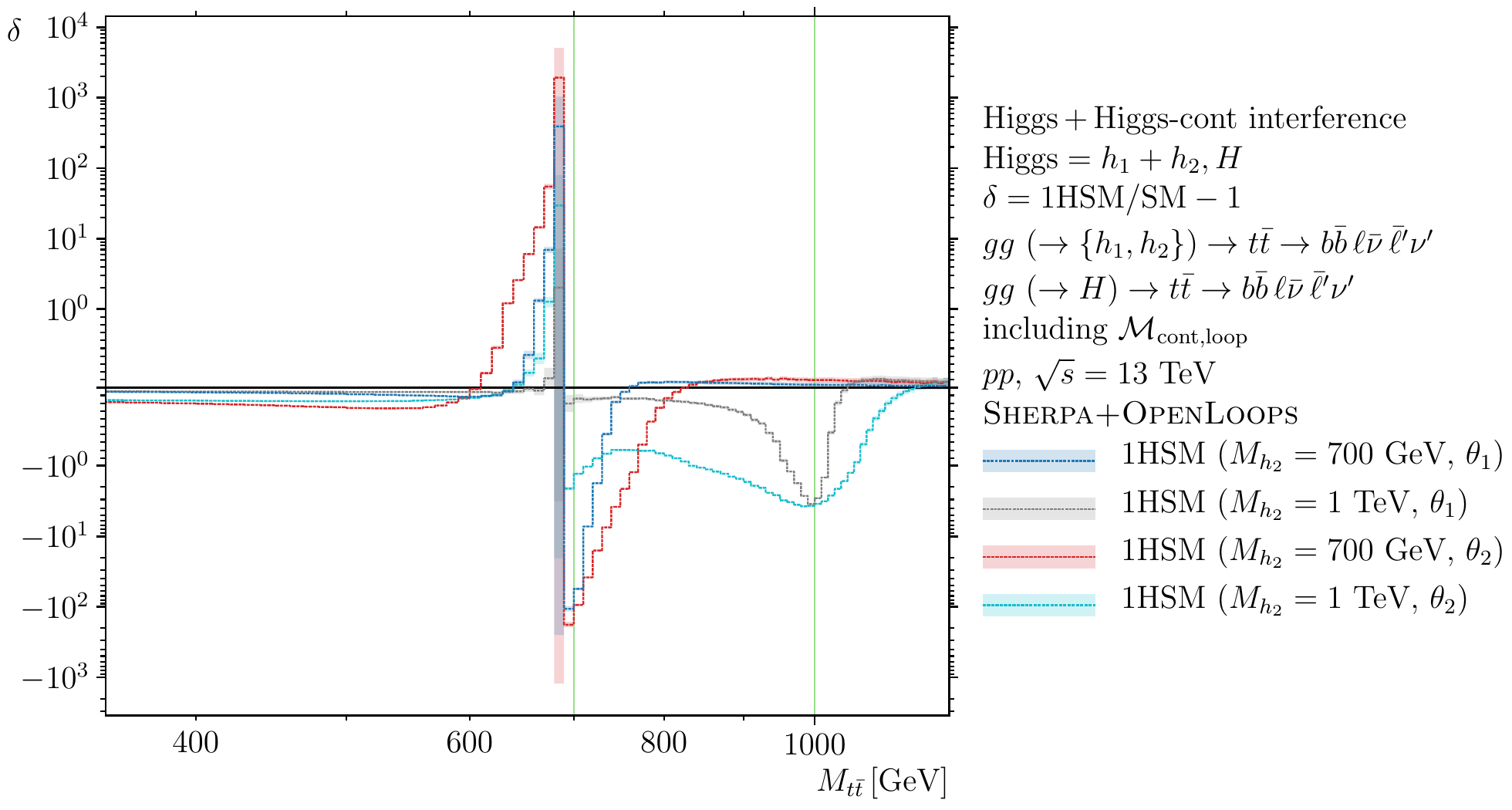}
\caption{\label{fig:1hsm_tt_si_h12bbv-1256-sm_tt_si_bbv-delta} $M_{t\bar{t}}$ distributions of the relative deviation $\delta=R-1$ of the Higgs cross section including its interference with the background in the 1HSM ($M_{h_2}=\{700,1000\}$~GeV, $\{\theta_1,\theta_2\}$) compared to the SM for $gg\ (\to \{h_1,h_2,H\}) \to t\bar{t} \to b\bar{b}\,\ell\bar{\nu}\,\bar{\ell}^\prime \nu^\prime$ in $pp$ collisions at $\sqrt{s}=13$~TeV.  $R$ is the ratio of $\sigma(\text{$h_{1+2}$+I(C$_{+\circlearrowleft}$)})$ to $\sigma(\text{$H$+I(C$_{+\circlearrowleft}$)})$.  Other details as in \rfig{fig:ww_1hsm3_mass}.}
\end{figure}
%
%
%
\begin{figure}[tbp]
\vspace{0.cm}
\centering
\includegraphics[width=\textwidth, clip=true]{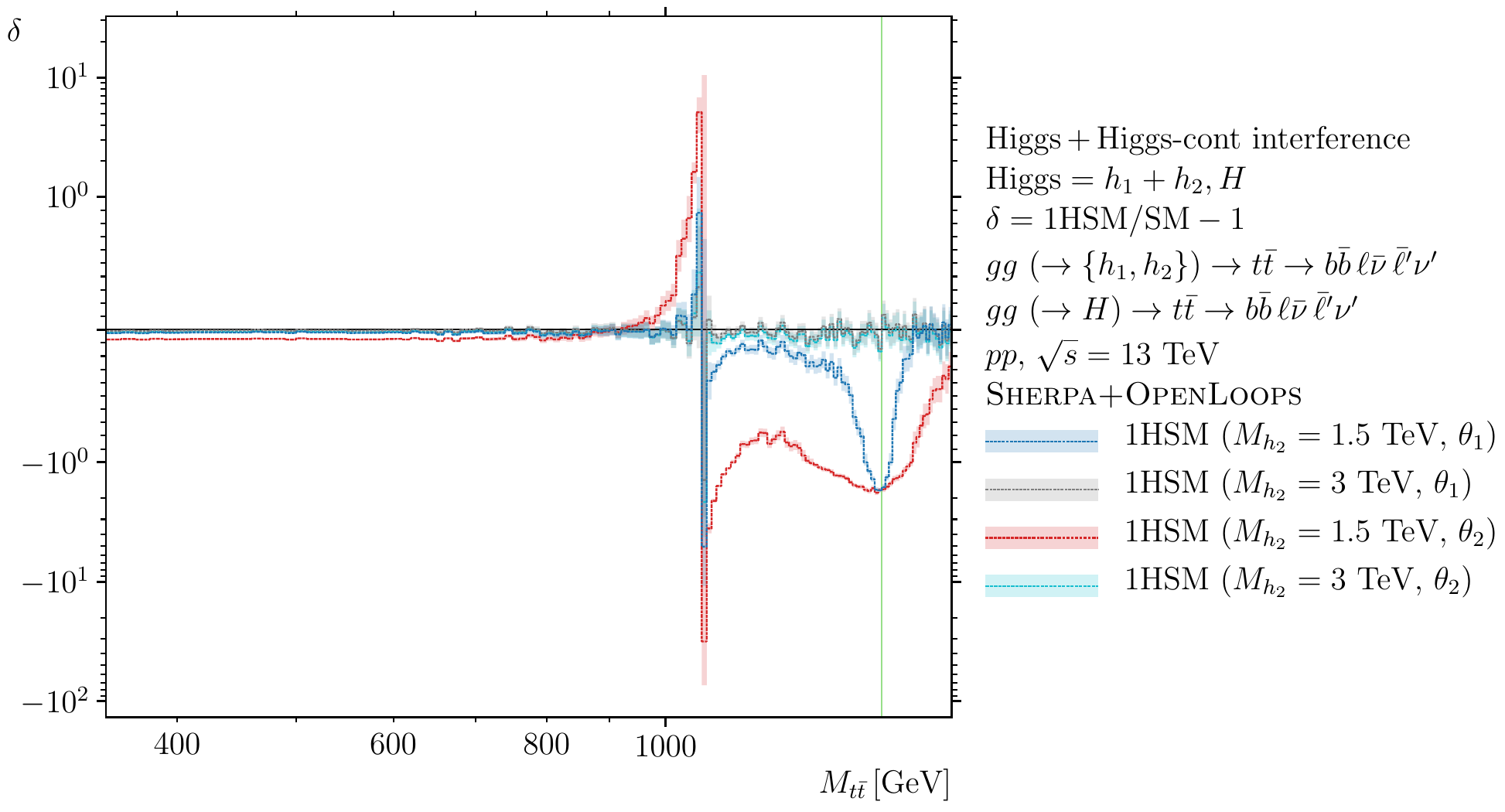}
\caption{\label{fig:1hsm_tt_si_h12-3478-sm_tt_si_-delta} $M_{t\bar{t}}$ distributions of the relative deviation $\delta=R-1$ of the Higgs cross section including its interference with the background in the 1HSM ($M_{h_2}=\{1.5,3\}$~TeV, $\{\theta_1,\theta_2\}$) compared to the SM for $gg\ (\to \{h_1,h_2,H\}) \to t\bar{t} \to b\bar{b}\,\ell\bar{\nu}\,\bar{\ell}^\prime \nu^\prime$ in $pp$ collisions at $\sqrt{s}=13$~TeV.  $R$ is the ratio of $\sigma(\text{$h_{1+2}$+I(C)})$ to $\sigma(\text{$H$+I(C)})$.  Other details as in \rfig{fig:ww_1hsm3_mass}.}
\end{figure}
%
%
%
\begin{figure}[tbp]
\vspace{0.cm}
\centering
\includegraphics[width=\textwidth, clip=true]{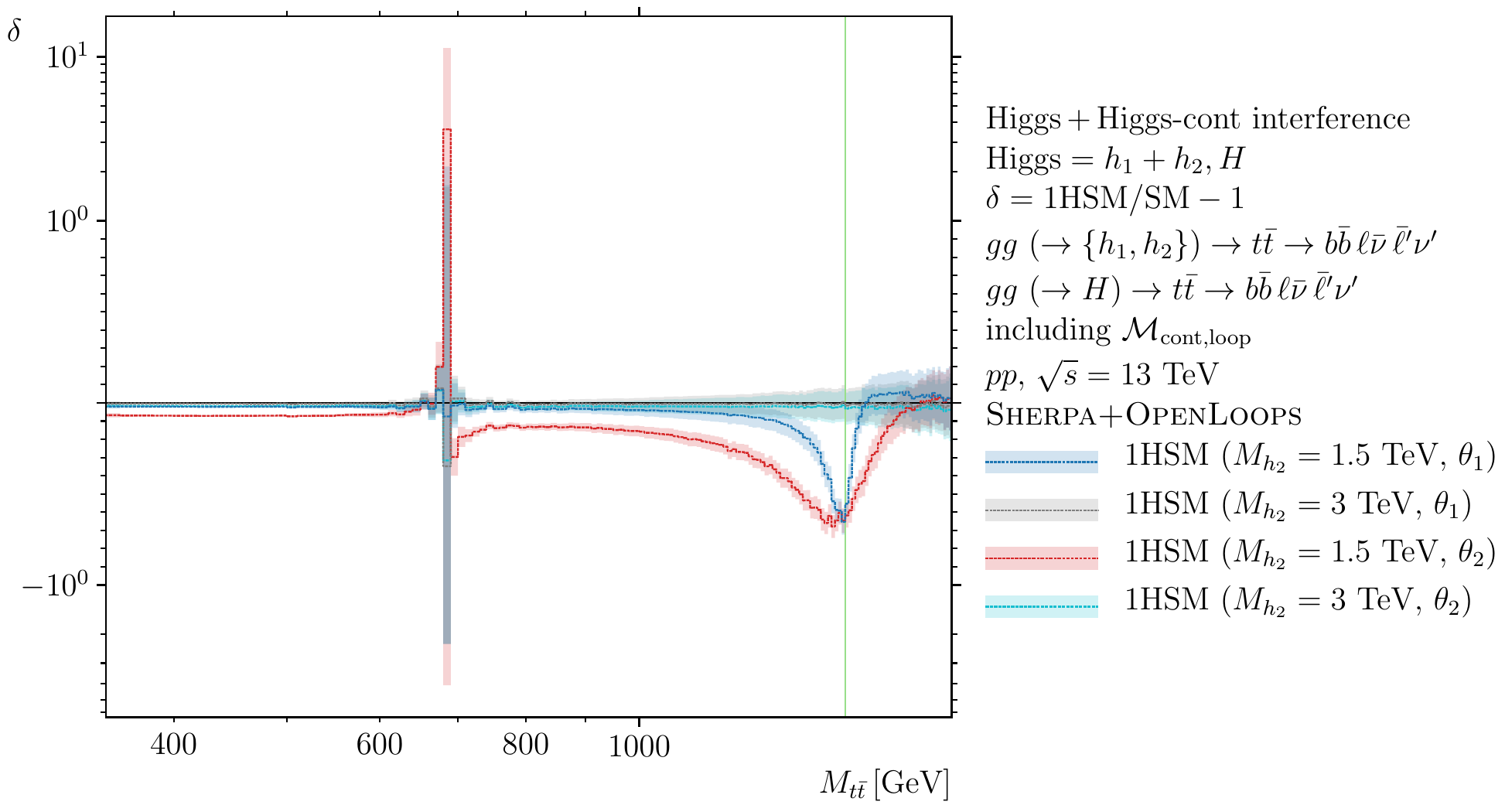}
\caption{\label{fig:1hsm_tt_si_h12bbv-3478-sm_tt_si_bbv-delta} $M_{t\bar{t}}$ distributions of the relative deviation $\delta=R-1$ of the Higgs cross section including its interference with the background in the 1HSM ($M_{h_2}=\{1.5,3\}$~TeV, $\{\theta_1,\theta_2\}$) compared to the SM for $gg\ (\to \{h_1,h_2,H\}) \to t\bar{t} \to b\bar{b}\,\ell\bar{\nu}\,\bar{\ell}^\prime \nu^\prime$ in $pp$ collisions at $\sqrt{s}=13$~TeV.  $R$ is the ratio of $\sigma(\text{$h_{1+2}$+I(C$_{+\circlearrowleft}$)})$ to $\sigma(\text{$H$+I(C$_{+\circlearrowleft}$)})$.  Other details as in \rfig{fig:ww_1hsm3_mass}.}
\end{figure}
%
%
%
\begin{figure}[tbp]
\vspace{0.cm}
\centering
\includegraphics[width=\textwidth, clip=true]{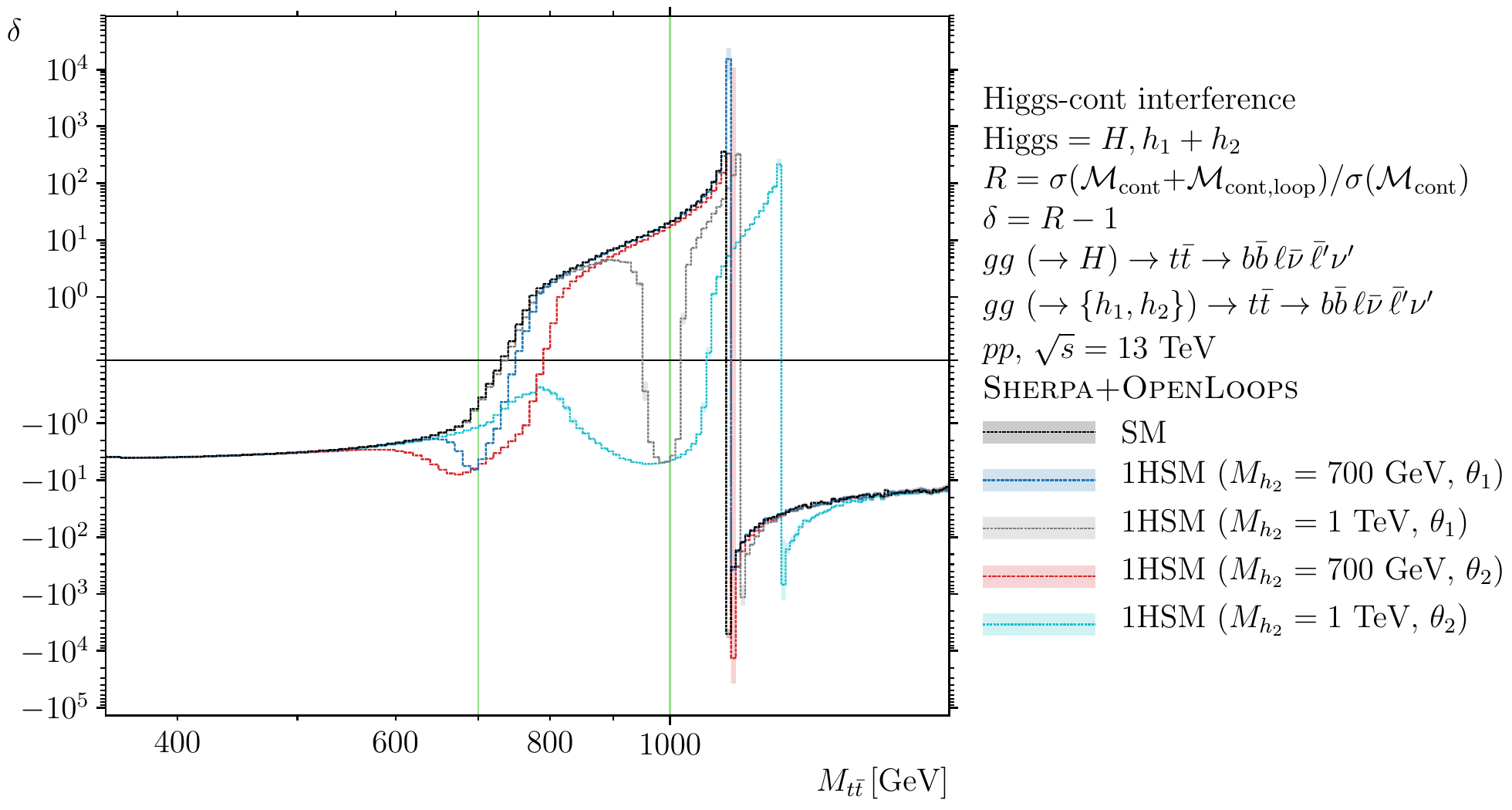}
\caption{\label{fig:tt_1hsm1256_mass-i_Hh12bv-div-i_Hh12b} $M_{t\bar{t}}$ distributions of the relative deviation $\delta=R-1$ of the Higgs interference with the background without and with the virtual corrections ($\mathcal{M}_\mathrm{cont,loop}$) in the SM and 1HSM ($M_{h_2}=\{700,1000\}$~GeV, $\{\theta_1,\theta_2\}$) for $gg\ (\to \{h_1,h_2,H\}) \to t\bar{t} \to b\bar{b}\,\ell\bar{\nu}\,\bar{\ell}^\prime \nu^\prime$ in $pp$ collisions at $\sqrt{s}=13$~TeV.  $R$ is the ratio of $\sigma(\text{I($H$,C$_{+\circlearrowleft}$)})$ to $\sigma(\text{I($H$,C)})$ and $\sigma(\text{I($h_{1+2}$,C$_{+\circlearrowleft}$)})$ to $\sigma(\text{I($h_{1+2}$,C)})$ in the SM and 1HSM, respectively.  Other details as in \rfig{fig:ww_1hsm3_mass}.}
\end{figure}
%
%
%
\begin{figure}[tbp]
\vspace{0.cm}
\centering
\includegraphics[width=\textwidth, clip=true]{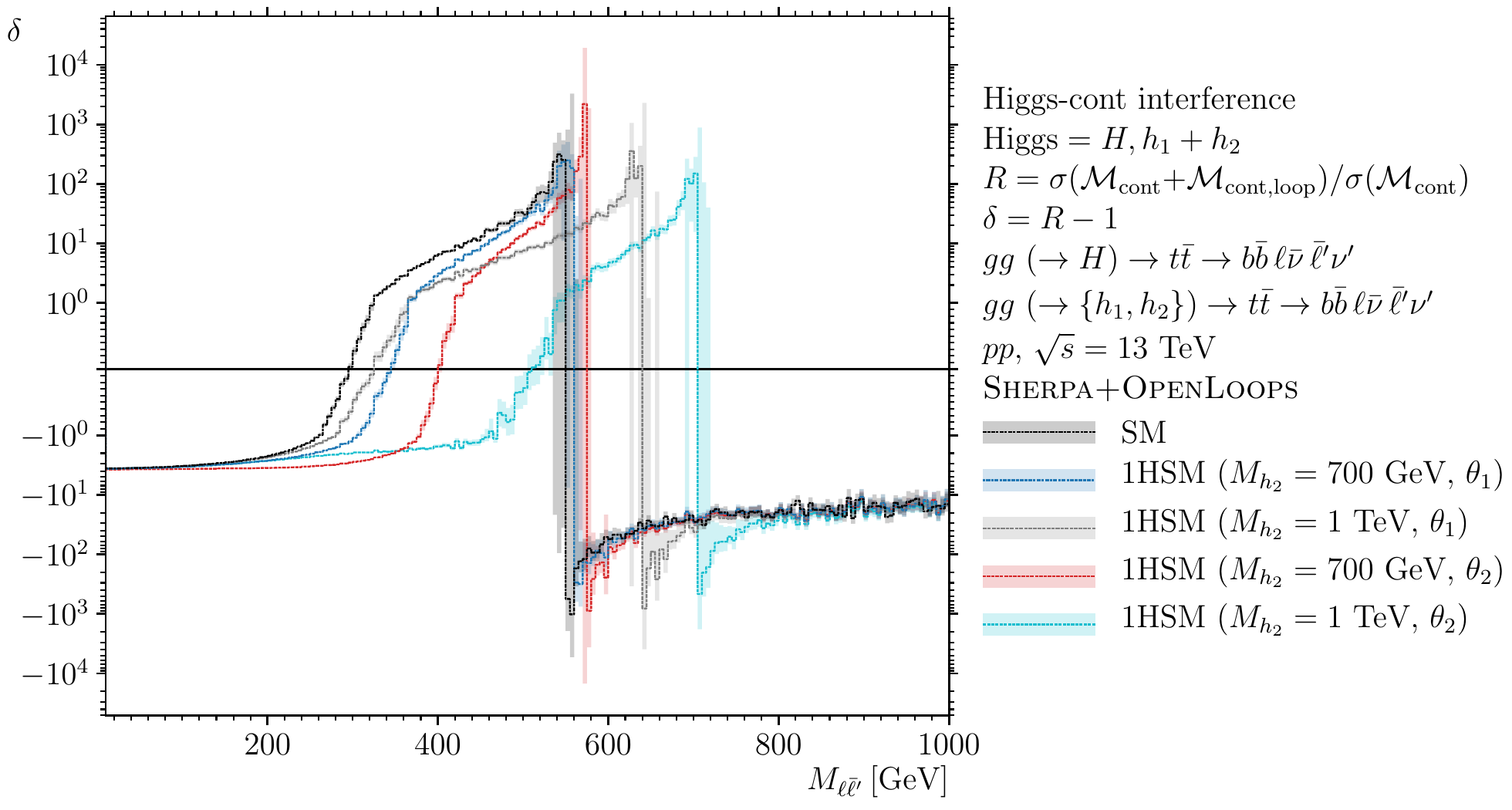}
\caption{\label{fig:tt_1hsm1256_mll-i_Hh12bv-div-i_Hh12b} $M_{\ell\bar{\ell}^\prime}$ distributions of the relative deviation $\delta=R-1$ of the Higgs interference with the background without and with the virtual corrections in the SM and 1HSM ($M_{h_2}=\{700,1000\}$~GeV, $\{\theta_1,\theta_2\}$) for $gg\ (\to \{H,h_1,h_2\}) \to t\bar{t} \to b\bar{b}\,\ell\bar{\nu}\,\bar{\ell}^\prime \nu^\prime$ in $pp$ collisions at $\sqrt{s}=13$~TeV.  Other details as in \rfig{fig:tt_1hsm1256_mass-i_Hh12bv-div-i_Hh12b}.}
\end{figure}
%
%
%
\begin{figure}[tbp]
\vspace{0.cm}
\centering
\includegraphics[width=\textwidth, clip=true]{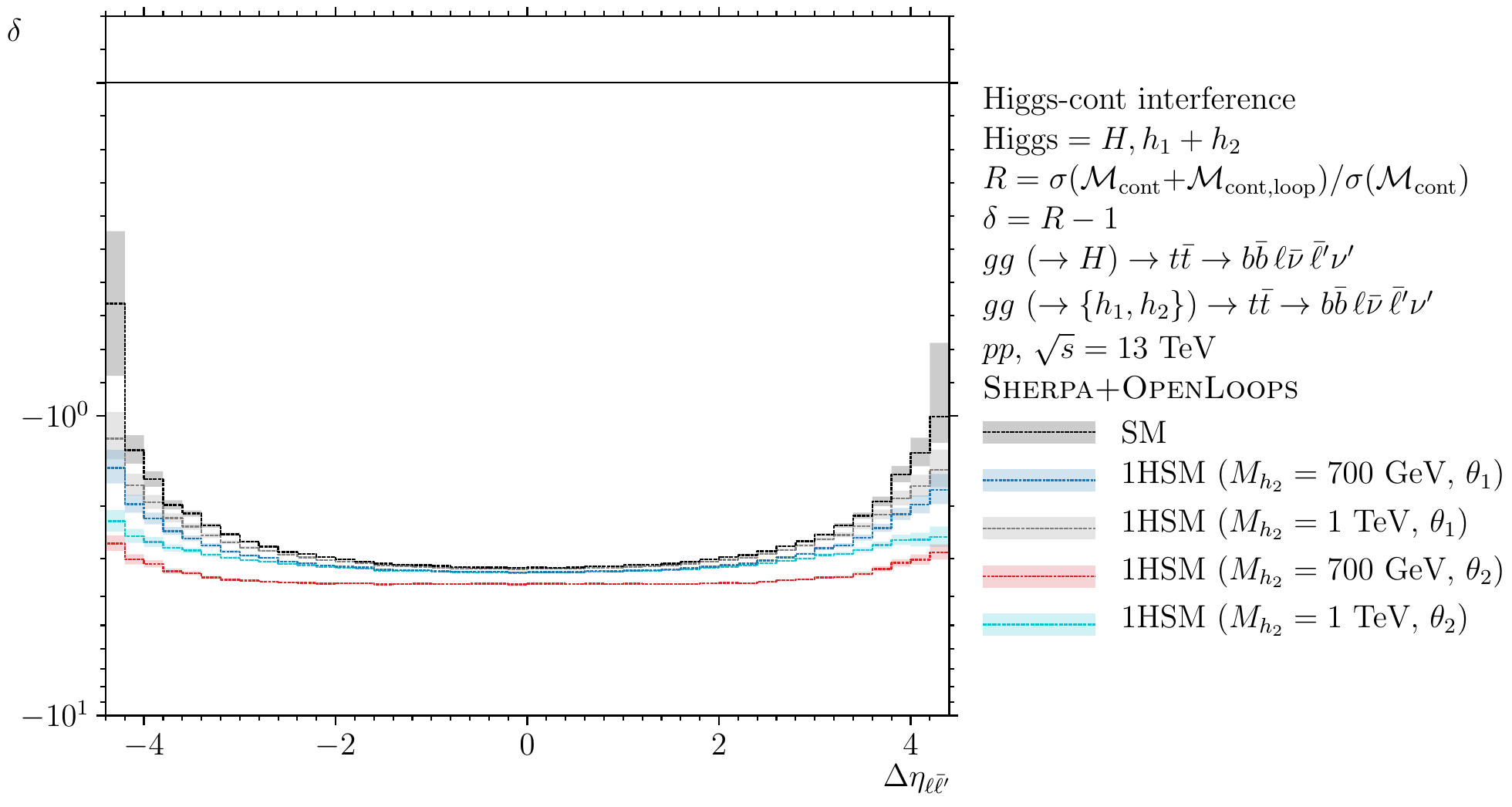}
\caption{\label{fig:tt_1hsm1256_deltaEtall-i_Hh12bv-div-i_Hh12b} $\Delta\eta_{\ell\bar{\ell}^\prime}$ distributions of the relative deviation $\delta=R-1$ of the Higgs interference with the background without and with the virtual corrections in the SM and 1HSM ($M_{h_2}=\{700,1000\}$~GeV, $\{\theta_1,\theta_2\}$) for $gg\ (\to \{H,h_1,h_2\}) \to t\bar{t} \to b\bar{b}\,\ell\bar{\nu}\,\bar{\ell}^\prime \nu^\prime$ in $pp$ collisions at $\sqrt{s}=13$~TeV.  Other details as in \rfig{fig:tt_1hsm1256_mass-i_Hh12bv-div-i_Hh12b}.}
\end{figure}
%
%
%
\begin{figure}[tbp]
\vspace{0.cm}
\centering
\includegraphics[width=\textwidth, clip=true]{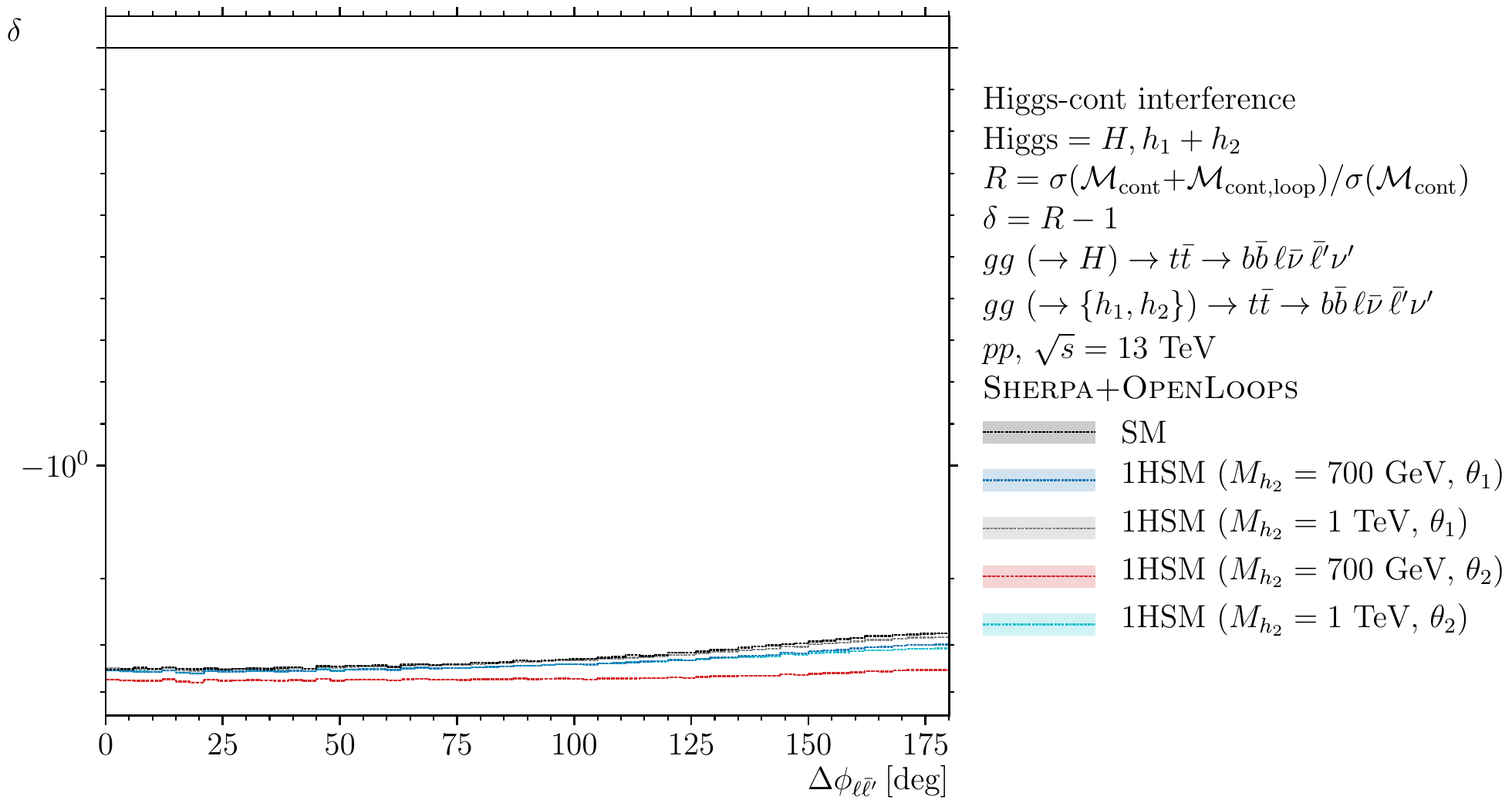}
\caption{\label{fig:tt_1hsm1256_deltaPhill-i_Hh12bv-div-i_Hh12b} $\Delta\phi_{\ell\bar{\ell}^\prime}$ distributions of the relative deviation $\delta=R-1$ of the Higgs interference with the background without and with the virtual corrections in the SM and 1HSM ($M_{h_2}=\{700,1000\}$~GeV, $\{\theta_1,\theta_2\}$) for $gg\ (\to \{H,h_1,h_2\}) \to t\bar{t} \to b\bar{b}\,\ell\bar{\nu}\,\bar{\ell}^\prime \nu^\prime$ in $pp$ collisions at $\sqrt{s}=13$~TeV.  Other details as in \rfig{fig:tt_1hsm1256_mass-i_Hh12bv-div-i_Hh12b}.}
\end{figure}


\section{Discussion\label{sec:discussion}}

The Sq($h_{1+2}$) integrated cross sections displayed in \rtab{tab:res_ww} for $WW$ production in the 1HSM exhibit a relatively small deviation from the SM cross section Sq($H$), which ranges from $-5\%$ to $-0.05\%$ for $M_{h_2}=1$~TeV and $M_{h_2}=3$~TeV, respectively, and the mixing angle $\theta_1$ ($\theta_1\lesssim 0.2$, see \rtab{tab:benchmark}).  Increasing, for illustration,\footnote{%
Note that $\theta_2$ is no longer compatible with experimental bounds.}
the mixing angle to $\theta_2\lesssim 0.4$, the cross section deviation range increases to $-16\%$ to $-2\%$ with corresponding heavy Higgs masses.  When adding the continuum background interference, in the 1HSM and SM the cross section is reduced uniformly by a factor close to $0.76$.  Since in \rtab{tab:res_ww} the $|\mathcal{M}_{h_1}+\mathcal{M}_{h_2}|^2$ and $|\mathcal{M}_H|^2$ Higgs cross sections are compared, due to unitarity constraints it is not surprising that cross section deviations are small and the impact of the interference is uniform.

In \rtab{tab:res_ww_h2}, we show for $WW$ production how interference affects the integrated heavy Higgs resonance cross section Sq($h_2$).  Due to the falling gluon PDF and the decreasing value of $\theta_{1,2}$ for $M_{h_2}=\{1.5,3\}$~TeV (see \rtab{tab:benchmark}), Sq($h_2$) decreases rapidly with increasing $M_{h_2}$ and, as expected, is roughly a factor $3$--$5$ higher for the mixing angle $\theta_2<\pi/4$, which is larger than $\theta_1>0$.  The heavy Higgs cross section Sq($h_2$) is drastically altered when taking into account the interference with the light Higgs I($h_1$,$h_2$), because the light Higgs cross section Sq($h_1$) is significantly larger than Sq($h_2$) throughout (see \rtab{tab:app_ww_1hsm} in \rapp{app:results}).  As seen in \rtab{tab:res_ww_h2}, the cross section ratio ($h_2$+I($h_1$))/Sq($h_2$) ranges from $0.527$ ($0.625$) to $-604$ ($-461$) when $M_{h_2}$ increases from $700$~GeV to $3$~TeV with mixing angle $\theta_1$ ($\theta_2$).  When comparing the integrated cross sections Sq($h_2$), $h_2$+I($h_1$) and $h_2$+I(C+$h_1$), where the heavy Higgs-continuum background interference has also been added in the third quantity, it is apparent that the heavy Higgs-light Higgs interference I($h_1$,$h_2$) and I($h_1$,C) always have opposite signs (see also \rtab{tab:app_ww_1hsm} in \rapp{app:results}), which results in a substantial reduction of the interference impact on the heavy resonance in $WW$ production.  This can be seen in \rtab{tab:res_ww_h2}: the cross section ratio ($h_2$+I(C+$h_1$))/Sq($h_2$) only ranges from $1.228$ ($1.261$) to $65$ ($66$) when $M_{h_2}$ increases from $700$~GeV to $3$~TeV with mixing angle $\theta_1$ ($\theta_2$).

Analogous results, but for $t\bar{t}$ production extended with the one-loop ($\circlearrowleft$) continuum background amplitude, are presented in \rtabs{tab:res_tt} and \ref{tab:res_tt_h2}.  The Sq($h_{1+2}$) integrated cross sections in the 1HSM displayed in \rtab{tab:res_tt} exhibit a deviation from the SM cross section Sq($H$) ranging from $-12\%$ ($-35\%$) to $-1\%$ ($-4\%$) for $M_{h_2}=1$~TeV and $M_{h_2}=3$~TeV, respectively, and the mixing angle $\theta_1$ ($\theta_2$).  We note that the extreme deviations occur for the same values of $M_{h_2}$ for $t\bar{t}$ and $WW$ production and that the deviations are larger in $t\bar{t}$ than in $WW$ production.  When adding the interference with the tree-level continuum background, in the 1HSM and SM the cross section changes by a factor of approximately $-40$, i.e.\ the interference at tree level is negative and about $40$ times larger than the heavy resonance cross section.  When the interference with the one-loop continuum background is included, the result changes sign and is at least twice as large.  This implies that the interference with the one-loop continuum background is at least three times larger than the tree-level interference, with opposite sign.  Already at the integrated cross section level, it is therefore important to include all one-loop contributions to obtain reliable signal plus interference results.

In \rtab{tab:res_tt_h2}, we show for $t\bar{t}$ production how interference affects the heavy Higgs cross section Sq($h_2$).  As before, Sq($h_2$) decreases rapidly with increasing $M_{h_2}$ and, as expected, is roughly a factor $3$--$10$ higher for $\theta_2$ than for $\theta_1$.  The heavy Higgs cross section Sq($h_2$) is substantially or even drastically altered when taking into account the interference with the light Higgs I($h_1$,$h_2$), because the $t\bar{t}$ light Higgs cross section Sq($h_1$) is much larger than Sq($h_2$) (see \rtab{tab:app_tt_1hsm} in \rapp{app:results}).
The cross section ratio ($h_2$+I($h_1$))/Sq($h_2$) ranges from $0.400$ ($0.527$) to $-5.92\times 10^{3}$ ($-1870$) when $M_{h_2}$ increases from $700$~GeV to $3$~TeV with mixing angle $\theta_1$ ($\theta_2$), where the ratio is negative for $M_{h_2}\gtrsim 1$~TeV.  When comparing the integrated cross sections Sq($h_2$), $h_2$+I($h_1$) and $h_2$+I(C+$h_1$), where the heavy Higgs-(tree-level-)continuum background interference has been added in the third quantity, it is apparent that the heavy Higgs-light Higgs interference I($h_1$,$h_2$) and I($h_2$,C) typically have opposite signs (see also \rtab{tab:app_tt_1hsm} in \rapp{app:results}). In contrast to $WW$ production, despite the opposite sign, the result is a strong increase of the interference impact on the heavy resonance for $M_{h_2}\gtrsim 1$~TeV.  As seen in \rtab{tab:res_tt_h2}, the cross section ratio ($h_2$+I(C+$h_1$))/Sq($h_2$) ranges from $26.3$ ($17.6$) to $7.36\times 10^{4}$ ($2.07\times 10^{4}$) when $M_{h_2}$ increases from $1$ to $3$~TeV with mixing angle $\theta_1$ ($\theta_2$).  Furthermore, the rightmost column of \rtab{tab:res_tt_h2} demonstrates that it is essential to take into account the interference with the one-loop continuum background.  The cross section ratio ($h_2$+I(C$_{+\circlearrowleft}$+$h_1$))/Sq($h_2$) ranges from $45.80$ ($45.2$) to $-9.7\times 10^{4}$ ($-1.47\times 10^{4}$) when $M_{h_2}$ increases from $1$ to $3$~TeV with mixing angle $\theta_1$ ($\theta_2$).  In all studied cases, the inclusion of the one-loop continuum background changes the cross section substantially or even drastically.

Additional insight is gained by studying differential distributions.  For $WW$ production, the $M_{WW}$ distribution shown in \rfig{fig:ww_1hsm3_mass} illustrates several characteristics.  First, as expected, $d\sigma(|\mathcal{M}_{h_1}|^2)$ resembles $d\sigma(|\mathcal{M}_H|^2)$ (see \rfig{fig:ww_sm_mass_low}).  Comparing the same figures, one also finds that, as expected, $d\sigma(2\mathrm{Re}(\mathcal{M}^\ast_{h_1}\mathcal{M}_\mathrm{cont}))$ and $d\sigma(2\mathrm{Re}(\mathcal{M}^\ast_H\mathcal{M}_\mathrm{cont}))$ have the same shape.  Secondly, \rfig{fig:ww_1hsm3_mass} illustrates that $d\sigma(2\mathrm{Re}(\mathcal{M}^\ast_{h_2}\mathcal{M}_\mathrm{cont}))$ and $d\sigma(2\mathrm{Re}(\mathcal{M}^\ast_{h_2}\mathcal{M}_{h_1}))$ have opposite sign behaviour with respect to $M_{WW}=M_{h_2}$.  Furthermore, the sign behaviour of $d\sigma(2\mathrm{Re}(\mathcal{M}^\ast_{h_{1/2}}\mathcal{M}_\mathrm{cont}))$ at $M_{WW}=M_{h_{1/2}}$ is identical to the sign behaviour of $d\sigma(2\mathrm{Re}(\mathcal{M}^\ast_H\mathcal{M}_\mathrm{cont}))$ at $M_{WW}=M_H$, which is dictated by unitarity cancellations at high energy.  $d\sigma(2\mathrm{Re}(\mathcal{M}^\ast_{h_2}(\mathcal{M}_\mathrm{cont}+\mathcal{M}_\mathrm{h_1})))$ illustrates the compensation between the two types of interference for the heavy resonance, which was discussed for \rtab{tab:res_ww_h2} above, at the differential level.  We note the strong interference reduction in the vicinity of the $h_2$ peak.  But, the mitigating effect of the heavy Higgs-light Higgs interference decreases steadily down to the $WW$ threshold.  The $M_{T,WW}$ distribution shown in \rfig{fig:ww_1hsm3_mass_trans1} is related to the $M_{WW}$ distribution discussed above by the fact that $M_{T,WW}\leq M_{WW}$ is guaranteed for the $M_{T,WW}$ definition used here.  
The $M_{\ell\bar{\ell}^\prime}$ distribution displayed in \rfig{fig:ww_1hsm3_mll} demonstrates that the interference impact for the heavy Higgs signal is largest for $M_{\ell\bar{\ell}^\prime}\lesssim 150$~GeV and decreases continuously for higher dilepton invariant masses.
The $\Delta\eta_{\ell\bar{\ell}^\prime}$, $\Delta\phi_{\ell\bar{\ell}^\prime}$ and $\Delta R_{\ell\bar{\ell}^\prime}$ distributions displayed in \rfigs{fig:ww_1hsm3_deltaEtall}, \ref{fig:ww_1hsm3_deltaPhill} and \ref{fig:ww_1hsm3_deltaRll}, respectively, illustrate that the interference impact for the heavy Higgs signal is large except for approximately back-to-back dilepton configurations.  As small dilepton opening angles are typically selected in Higgs $\to WW$ searches \cite{Dittmar:1996ss}, this implies that the angular dependence of interference effects is important and should be taken into account in such studies.

For $t\bar{t}$ production, the $M_{t\bar{t}}$ and $M_{T,WW}$ distributions are shown in \rfigs{fig:tt_1hsm1_mass} and \ref{fig:tt_1hsm1_mass_trans1}, respectively.  Comparing the $M_{t\bar{t}}$ distributions in the 1HSM (\rfig{fig:tt_1hsm1_mass}) and the SM (\rfig{fig:tt_sm_mass_low}, see also \rfig{fig:tt_sm_mass_trans1_low}) yields: First, in analogy to $WW$ production, shape agreement is found when $h_1$-dependent 1HSM cross sections are compared with the corresponding $H$-dependent SM cross sections.  Secondly, the same pattern for the sign behaviour of $d\sigma(2\mathrm{Re}(\mathcal{M}^\ast_{h_2}\mathcal{M}_\mathrm{cont}))$ and $d\sigma(2\mathrm{Re}(\mathcal{M}^\ast_{h_2}\mathcal{M}_{h_1}))$ is found as in $WW$ production.  As new feature, the typically dominant impact of the one-loop continuum background amplitude on the $\mathcal{M}_\mathrm{cont}$-dependent distributions is clearly demonstrated in \rfigs{fig:tt_1hsm1_mass} and \ref{fig:tt_1hsm1_mass_trans1}.  (In \rfigs{fig:tt_sm_mass_low} and \ref{fig:tt_sm_mass_trans1_low}, the same is demonstrated for the corresponding SM distributions.)  In these figures and all other $t\bar{t}$ distributions, it is apparent that interference is the leading cross section contribution and the Higgs resonance cross section is subleading.  The $M_{\ell\bar{\ell}^\prime}$, $\Delta\eta_{\ell\bar{\ell}^\prime}$, $\Delta\phi_{\ell\bar{\ell}^\prime}$ and $\Delta R_{\ell\bar{\ell}^\prime}$ distributions displayed in \rfigs{fig:tt_1hsm1_mll}, \ref{fig:tt_1hsm1_deltaEtall}, \ref{fig:tt_1hsm1_deltaPhill} and \ref{fig:tt_1hsm1_deltaRll}, respectively, 
confirm both statements for the dilepton invariant mass and angular observables.

Employing the Higgs invariant mass distribution and considering all benchmark points, for $WW$ production in \rfigs{fig:1hsm_ww_si_h12-1256-sm_ww_si_-delta} and \ref{fig:1hsm_ww_si_h12-3478-sm_ww_si_-delta} and for $t\bar{t}$ production without and with one-loop continuum background amplitude in \rfigs{fig:1hsm_tt_si_h12-1256-sm_tt_si_-delta} and \ref{fig:1hsm_tt_si_h12-3478-sm_tt_si_-delta} and \rfigs{fig:1hsm_tt_si_h12bbv-1256-sm_tt_si_bbv-delta} and \ref{fig:1hsm_tt_si_h12bbv-3478-sm_tt_si_bbv-delta}, respectively, the relative deviation of the Higgs cross section in the 1HSM and SM ($|\mathcal{M}_{h_1}+\mathcal{M}_{h_2}|^2$, $|\mathcal{M}_H|^2$) including interference with the continuum background is shown.  Near the heavy Higgs resonance and in extended neighbouring regions the deviation can be $\calO(100\%)$ or even significantly larger.  Elsewhere, the deviation is $\calO(10\%)$.

Similarly, for $t\bar{t}$ production in the SM and 1HSM various distributions shown in \rfigs{fig:tt_1hsm1256_mass-i_Hh12bv-div-i_Hh12b}, \ref{fig:tt_1hsm1256_mll-i_Hh12bv-div-i_Hh12b}, \ref{fig:tt_1hsm1256_deltaEtall-i_Hh12bv-div-i_Hh12b} and \ref{fig:tt_1hsm1256_deltaPhill-i_Hh12bv-div-i_Hh12b} illustrate the relative deviation of the interference cross section without and with one-loop continuum background amplitude at the differential level.  
For $M_{t\bar{t}}$ and $M_{\ell\bar{\ell}^\prime}$ the deviation significantly exceeds $100\%$ in large invariant mass regions.  For $\Delta\eta_{\ell\bar{\ell}^\prime}$ and $\Delta\phi_{\ell\bar{\ell}^\prime}$ the deviation is $\calO(2$--$4)$ and its differential variation is non-negligible, but less pronounced.


\section{Conclusions\label{sec:sum}}

A detailed study of Higgs interference effects at the one-loop level in the 1HSM was presented for the $WW$ and $t\bar{t}$ decay modes with fully leptonic $WW$ decay.  We calculated with massive top and bottom quarks and explored interference effects for benchmark points with a heavy Higgs mass that significantly exceeds $2m_t$.  More specifically, the $M_{h_2}$ range 700--3000~GeV was studied with corresponding mixing angles compatible with current limits as well as a second set of mixing angles, roughly twice as large, to illustrate the dependence on the mixing angle.  In the $WW$ channel, the Higgs signal and the interfering continuum background are loop induced.  In the $t\bar{t}$ channel, which features a tree-level background, we also calculated the interference with the one-loop background (applying the NWA to $t$ and $\bar{t}$), which, due to the appearance of the absorptive part, was found to dominate the studied distributions.  More generally, our results indicate that NLO interference contributions substantially change the normalisation and shape of BSM and SM differential Higgs cross section distributions in invariant-mass as well as angular kinematic variables.  This can be understood via the appearance of a non-trivial phase that is caused by loop corrections to the continuum background.  Full NLO corrections are therefore essential and, when available, should be taken into account in all interference-affected experimental searches for heavy Higgs resonances.  We conjecture that the same applies to searches for other heavy resonances.  As corollary, we find that the commonly used geometric average $K$-factor approximation $K_\text{interference}\approx (K_\text{Higgs}K_\text{background})^{1/2}$ is not appropriate.

Finally, we note that our 1HSM and SM implementation in \p{Sherpa+OpenLoops}, which can be used as parton-level integrator or event generator, is included in the \href{http://arxiv.org/}{arXiv} submission as ancillary file.  Supplementary figures with distributions for all studied quantities, models and benchmark points are available as Web download.


\acknowledgments
We are grateful to Silvan Kuttimalai and Stefan H\"oche for useful suggestions and \p{Sherpa} and \p{Matplotlib} support.
N.K.\ would like to thank CERN for hospitality and partial financial support through the CERN Theory Institute on \emph{LHC and the Standard Model: Physics and Tools} 
and the general assembly meetings of the LHC Higgs Cross Section Working Group.  N.K.\ is also grateful to C.\ Anastasiou and V.\ Hirschi for hospitality at ETH Zurich as well as access to the RHUL Physics HPC cluster.  A.L.\ would like to thank the Department of Physics, Royal Holloway, University of London, for supplementary financial support and the Theoretical Particle Physics Group, University of Sussex, for hospitality and access to their computer cluster. This work was supported in part by STFC grant ST/P000738/1 and by the U.S.\ Department of Energy under contract number DE-AC02-76SF00515.  W.S.\ has received funding from the European Union's Horizon 2020 research and innovation programme under the Marie Sk\l{}odowska-Curie grant agreement (GA) No 665593 awarded to the Science and Technology Facilities Council.  Feynman graphs were drawn with \p{JaxoDraw} \cite{Binosi:2003yf}.  We are grateful to J.~Collins for providing a remedy for a known JaxoDraw flaw.  Plots were created with \p{Rivet} \cite{Buckley:2010ar} and \p{Matplotlib} \cite{Hunter:2007}.


\clearpage

\appendix


\section{\boldmath Partial decay widths for\texorpdfstring{\ \,$h_2\:\to\:h_1 h_1,\ h_1 h_1 h_1,\ h_1 h_1 h_1 h_1$}{ h_2 --> h_1 h_1, h_1 h_1 h_1, h_1 h_1 h_1 h_1} \label{app:partwidths}}

The partial decay widths for $h_2\to h_1 h_1$, $h_2\to h_1 h_1 h_1$ and $h_2\to h_1 h_1 h_1 h_1$ for the considered benchmark points are given in \rtab{tab:partwidths}.%
\footnote{The absorptive part of the occurring $h_2$ propagators, which due to the kinematic constraints cannot be on-shell, was neglected.  We checked, by iterating once with the obtained values for $\Gamma_{h_2}$, that the resulting uncertainty ranges from 0.05\% to 5\%, depending on the benchmark point, and is hence generally small.}
\begin{table}[tbp]
\renewcommand{\arraystretch}{1.2}
\centering 
\begin{tabular}{|l|c|c|c|}
\hline 
\multicolumn{4}{|c|}{$\theta=\theta_1$} \\
\hline
$M_{h_2}$  [GeV]  & $\Gamma(h_2 \to h_1 h_1)$ [GeV] & $\Gamma(h_2 \to h_1 h_1 h_1)$ [GeV] & $\Gamma(h_2 \to h_1 h_1 h_1 h_1)$ [GeV] \\ \hline
700   &  2.1556(1)  &  0.00468(2)  &  $6.24(4) \times 10^{-7}$  \\ \hline
1000  &  6.0953(1)  &  0.1692(7)  &  0.001718(9)  \\ \hline
1500  &  9.8911(1)  &  0.218(2)  &  0.001632(8)  \\ \hline
3000  &  20.658(1)  &  0.306(2)  &  0.001060(7)  \\ \hline
\multicolumn{4}{c}{} \\
\hline 
\multicolumn{4}{|c|}{$\theta=\theta_2$} \\
\hline 
$M_{h_2}$  [GeV]  & $\Gamma(h_2 \to h_1 h_1)$ [GeV] & $\Gamma(h_2 \to h_1 h_1 h_1)$ [GeV] & $\Gamma(h_2 \to h_1 h_1 h_1 h_1)$ [GeV] \\ \hline
700   &  4.1798(1)  &  0.507(2)  &  0.01451(8)  \\ \hline
1000  &  11.604(1)  &  7.34(4)  &  2.46(2)  \\ \hline
1500  &  27.26(1)  &  12.9(2)  &  3.91(2)  \\ \hline
3000  &  66.8(1)  &  21.4(2)  &  4.17(2)  \\
\hline
\end{tabular}
\caption{\label{tab:partwidths} 
Partial decay widths for $h_2\to h_1 h_1$, $h_2\to h_1 h_1 h_1$ and $h_2\to h_1 h_1 h_1 h_1$. Other details as in \protect\rtabs{tab:benchmark} and \ref{tab:widths}.}
\end{table}


\section{\boldmath Nonredundant complete set of integrated results\label{app:results}}

In \rtabs{tab:app_ww_sm}--\ref{tab:app_tt_1hsm}, a nonredundant complete set of integrated results is given.

\begin{table}[tbp]
\vspace{0.cm}
\centering
\renewcommand{\arraystretch}{1.2}
\begin{tabular}{|c|c|}
\hline
\multicolumn{2}{|c|}{$gg\ (\to H) \to W^-W^+ \!\to \ell\bar{\nu}\,\bar{\ell}^\prime \nu^\prime$} \\ 
\multicolumn{2}{|c|}{SM, $pp$, $\sqrt{s}=13$ TeV} \\ 
\hline
$|\mathcal{M}|^2$ & $\sigma$ [fb] \\ 
\hline
Sq(C) & $27.616(7)$ \\
Sq($H$) & $13.689(4)$ \\
I($H$,C) & $-3.269(4)$ \\
\hline
\end{tabular}\\[0cm] 
\caption{\label{tab:app_ww_sm}
Cross sections for $gg\ (\to H) \to W^-W^+ \!\to \ell\bar{\nu}\,\bar{\ell}^\prime \nu^\prime$ in $pp$ collisions at $\sqrt{s}=13$~TeV in the Standard Model.  Mod-squared amplitude contributions are specified using the abbreviations defined in \rtab{tab:abbrev}. The selection cuts in \eqref{cuts} are applied.  Cross sections are given for a single lepton flavour combination.  The integration error is displayed in brackets.
} 
\end{table}
%
%
%
\begin{table}[tbp]
\vspace{0.cm}
\centering
\renewcommand{\arraystretch}{1.2}
\begin{tabular}{|c|c|c|c|c|c|}
\hline
\multicolumn{6}{|c|}{$gg\ (\to \{h_1,h_2\}) \to W^-W^+ \!\to \ell\bar{\nu}\,\bar{\ell}^\prime \nu^\prime$} \\ 
\multicolumn{6}{|c|}{$\sigma$ [fb], $pp$, $\sqrt{s}=13$ TeV} \\ 
\multicolumn{6}{|c|}{1HSM (see \rtabs{tab:benchmark} and \ref{tab:abbrev})} \\ 
\hline
\multirow{2}{*}{$|\mathcal{M}|^2$} & \multirow{2}{*}{$\theta$} & \multicolumn{4}{|c|}{$ M_{h_2}$  [GeV]} \\ \cline{3-6}
 &  & 700 & 1000 & 1500 & 3000 \\
\hline
\multirow{2}{*}{Sq(C)} & $\theta_1$ & $27.616(7)$ & $27.616(7)$ & $27.616(7)$ & $27.616(7)$ \\
 & $\theta_2$ & $27.616(7)$ & $27.616(7)$ & $27.616(7)$ & $27.616(7)$ \\
\hline
\multirow{2}{*}{Sq($h_1$)} & $\theta_1$ & $13.048(4)$ & $13.048(4)$ & $13.393(4)$ & $13.619(4)$ \\
 & $\theta_2$ & $11.541(4)$ & $11.541(4)$ & $12.697(4)$ & $13.436(4)$ \\
\hline
\multirow{2}{*}{Sq($h_2$)} & $\theta_1$ & $0.07810(2)$ & $0.010824(2)$ & $0.00027818(5)$ & $5.3026(9)\times 10^{-7}$ \\
 & $\theta_2$ & $0.27776(5)$ & $0.035182(6)$ & $0.0008885(2)$ & $2.3605(4)\times 10^{-6}$ \\
\hline
\multirow{2}{*}{I($h_1$,$h_2$)} & $\theta_1$ & $-0.03697(4)$ & $-0.02704(2)$ & $-0.006028(2)$ & $-0.00032061(7)$ \\
 & $\theta_2$ & $-0.1041(1)$ & $-0.07363(3)$ & $-0.017115(5)$ & $-0.0010893(3)$ \\
\hline
\multirow{2}{*}{I($h_1$,C)} & $\theta_1$ & $-3.132(3)$ & $-3.132(3)$ & $-3.205(4)$ & $-3.251(4)$ \\
 & $\theta_2$ & $-2.796(3)$ & $-2.796(3)$ & $-3.051(2)$ & $-3.221(4)$ \\
\hline
\multirow{2}{*}{I($h_2$,C)} & $\theta_1$ & $0.05478(5)$ & $0.03401(2)$ & $0.006963(3)$ & $0.00035468(8)$ \\
 & $\theta_2$ & $0.1765(2)$ & $0.10678(4)$ & $0.021519(6)$ & $0.0012430(3)$ \\
\hline
\end{tabular}\\[0cm] 
\caption{\label{tab:app_ww_1hsm}
Cross sections for $gg\ (\to \{h_1,h_2\}) \to W^-W^+ \!\to \ell\bar{\nu}\,\bar{\ell}^\prime \nu^\prime$ in $pp$ collisions at $\sqrt{s}=13$~TeV in the 1-Higgs-Singlet Extension of the SM with $M_{h_1} = 125$~GeV, $M_{h_2} = 700, 1000, 1500, 3000$~GeV and mixing angles $\theta_1$ and $\theta_2$ (see \rtab{tab:benchmark}).  Other details as in \rtab{tab:app_ww_sm}.
} 
\end{table}
%
%
%
\begin{table}[tbp]
\vspace{0.cm}
\centering
\renewcommand{\arraystretch}{1.2}
\begin{tabular}{|c|c|}
\hline
\multicolumn{2}{|c|}{$gg\ (\to H) \to t\bar{t} \to b\bar{b}\,\ell\bar{\nu}\,\bar{\ell}^\prime \nu^\prime$} \\ 
\multicolumn{2}{|c|}{SM, $pp$, $\sqrt{s}=13$ TeV} \\ 
\hline
$|\mathcal{M}|^2$ & $\sigma$ [fb] \\ 
\hline
Sq(C) & $2535.5(6)$ \\
Sq($H$) & $0.13367(4)$ \\
I($H$,C) & $-5.117(2)$ \\
I($H$,C$_\circlearrowleft$) & $15.967(5)$ \\
\hline
\end{tabular}\\[0cm] 
\caption{\label{tab:app_tt_sm}
Cross sections for $gg\ (\to H) \to t\bar{t} \to b\bar{b}\,\ell\bar{\nu}\,\bar{\ell}^\prime \nu^\prime$ in $pp$ collisions at $\sqrt{s}=13$~TeV in the Standard Model.  Other details as in \rtab{tab:app_ww_sm}.
} 
\end{table}
%
%
%
\begin{table}[tbp]
\vspace{0.cm}
\centering

\renewcommand{\arraystretch}{1.2}
\begin{tabular}{|c|c|c|c|c|c|}
\hline
\multicolumn{6}{|c|}{$gg\ (\to \{h_1,h_2\}) \to t\bar{t} \to b\bar{b}\,\ell\bar{\nu}\,\bar{\ell}^\prime \nu^\prime$} \\ 
\multicolumn{6}{|c|}{$\sigma$ [fb], $pp$, $\sqrt{s}=13$ TeV} \\ 
\multicolumn{6}{|c|}{1HSM (see \rtabs{tab:benchmark} and \ref{tab:abbrev})} \\ 
\hline
\multirow{2}{*}{$|\mathcal{M}|^2$} & \multirow{2}{*}{$\theta$} & \multicolumn{4}{|c|}{$ M_{h_2}$  [GeV]} \\ \cline{3-6}
 &  & 700 & 1000 & 1500 & 3000 \\
\hline
\multirow{2}{*}{Sq(C)} & $\theta_1$ & $2535.2(6)$ & $2535.2(6)$ & $2535.2(6)$ & $2535.2(6)$ \\
 & $\theta_2$ & $2535.2(6)$ & $2535.2(6)$ & $2535.2(6)$ & $2535.2(6)$ \\
\hline
\multirow{2}{*}{Sq($h_1$)} & $\theta_1$ & $0.12228(4)$ & $0.12228(4)$ & $0.12827(4)$ & $0.13233(4)$ \\
 & $\theta_2$ & $0.09734(3)$ & $0.09734(3)$ & $0.11631(3)$ & $0.12911(4)$ \\
\hline
\multirow{2}{*}{Sq($h_2$)} & $\theta_1$ & $0.015207(4)$ & $0.0012148(4)$ & $1.2910(4)\times 10^{-5}$ & $7.858(3)\times 10^{-9}$ \\
 & $\theta_2$ & $0.05395(2)$ & $0.004151(2)$ & $5.566(2)\times 10^{-5}$ & $8.503(3)\times 10^{-8}$ \\
\hline
\multirow{2}{*}{I($h_1$,$h_2$)} & $\theta_1$ & $-0.009140(9)$ & $-0.005293(3)$ & $-0.0009301(3)$ & $-4.656(2)\times 10^{-5}$ \\
 & $\theta_2$ & $-0.02553(2)$ & $-0.014530(5)$ & $-0.0027239(8)$ & $-0.00015904(5)$ \\
\hline
\multirow{2}{*}{I($h_1$,C)} & $\theta_1$ & $-4.893(2)$ & $-4.893(2)$ & $-5.011(2)$ & $-5.090(2)$ \\
 & $\theta_2$ & $-4.365(2)$ & $-4.365(2)$ & $-4.772(2)$ & $-5.027(2)$ \\
\hline
\multirow{2}{*}{I($h_2$,C)} & $\theta_1$ & $-0.01350(2)$ & $0.03602(3)$ & $0.009967(4)$ & $0.0006248(2)$ \\
 & $\theta_2$ & $-0.07772(8)$ & $0.08367(9)$ & $0.02335(1)$ & $0.0019221(6)$ \\
\hline
\multirow{2}{*}{I($h_1$,C$_\circlearrowleft$)} & $\theta_1$ & $15.277(5)$ & $15.277(5)$ & $15.644(5)$ & $15.890(5)$ \\
 & $\theta_2$ & $13.629(4)$ & $13.629(4)$ & $14.898(5)$ & $15.695(5)$ \\
\hline
\multirow{2}{*}{I($h_2$,C$_\circlearrowleft$)} & $\theta_1$ & $0.7040(6)$ & $0.06697(7)$ & $-0.01183(2)$ & $-0.0013431(7)$ \\
 & $\theta_2$ & $2.485(2)$ & $0.4122(4)$ & $0.01486(2)$ & $-0.003009(3)$ \\
\hline
\end{tabular}\\[0cm] 
\caption{\label{tab:app_tt_1hsm}
Cross sections for $gg\ (\to \{h_1,h_2\}) \to t\bar{t} \to b\bar{b}\,\ell\bar{\nu}\,\bar{\ell}^\prime \nu^\prime$ in $pp$ collisions at $\sqrt{s}=13$~TeV in the 1HSM.  Other details as in \rtabs{tab:app_ww_sm} and \ref{tab:app_ww_1hsm}.
} 
\end{table}


\section{\boldmath Standard Model distributions\label{app:sm_figs}}

Invariant mass and transverse invariant mass distributions for $WW$ and $t\bar{t}$ production in the SM are displayed in \rfigs{fig:ww_sm_mass_low}--\ref{fig:tt_sm_mass_trans1_low}.

\begin{figure}[tbp]
\vspace{0.cm}
\centering
\includegraphics[width=\textwidth, clip=true]{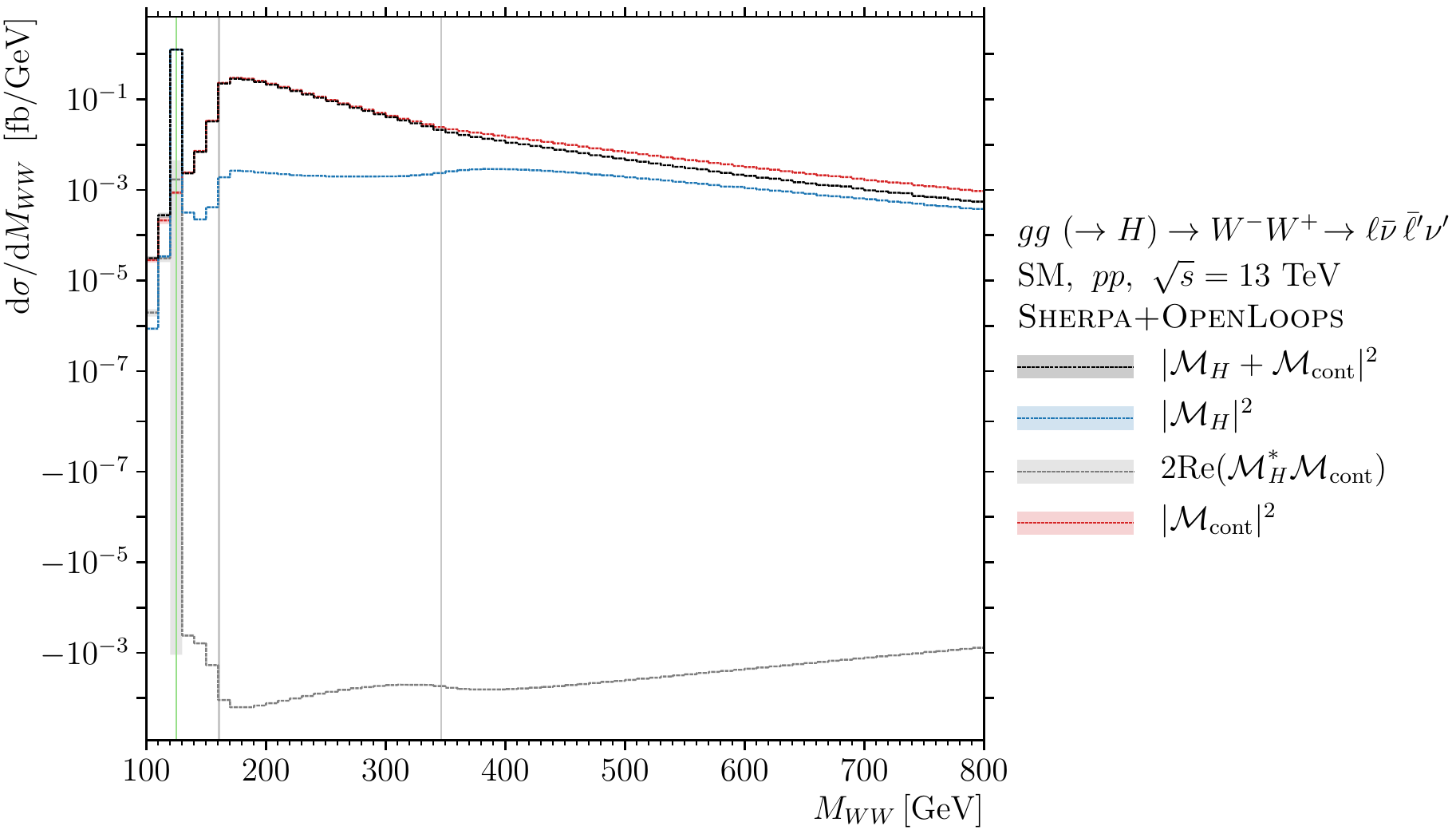}
\caption{\label{fig:ww_sm_mass_low} $M_{WW}$ distributions for the process $gg\ (\to H) \to W^-W^+ \!\to \ell\bar{\nu}\,\bar{\ell}^\prime \nu^\prime$ in the SM including its interference with the background in $pp$ collisions at $\sqrt{s}=13$~TeV.  Other details as in \rfig{fig:ww_1hsm3_mass}.}
\end{figure}
%
%
%
\begin{figure}[tbp]
\vspace{0.cm}
\centering
\includegraphics[width=\textwidth, clip=true]{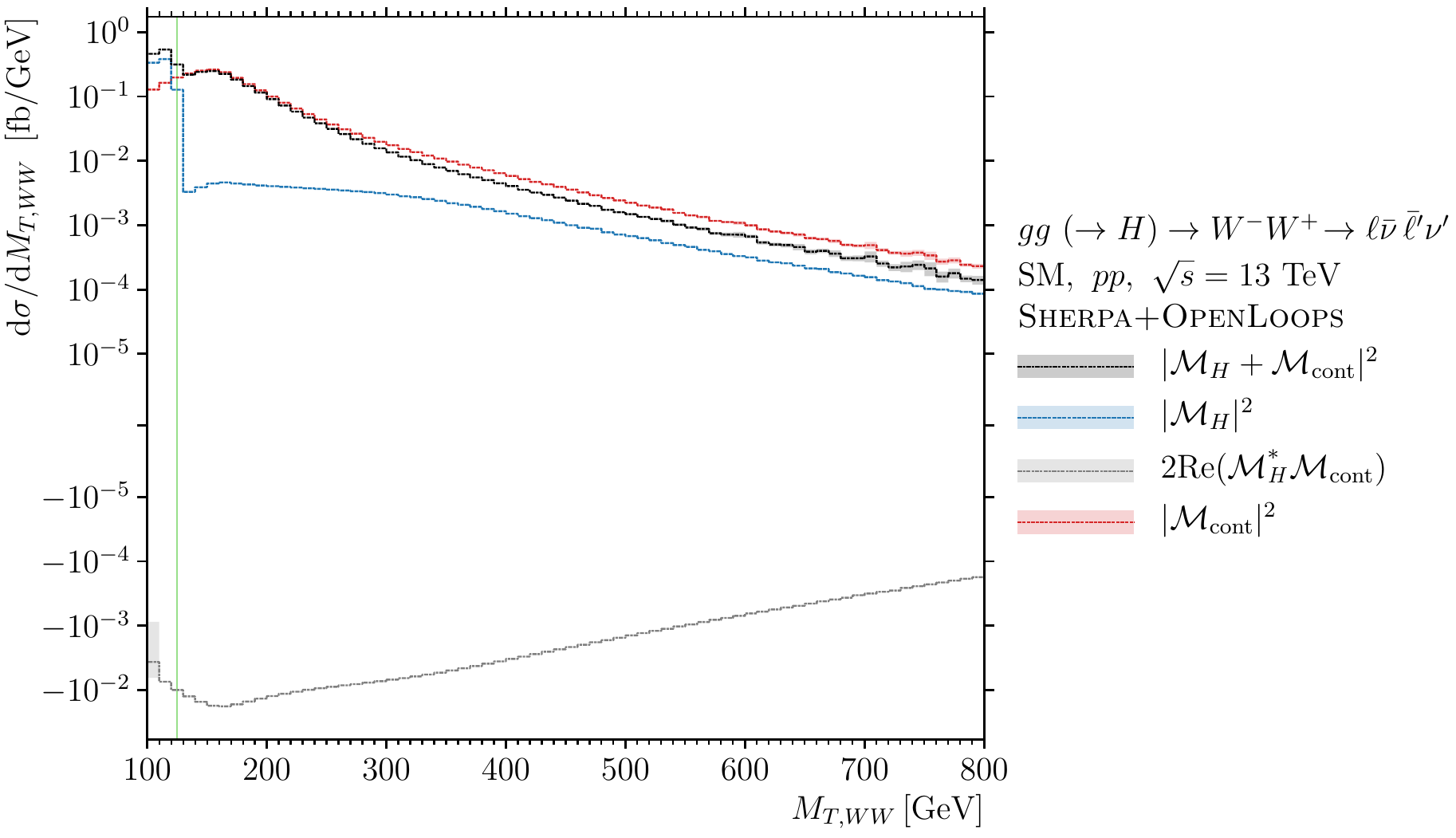}
\caption{\label{fig:ww_sm_mass_trans1_low} $M_{T,WW}$ distributions for the process $gg\ (\to H) \to W^-W^+ \!\to \ell\bar{\nu}\,\bar{\ell}^\prime \nu^\prime$ in the SM including its interference with the background in $pp$ collisions at $\sqrt{s}=13$~TeV.  Other details as in \rfig{fig:ww_1hsm3_mass}.}
\end{figure}
%
%
%
\begin{figure}[tbp]
\vspace{0.cm}
\centering
\includegraphics[width=\textwidth, clip=true]{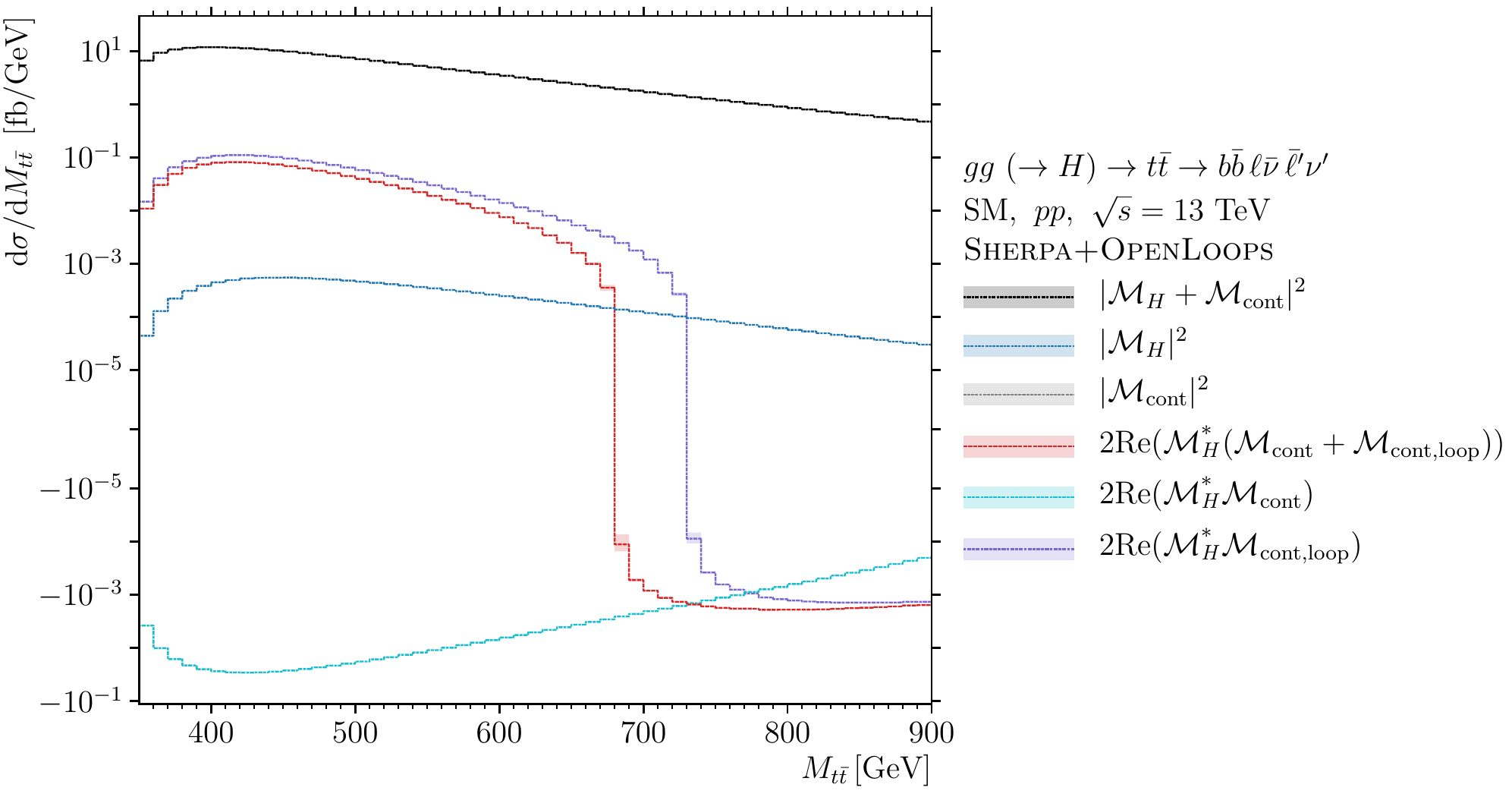}
\caption{\label{fig:tt_sm_mass_low} $M_{t\bar{t}}$ distributions for the process $gg\ (\to H) \to t\bar{t} \to b\bar{b}\,\ell\bar{\nu}\,\bar{\ell}^\prime \nu^\prime$ in the SM including its interference with the background in $pp$ collisions at $\sqrt{s}=13$~TeV.  Other details as in \rfig{fig:ww_1hsm3_mass}.}
\end{figure}
%
%
%
\begin{figure}[tbp]
\vspace{0.cm}
\centering
\includegraphics[width=\textwidth, clip=true]{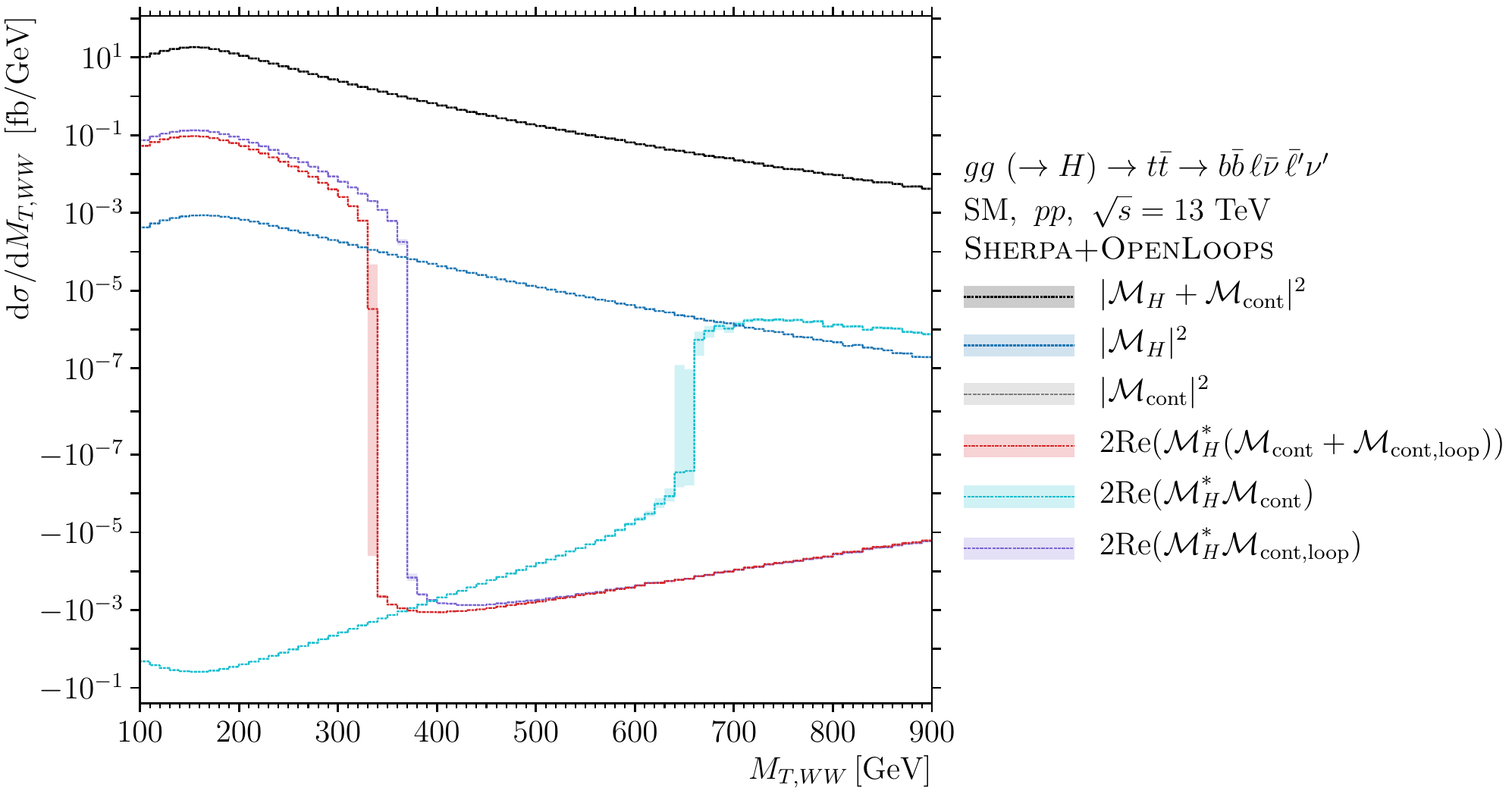}
\caption{\label{fig:tt_sm_mass_trans1_low} $M_{T,WW}$ distributions for the process $gg\ (\to H) \to t\bar{t} \to b\bar{b}\,\ell\bar{\nu}\,\bar{\ell}^\prime \nu^\prime$ in the SM including its interference with the background in $pp$ collisions at $\sqrt{s}=13$~TeV.  Other details as in \rfig{fig:ww_1hsm3_mass}.}
\end{figure}


\clearpage

\bibliography{kauer}
\bibliographystyle{JHEP}



\end{document}